\documentclass[aps,twocolumn,english,superscriptaddress,floatfix,a4paper]{revtex4}
\usepackage{placeins}

\usepackage{graphicx}
\usepackage{dcolumn}
\usepackage{bm}
\usepackage{amsmath} 

\usepackage{pifont}
\usepackage{siunitx}
\usepackage{tabularx}
\usepackage{booktabs}
\usepackage{verbatim}
\usepackage{placeins}
\usepackage{float}
\usepackage[utf8]{inputenc}  

\usepackage[cmyk,dvipsnames]{xcolor}
\usepackage{array} 

\definecolor{zn_color}{rgb}{0.494,0.498,0.733}
\definecolor{mg_color}{rgb}{0.890,0.533,0.286}

\usepackage{float}
\usepackage{hyperref}
\hypersetup{
colorlinks = true,
urlcolor   = blue,
linkcolor  = blue,
citecolor  = blue
}

\usepackage[cmyk,dvipsnames]{xcolor}
\definecolor{pacificb}{HTML}{1CA9C9}
\definecolor{magenta}{HTML}{FF00FF}
\definecolor{goldbrown}{HTML}{994C00}
\definecolor{darkblue}{HTML}{0000CC}
\definecolor{ocre}{HTML}{923232}
\definecolor{purple}{HTML}{4B0082}
\definecolor{indigo}{HTML}{1500FF}
\definecolor{darkteal}{HTML}{0060CC}

\newcommand{\SeRC}{SeRC (Swedish e-Science Research Center), KTH Royal Institute of Technology, SE-10044 Stockholm, Sweden}

\newcommand{\WISEKTH}{Wallenberg Initiative Materials Science for Sustainability (WISE), KTH Royal Institute of Technology, SE-10044 Stockholm, Sweden}

\definecolor{GanapathyPurple}{HTML}{7030A0}

\begin{document}

\title{Evidence for crystal-field-selected octupolar order in wolframite oxides}

\author{Ugne~Miniotaite}
 \email{ugnem@kth.se}
\affiliation{Department of Applied Physics, KTH Royal Institute of Technology, SE-106 91 Stockholm, Sweden}
\author{Leonid V. Pourovskii }
\affiliation{CPHT, CNRS, \'Ecole polytechnique, Institut Polytechnique de Paris, 91120 Palaiseau, France}
\affiliation{Coll\`ege de France, Universit\'e PSL, 11 place Marcelin Berthelot, 75005 Paris, France}
\author{Ola~K.~Forslund}
\affiliation{Physik-Institut, Universit\textnormal{\"a}t Z\textnormal{\"u}rich, Z\textnormal{\"u}rich, Switzerland}
\affiliation{Department of Physics and Astronomy, Uppsala University, Uppsala, Sweden }
\author{Elisabetta~Nocerino}
\affiliation{Department of Chemistry, Stockholm University, SE-106 91 Stockholm, Sweden}
\affiliation{PSI Center for Neutron and Muon Sciences, Paul Scherrer Institute, CH-5232 Villigen PSI, Switzerland}
\author{Yuqing~Ge}
\author{Frank~Elson}
\affiliation{Department of Applied Physics, KTH Royal Institute of Technology, SE-106 91 Stockholm, Sweden}
\author{Michael~Sannemo}
\affiliation{Department of Materials and Environmental Chemistry, Stockholm University, Stockholm, Sweden}
\author{Hiroya~Sakurai}
\affiliation{National Institute for Materials Science (NIMS), Namiki Tsukuba Ibaraki, Japan}
\author{Jun~Sugiyama}
\affiliation{Neutron Science and Technology Center, Comprehensive Research Organization for Science and Society (CROSS),
Tokai, Japan}
\author{Hubertus~Luetkens}
\affiliation{PSI Center for Neutron and Muon Sciences CNM, Villigen PSI, Switzerland}
\author{Toru~Ishigaki}
\affiliation{Neutron Industrial Application Promotion Center, Comprehensive Research Organization for Science and Society (CROSS), Tokai, Japan}
\author{Rasmus~Palm}
\affiliation{Institute of Chemistry, University of Tartu, Tartu,  Estonia}
\author{Chennan Wang}
\affiliation{PSI Center for Neutron and Muon Sciences CNM, Villigen PSI, Switzerland}
\author{David~W.~Tam}
\affiliation{Department of Applied Physics, KTH Royal Institute of Technology, SE-106 91 Stockholm, Sweden}
\author{
Sonali S. Pradhan}
\affiliation{Department of Physics, Indian Institute of Technology Hyderabad, Kandi, Sangareddy 502285,
Telangana, India.}
\author{
V. Kanchana}
\affiliation{Department of Physics, Indian Institute of Technology Hyderabad, Kandi, Sangareddy 502285,
Telangana, India.}
\author{Ganapathy Vaitheeswaran}
\affiliation{School of Physics, University of Hyderabad, Prof. C. R. Rao Road, Gachibowli,
Hyderabad-500046, Telangana, India.}
\author{Anna Delin}
\affiliation{Department of Applied Physics, KTH Royal Institute of Technology, SE-106 91 Stockholm, Sweden}
    \affiliation{\SeRC}
    \affiliation{\WISEKTH}
\author{Martin~M{\aa}nsson}
\email{condmat@kth.se}
\author{Yasmine Sassa}
 \email{sassa@kth.se}
\affiliation{Department of Applied Physics, KTH Royal Institute of Technology, SE-106 91 Stockholm, Sweden}

\date{\today}

{\bf \noindent
\begin{abstract}
{\bf \noindent
Magnetic order can involve higher-rank moments such as octupoles, which are difficult to identify in strongly distorted crystals. Here we present evidence for predominantly octupolar order in the monoclinic wolframites MgReO$_4$ and ZnReO$_4$. Muon spin spectroscopy detects magnetic transitions near 83 and 126~K, respectively, with similar precession frequencies in both compounds. An ab initio effective Hamiltonian shows that strong crystal fields split the spin--orbit quartet, isolating low-energy doublets with weak dipolar character. Exchange selects octupole-dominated order in distinct collinear antiferromagnetic structures, with moments along $b$ in MgReO$_4$ and tilted within the $ac$ plane in ZnReO$_4$. The estimated dipole moments of 0.06 and 0.12~$\mu_{\mathrm{B}}$ per Re, respectively, are consistent with the absence of a resolved magnetic contribution in neutron powder diffraction. A single-ion model illustrates how octupolar states generate local fields with a vanishing net dipole moment. These findings support a crystal-field route to hidden magnetism beyond double perovskites.}
\end{abstract}
}

\maketitle

\section*{Introduction}
The alignment of dipole moments does not always fully describe the magnetic order in a material. In Re$^{6+}$ ($5d^1$) oxides, identifying the ordered state is particularly challenging because spin--orbit coupling (SOC) both enables higher-rank magnetic multipoles, such as octupoles, and can suppress the physical dipole through spin--orbital compensation~\cite{Chen2010,FioreMosca2025}. A small magnetic moment therefore does not, by itself, distinguish compensated dipolar order from a predominantly octupolar state. Higher-order spin and orbital multipoles can profoundly influence material properties even when conventional dipolar moments remain small or unchanged, making identification of the complete symmetry-breaking order parameter essential. In the ideal octahedral limit, SOC acts within the $t_{2g}$ orbital manifold to produce a low-energy $J_{\mathrm{eff}}=3/2$ quartet supporting dipolar, quadrupolar and octupolar degrees of freedom~\cite{Chen2010}. The problem is to determine which of these degrees of freedom actually orders in a real crystal.

Microscopic models of $4d^1$ and $5d^1$ double perovskites predicted octupole-dominated antiferromagnetism already in 2010~\cite{Chen2010}. Subsequent experiments on Ba$_2$MgReO$_6$ established coupled structural and magnetic transitions, followed by direct spectroscopic identification of charge-quadrupolar order~\cite{Hirai2020,Soh2024}. Spectroscopic and first-principles studies further revealed the importance of coupling between electronic and lattice degrees of freedom in selecting the observed states~\cite{Frontini2024,FioreMosca2024,Martinelli2026}. However, charge quadrupoles are even under time reversal, whereas magnetic octupoles are odd: identifying the former does not establish the latter. The absence of resolved magnetic reflections in a recent neutron diffraction study of Ba$_2$MgReO$_6$ has also motivated a proposal of octupolar magnetism~\cite{Barbosa2024}. These developments underline the need to distinguish experimentally established symmetry breaking from the microscopic multipolar interpretation of the ordered state.

A second challenge is to determine how the magnetism is affected by the local environment deviating from cubic symmetry. Distortions split and mix the spin--orbit states and modify the exchange interactions, producing different magnetic behaviour even among compounds with the same $5d^1$ configuration~\cite{Barbosa2022}. When the quartet splits into doublets, octupolar degrees of freedom can remain active within the lowest doublet. Its two states can be represented by an effective spin one-half, while their dipolar and octupolar character depends on the underlying crystal-field wavefunctions. Building on the crystal-field limit considered by Chen \textit{et al.}~\cite{Chen2010}, an ab initio effective Hamiltonian for tetragonal Sr$_2$MgReO$_6$ predicts antiferro-octupolar order within an isolated low-energy doublet, accompanied by weak physical dipoles~\cite{FioreMosca2025}. This provides a microscopic interpretation of its experimentally established antiferromagnetic phase~\cite{Gao2020}. Lower symmetry can further mix dipolar and octupolar components, making their relative weights a material-specific question. These findings motivate asking whether predominantly octupolar order can also emerge in lower-symmetry structures, particularly strongly distorted monoclinic oxides outside the double-perovskite family. Answering this question requires a material-specific analysis of the crystal-field wavefunctions and exchange couplings, since crystal symmetry alone does not determine the character of the magnetic order.

The $5d^1$ wolframite-type rhenium oxides MgReO$_4$ and ZnReO$_4$ provide a direct test of this question. First reported by Sleight~\cite{Sleight1975}, they crystallize in the monoclinic space group $P2/c$ and contain strongly distorted, edge-sharing ReO$_6$ octahedra, providing a different local environment and Re--Re exchange paths from double perovskites. Their closed-shell Mg$^{2+}$ and Zn$^{2+}$ cations avoid the additional magnetic sublattice present in MnReO$_4$~\cite{Bramnik2005}, allowing the Re $5d^1$ magnetism to be investigated without a second magnetic cation contribution. Earlier muon spin spectroscopy established magnetic ordering in MgReO$_4$ but did not determine its microscopic order parameter~\cite{Nocerino2023}. Comparing the two closely related wolframites therefore offers an opportunity to determine how local crystal-field states and intersite interactions together control the character and spatial arrangement of the magnetic order.

Here, we combine neutron powder diffraction, ac magnetic susceptibility and muon spin spectroscopy with muon-site and local-field calculations and first-principles modelling of the electronic and magnetic states. Magnetic transitions near 83 and 126~K in MgReO$_4$ and ZnReO$_4$, respectively, are accompanied by spontaneous local fields without a resolved magnetic diffraction contribution. An ab initio effective Hamiltonian constructed from the refined crystal structures predicts octupole-dominated ground states. Strong crystal fields isolate low-energy doublets whose ordered components have weak dipolar content. The calculated exchange couplings select a collinear antiferromagnetic structure in MgReO$_4$ and a collinear one with a canted axis in the $ac$ plane in ZnReO$_4$, despite similar observed muon precession frequencies. Complementary hybrid-functional calculations including SOC predict semiconducting states and strong spin--orbital compensation. A single-ion model including spin magnetization density
and orbital currents illustrates how an octupolar state can generate a local magnetic field even when its integrated magnetic dipole moment vanishes. Together, these results provide evidence for crystal-field-selected octupolar order beyond double perovskites and identify the microscopic content of the low-energy states as a guide to understanding weak-moment magnetism in low-symmetry materials.

\begin{figure*}
\centering
\includegraphics[width=\textwidth]{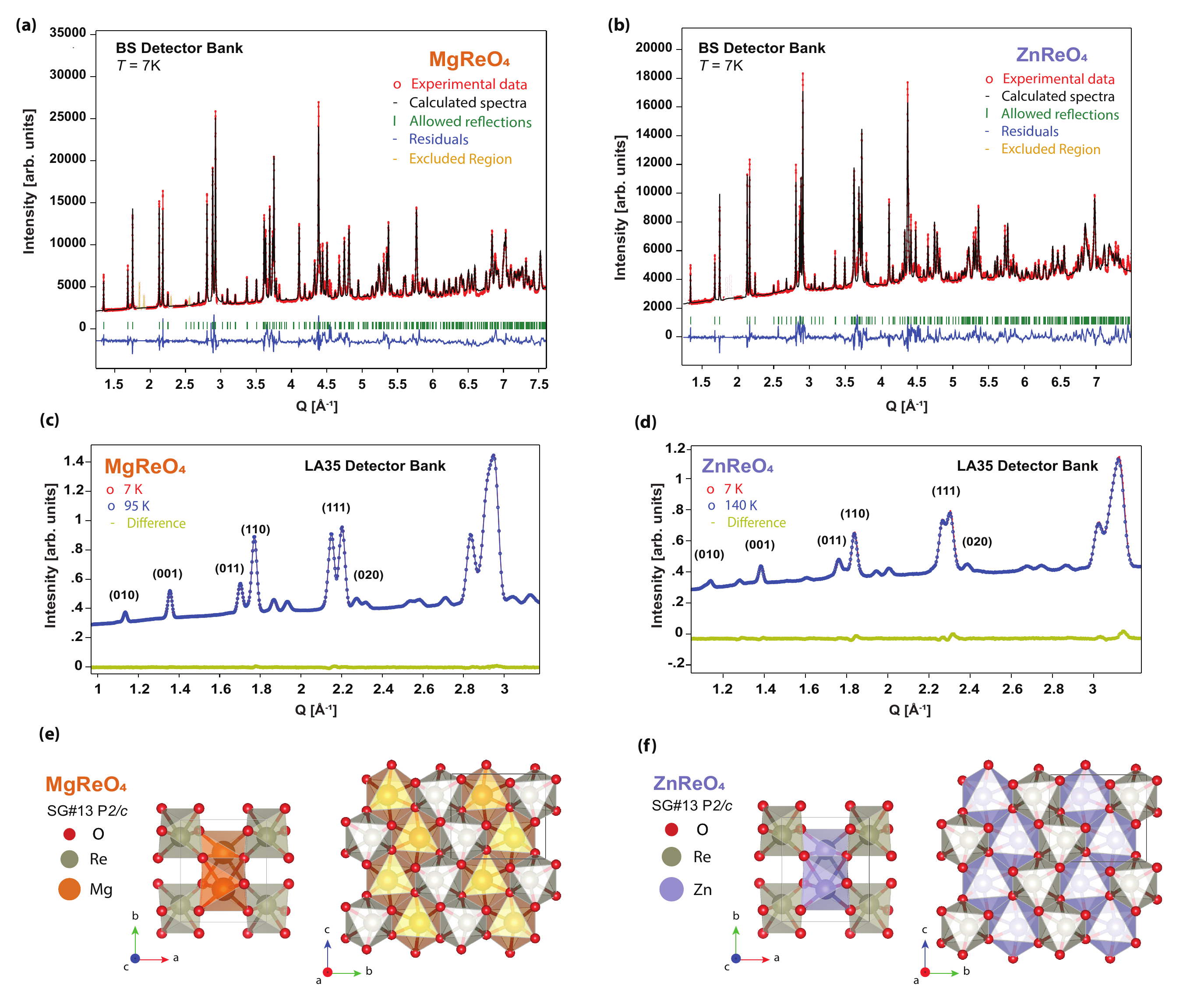}
\caption{\textbf{Distorted wolframite structures and magnetic diffraction constraints.}
\textbf{a,b}, Neutron powder diffraction patterns of MgReO$_4$ and ZnReO$_4$, respectively, measured at $7$~K using the backscattering (BS) detector bank of the iMATERIA time-of-flight diffractometer at J-PARC. Excluded regions contain unindexed reflections.\textbf{c,d}, Low-$Q$ patterns measured using the
low-angle (LA35) detector bank at $7$~K and above the magnetic transitions: $95$~K for MgReO$_4$ and $140$~K for ZnReO$_4$. The difference of the patterns show no additional magnetic contribution. \textbf{e,f}, Two views of the refined monoclinic $P2/c$ structures of MgReO$_4$ and ZnReO$_4$, showing the distorted, edge-sharing ReO$_6$ octahedra.}
\label{fig:npd}
\end{figure*}

\section*{Results}
\subsection*{Distorted wolframite structures and magnetic diffraction constraints}
\label{sec:structure}
We first determine the low-temperature crystal structures that define the local Re environment and provide the input for the magnetic calculations. The 7~K neutron powder diffraction patterns of both compounds are described by the monoclinic wolframite structure, space group $P2/c$ (Fig.~\ref{fig:npd}). Edge-sharing ReO$_6$ octahedra form zigzag chains along $c$, separated by Mg or Zn sites. The local coordination is strongly noncubic in both materials: the three pairs of Re--O distances span approximately 1.79--2.08~\AA, and the O--Re--O angles in the ReO$_6$ octahedra span approximately $75$--$104^{\circ}$. The similar crystal distortion contrast with the different magnetic ordering temperatures discussed below. Full lattice parameters, atomic coordinates and local coordination are given in Supplementary Section~S2.

Comparing the $7$~K patterns with those measured above the magnetic transitions, at $95$~K for MgReO$_4$ and
$140$~K for ZnReO$_4$, reveals no resolved change of crystallographic symmetry or additional magnetic diffraction contribution. Simulations of the collinear $\mathbf{k}=(1/2,1/2,0)$ MgReO$_4$ structure predicted by first-principles calculations show that an ordered moment of approximately $0.3\,\mu_{\mathrm B}$ per Re would produce a detectable magnetic reflection (Supplementary Section~S3). For the predicted
$\mathbf{k}=(0,0,0)$ ZnReO$_4$ structure, magnetic reflections can overlap nuclear reflections, making a weak magnetic contribution more difficult to detect.

Unindexed peaks were excluded from the nuclear refinements. These peaks are also present above the magnetic ordering temperatures. Tests of lower-symmetry structural models could not account for them, so we attribute them to an impurity phase. Refinement details are given in Supplementary Section~S2, and sample-handling assessments are presented in Supplementary Sections~S1 and S4.

\subsection*{Magnetic transitions with spontaneous local fields}
\label{sec:muon}
\begin{figure*}
\centering
\includegraphics[width=\textwidth]{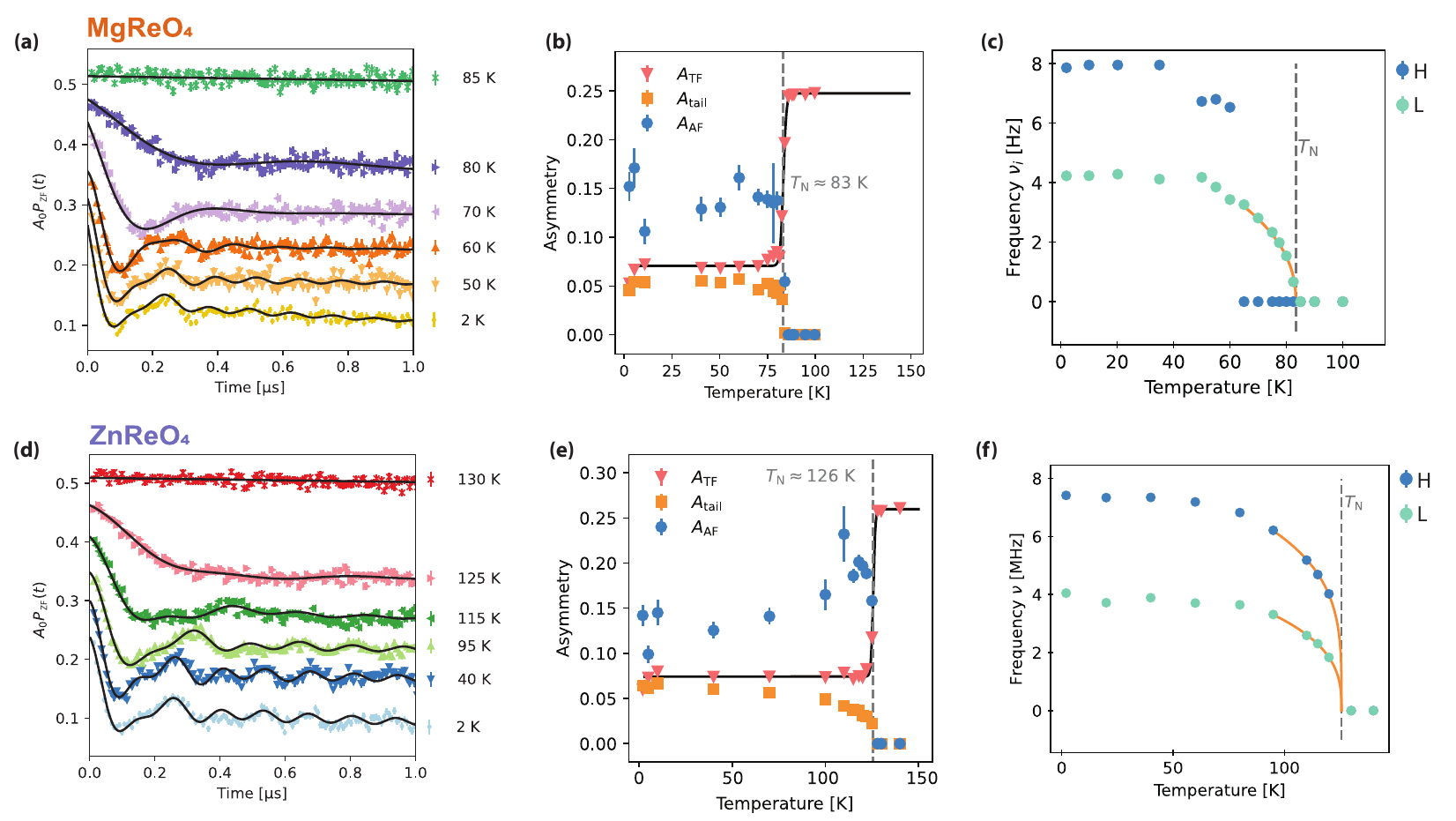}
\caption{\textbf{Spontaneous local fields and magnetic transition temperatures.}
\textbf{a,d}, Representative zero-field muon spectra for MgReO$_4$ and ZnReO$_4$ at a set of temperatures; black lines show the fitted polarization function. Spectra are vertically offset for clarity. \textbf{b,e}, Fitted amplitudes in a weak transverse field of 5~mT for MgReO$_4$ and ZnReO$_4$: the applied-field component $A_{\mathrm{TF}}$, internal-field component $A_{\mathrm{AF}}$ and slowly relaxing tail $A_{\mathrm{tail}}$. The black sigmoid fits to $A_{\mathrm{TF}}$ give $T_{\mathrm N}=83.35(2)$ and $125.67(10)$~K, respectively; vertical dashed lines mark the magnetic ordering N\'eel temperatures $T_\mathrm{N}$. Error bars show fit uncertainties.\textbf{c,f}, Temperature dependence of the frequency components, corresponding to the high and low frequency components H and L. Orange curves show power-law fits to the L component of MgReO$_4$ and to both H and L components of ZnReO$_4$. The fitted exponents are $\beta_{\rm L}^{\rm Mg}=0.440(25)$, $\beta_{\rm H}^{\rm Zn}=0.260(5)$ and $\beta_{\rm L}^{\rm Zn}=0.349(8)$. The zero-valued points in the frequency panels denote unresolved fit components not measured zero local fields. Full muon spin spectroscopy fit and analysis are given in Supplementary Sections~S6--S7, and details of the power-law fits in Section~S8.}
\label{fig:muon}
\end{figure*}

Muon spin spectroscopy establishes magnetic transitions despite the absence of a resolved magnetic diffraction contribution. At low temperature, both compounds exhibit two zero-field oscillations (Fig.~\ref{fig:muon}~(a)), corresponding to quasistatic internal magnetic fields. We denote the higher- and lower-frequency components by H and L, respectively. Their frequencies at base temperature are $7.86(7)$ and $4.229(9)$~MHz in MgReO$_4$, and $7.42(3)$ and $4.05(5)$~MHz in ZnReO$_4$. Using $f=\gamma_{\mu}|\mathbf{B}|/(2\pi)$, these correspond to approximately 58.0 and 31.2~mT, and 54.7 and 29.9~mT, respectively. Thus, the two compounds have closely similar local-field scales. The MgReO$_4$ spectra are reanalysed from the previously reported data set \cite{Nocerino2023}.

The temperature dependence of the weak-transverse-field asymmetry, measured in 5~mT, gives transition temperatures of $83.35(2)$~K for MgReO$_4$ and $125.67(10)$~K for ZnReO$_4$ (Fig.~\ref{fig:muon}~(b)). The uncertainties are statistical errors from the sigmoid fits. A weak anomaly in the real AC susceptibility of ZnReO$_4$ near 126~K corroborates the transition, although a substantial low-temperature upturn prevents extraction of an intrinsic moment from this data set (Supplementary Section~S1). Approximately 20\% of the high-temperature weak-field asymmetry remains at low temperature. This residual is treated as a non-ordering contribution, with sample background and secondary phases as possible origins, rather than as an independently measured impurity mass fraction.

Above the transitions, the weak relaxation is substantially decoupled by a $2$~mT longitudinal field, consistent with predominantly quasistatic nuclear fields (Supplementary Section~S6). The L component has the larger asymmetry and stronger damping in both compounds. The temperature evolution of the oscillating components differs between the compounds: in MgReO$_4$, the weaker H component becomes unresolved above approximately $60$~K (see Fig.~\ref{fig:muon}~(c), whereas both components remain resolvable throughout the ordered phase of ZnReO$_4$. Full spectra, fitting functions and temperature-dependent fit parameters are given in Supplementary Sections~S7--S8.

\subsection*{Crystal-field isolation of the low-energy doublets}
\label{sec:electronic}
\begin{figure*}
\centering
\includegraphics[width=\textwidth]{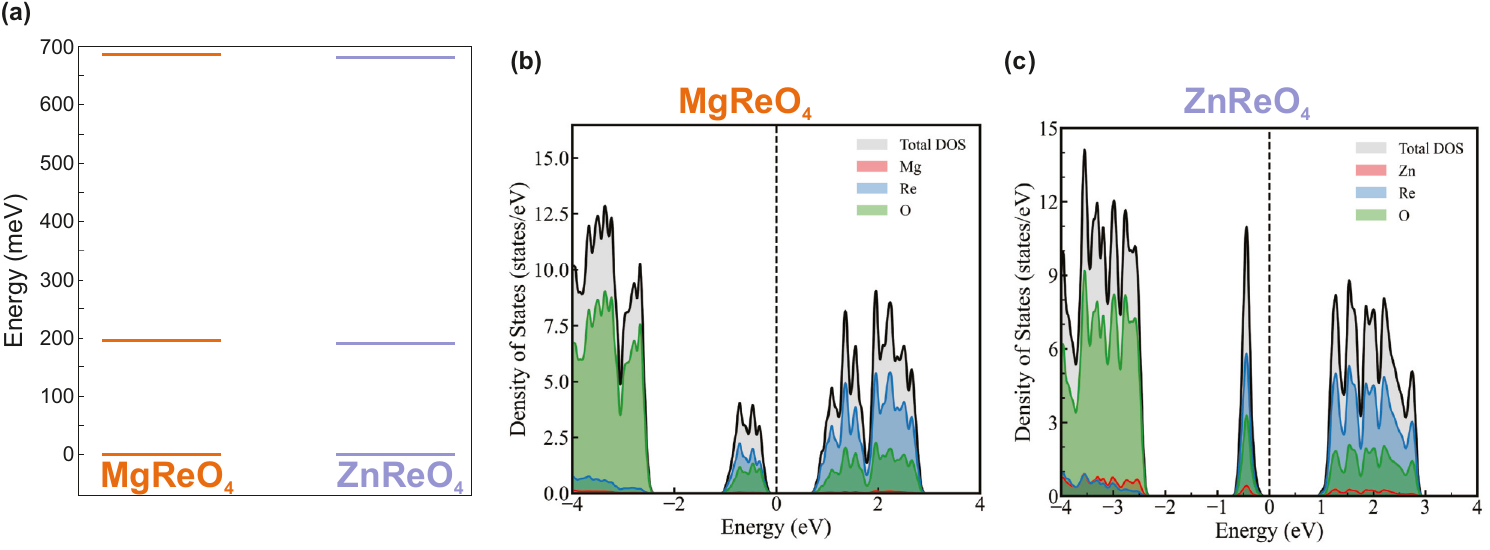}
\caption{\textbf{An isolated crystal-field doublet and a gapped calculated electronic structure.}
\textbf{a}, Spin--orbit and crystal-field levels obtained from the DFT+Hubbard-I description of MgReO$_4$ and ZnReO$_4$. The ground-state energy is set to zero for each compound. The two $J=3/2$ doublets are separated by 196 and 190~meV, respectively; the $J=1/2$ doublet lies approximately 680~meV above the ground doublet. These calculated levels provide the local-state basis of the effective Hamiltonian.
\textbf{b,c}, HSE06+SOC densities of states calculated for the relaxed MgReO$_4$ and ZnReO$_4$ structures in two-Re crystallographic cells, respectively. The dashed lines mark zero energy. The reported gaps are approximately 0.99 and 1.22~eV. The HSE magnetic solutions and the paramagnetic DFT+Hubbard-I construction are distinct calculations. Full spectra and computational details are given in Supplementary Sections~S11--S13.}
\label{fig:electronic}
\end{figure*}

To identify the local degrees of freedom underlying the observed magnetic transitions, we calculate the spin--orbit and crystal-field states of Re from the neutron-refined $7$~K structures using charge-self-consistent density functional theory combined with the Hubbard-I approximation (DFT+Hubbard-I). The $J=3/2$ quartet splits into two Kramers doublets separated by 196~meV in MgReO$_4$ and 190~meV in ZnReO$_4$ (Fig.~\ref{fig:electronic}). These separations greatly exceed the thermal energies at the transitions, $k_{\mathrm B}T_{\mathrm N}\simeq7.2$ and 10.8~meV. The lowest doublet therefore supplies the thermally active local states near ordering, while the four-state effective Hamiltonian retains the excited crystal-field doublet. The predominantly $J=1/2$ spin--orbit doublet lies about 680~meV above the ground doublet. Crystal-field mixing with this higher manifold distinguishes the physical states from pure atomic $J=3/2$ eigenstates.

The electronic spectra provide a complementary check on this local-state description. Calculations using the screened Heyd--Scuseria--Ernzerhof hybrid functional (HSE06), with and without SOC, describe magnetic states using structures derived from the experimental $7$~K structures and relaxed with the Perdew--Burke--Ernzerhof (PBE) functional. They use a two-Re crystallographic cell with magnetic propagation vector $\mathbf{k}=(0,0,0)$, chosen based on a candidate structure from the muon dipolar-field analysis described below (Supplementary Section~S10). The HSE06+SOC calculations predict semiconducting gaps of approximately $0.99$~eV and $1.22$~eV for MgReO$_4$ and ZnReO$_4$, respectively (Fig.~\ref{fig:electronic}b,c). HSE06 calculations without SOC are also gapped, whereas the semilocal PBE calculations are metallic. The comparison therefore reflects the treatment of electronic interactions rather than demonstrating that SOC alone opens the gap. The occupied and low-lying unoccupied spectral weight is predominantly O- and Re-derived, respectively; the complete orbital-resolved spectra are given in Supplementary Section~S11.

The DFT+Hubbard-I one-electron spectral functions (Supplementary Section~S12, Fig.~S16) agree closely with the HSE06+SOC densities of states in their relative band positions and gap magnitudes. Some differences are expected because the DFT+Hubbard-I calculation is paramagnetic whereas the HSE06+SOC calculation is magnetically ordered, and the approaches treat correlations and multiplet effects differently.

The HSE06+SOC solutions exhibit opposing site-projected Re spin and orbital contributions, with nearly cancelling local contributions in MgReO$_4$ and partial compensation in ZnReO$_4$. These results establish complementary features of the electronic structure; they are not used to determine the multipolar rank of the ordered state. That identification requires the effective-Hamiltonian and operator analysis below. The full moment table, structural inputs and computational conventions are found in Supplementary Sections~S11--S13.

A separate PBE comparison between relaxed and unrelaxed experimental structures changes the detailed spectral profiles but leaves the electronic structure near the Fermi level qualitatively similar in both compounds (Supplementary Section~S16). HSE and HSE06+SOC were evaluated only for the relaxed structures, so this comparison does not determine the relaxation dependence of the hybrid-functional gaps.

\subsection*{Octupole-dominated order and distinct magnetic structures}
\label{sec:order}
We determine the ordering of these local degrees of freedom from the ab initio effective Hamiltonian
\begin{equation}
H_{\mathrm{eff}}=H_{\mathrm{cf}}+
\sum_{\langle ij\rangle}\sum_{KQK'Q'}
V_{KK'}^{QQ'}(ij)O_{KQ}^{i}O_{K'Q'}^{j},
\label{eq:Heff}
\end{equation}
where $H_{\mathrm{cf}}$ describes the on-site crystal field and $O_{KQ}$ are multipolar operators in the specified $J=3/2$ pseudospin basis. Their ranks $K=1,2,3$ correspond to dipoles, quadrupoles and octupoles, respectively, with  $Q=-K,-K+1,\ldots,K$ labeling their projection. $V_{KK'}^{QQ'}(ij)$ is the intersite exchange coupling between corresponding multipoles on the sites $i$ and $j$.  The exchange couplings  are evaluated by the force theorem in Hubbard-I~\cite{Pourovskii2016} from the paramagnetic electronic structure. The operator conventions and exchange couplings are given in Supplementary Sections~S13 and S14. The calculated intersite exchange interactions depend strongly on the Re–Re bond direction and decay rapidly with increasing bond length. The strongest interaction in each material couples Re sites in edge-sharing octahedra. In MgReO$_4$ its scale is approximately four times that of the second-neighbour interaction, whereas longer-distance interactions are relatively more important in ZnReO$_4$. 

\begin{figure*}
\centering\includegraphics[width=0.9\textwidth]{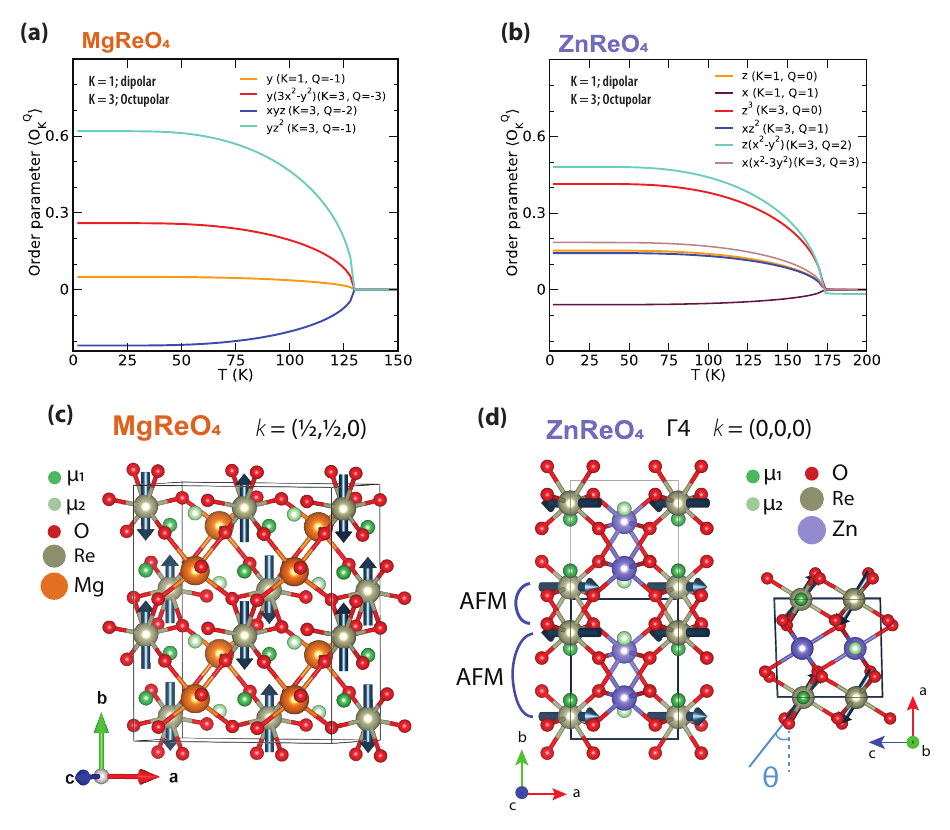}
\caption{\textbf{Octupole-dominated order and predicted magnetic structures.}
\textbf{a,b}, Mean-field expectation values of the leading dipolar ($K=1$) and octupolar ($K=3$) operators for MgReO$_4$ and ZnReO$_4$, respectively. All operators are normalized to $\operatorname{Tr}(O_{KQ}^{2})=1$. Quadrupolar moments induced by the crystal field and additional Zn octupolar components with magnitudes below 0.01 are not shown. \textbf{c,d}, Collinear antiferromagnetic structure types
predicted by the first-principles effective Hamiltonians: $\mathbf{k}=(1/2,1/2,0)$ for MgReO$_4$, with Re dipole moments along $b$, and $\Gamma_4$ with $\mathbf{k}=0$ for ZnReO$_4$, with moments in the $ac$ plane.
In \textbf{d}, $\theta$ denotes the canting angle of the antiferromagnetic axis away from $a$ within the $ac$ plane. $\mu_1$ and $\mu_2$ denote the calculated muon stopping sites of each compound (Supplementary Section~S9).}
\label{fig:order}
\end{figure*}

Solving the effective Hamiltonian (\ref{eq:Heff}) in mean-field we find collinear dipole-octupolar magnetic orders in both compounds. The dipole and octupole temperature-dependent order parameters are shown in Fig.~\ref{fig:order}. Quadrupolar moments are already constrained by the strong crystal field and do not exhibit any anomaly at the transition point. 

With all multipolar operators uniformly normalized to $\operatorname{Tr}(O_{KQ}^{2})=1$, the relative importance of dipolar and octupolar order parameters can  be quantified by $m_1/m_3$, where
\begin{equation}
m_K=\left[\sum_{Q=-K}^{K}\langle O_{KQ}\rangle^2\right]^{1/2}.
\label{eq:rank}
\end{equation}
We find $m_1/m_3=0.07$ for MgReO$_4$ and 0.24 for ZnReO$_4$. The order in MgReO$_4$ is thus almost purely octupolar within the defined $J=3/2$ manifold, whereas ZnReO$_4$ has a larger dipole component but remains octupole dominated. 

This octupolar dominance is not a consequence of dipolar exchange being very weak, since the overall magnitudes of the dipolar and octupolar exchange blocks are comparable (see Supplementary Section S14). Instead, the predominantly octupolar character originates from the crystal-field-constrained low-energy states. When the inter-doublet crystal-field splitting is much larger than the exchange field, as in the present case, the $J=3/2$ order can be faithfully mapped onto the ground-state doublet subspace and described in terms of effective spin-one-half operators. For MgReO$_4$, the calculated $J=3/2$ order maps onto an $\langle s_y\rangle$ order within the lowest doublet. Embedding of the $s_y$ operator into the  $J=3/2$ basis is
\begin{equation}
\begin{split}
\sqrt{2}P^{\dagger}s_yP={}&
-0.08O_{1,-1}-0.38O_{3,-3}\\
&+0.31O_{3,-2}-0.87O_{3,-1},
\end{split}
\label{eq:projection-main}
\end{equation}
where the rows of the $2\times4$ matrix $P$ are the ground-doublet bra states. The four terms on the right are precisely the order parameters found for MgReO$_4$ (Fig.~\ref{fig:order}a).  This projection explains the  hierarchy of order parameters in MgReO$_4$ by  revealing that the relevant spin-one-half operator is mainly octupolar in its microscopic content, with a small dipolar admixture. An analogous analysis applies also to ZnReO$_4$.

Now we turn to the analysis of the predicted dipolar order and the corresponding magnetic dipole moments. The predicted MgReO$_4$ dipolar structure has propagation vector $\mathbf{k}=(1/2,1/2,0)$ with its weak dipole magnetic moment of approximately $0.06\,\mu_{\mathrm B}$ directed along the $b$ (or $y$) axis. For ZnReO$_4$ we instead predict a collinear antiferromagnetic structure described by the $\Gamma_4$ irreducible representation, $\mathbf{k}=(0,0,0)$ structure, with a dipole moment of approximately $0.12\,\mu_{\mathrm B}$ lying in the $ac$ plane (Fig.~\ref{fig:order} and Table~\ref{tab:summary}). 

In the ideal atomic $t_{2g}$, $J=3/2$ limit, spin and orbital contributions to the physical dipole cancel. Covalency and $e_g$ admixture can reduce the orbital contribution~\cite{Ahn2017,Agrestini2024}; crystal-field mixing with the excited spin--orbit manifold also contributes to the physical moment. 
In particular, 5$d$ covalency was theoretically shown to significantly reduce the orbital moment in 5$d$ double perovskites~\cite{Ahn2017}. Our Wannier orbitals constructed from the manifold of mainly 5$d$ Kohn-Sham bands do include the effect of $p-d$ mixing~\cite{Aichhorn2009}, which affects intersite exchange and crystal field. However, calculating the impact of this mixing on the orbital moment is a complex procedure~\cite{Ahn2017} that we do not employ here. Instead, 
to account for the covalency effect we reduce the orbital moment contribution by the orbital-reduction (covalency) factor $\gamma$.  
We find that the moment magnitude in ZnReO$_4$ remains close to $0.12\,\mu_{\mathrm B}$, but its angle with respect to $a$ changes from about $60^{\circ}$ to $10^{\circ}$ as $\gamma$ increases from 0.5 to 0.8. The $30^{\circ}$ direction is obtained for $\gamma=0.67$, a value within the range experimentally found for $\gamma$ in 5$d$ oxides and halides~\cite{Ahn2017,Ishikawa2019,Gao2020}.

Mean-field transition temperatures of 130 and 170~K reproduce the higher ordering temperature of ZnReO$_4$, while overestimating both experimental values. Such an overestimate is consistent with the limitations of single-site mean field and previous applications of the force-theorem Hubbard-I (FT-HI) approach~\cite{Pourovskii2023,FioreMosca2024,FioreMosca2025}. The relatively smaller discrepancy for ZnReO$_4$ may reflect its more substantial longer-distance interactions and higher effective coordination.

\begin{table}[htbp]
\centering\small
\caption{\textbf{Experimental transitions and theoretical magnetic states.}
Experimental transition temperatures are obtained from weak-transverse-field muon spin spectroscopy fits. The remaining quantities are obtained from the first-principles effective-Hamiltonian analysis. The dipole direction in ZnReO$_4$ depends on the orbital-reduction (covalency) factor $\gamma$.}
\label{tab:summary}
\begin{tabular}{lcc}\toprule
Quantity & MgReO$_4$ & ZnReO$_4$\\\midrule
Measured $T_{\mathrm N}$ (K) & 83.35(2) & 125.67(10)\\
Mean-field $T_{\mathrm N}$ (K) & 130 & 170\\
Crystal-field splitting (meV) & 196 & 190\\
Normalized $m_1/m_3$ & 0.07 & 0.24\\
Predicted $\mathbf{k}$ & $(1/2,1/2,0)$ & $(0,0,0)$\\
Dipole magnetic moment ($\mu_{\mathrm B}$/Re) & $\simeq0.06$ & $\simeq0.12$\\
Dipole direction & $b$ & $ac$ plane\\\bottomrule
\end{tabular}
\end{table}

\subsection*{Local fields beyond an integrated dipole}
\label{sec:fields}
Electrostatic-potential calculations identify two candidate muon stopping sites in each compound (Supplementary Section~S9). We used irreducible-representation analysis to examine dipolar magnetic structures with $\mathbf{k}=(0,0,0)$, $(1/2,0,0)$ and $(0,1/2,0)$, and calculated their point-dipole fields at these muon sites. Among the tested structures, $\Gamma_3$ with $\mathbf{k}=(0,1/2,0)$ and $\Gamma_4$ with $\mathbf{k}=0$, both with moments in the $ac$ plane, approximately reproduce the two low-temperature frequencies with moment magnitudes of approximately $0.22$--$0.26\,\mu_{\mathrm B}$ per Re. Several dipolar structures are therefore compatible with the measured frequencies.

We tested whether a rotation of the Re moments could explain why the H oscillation in MgReO$_4$ becomes unresolved near $60$~K. For the $\Gamma_3$ and $\Gamma_4$ candidates, rotations within the $ac$ plane can bring the two calculated site frequencies together. However, these rotations also predict
an increase in the L frequency that is not observed. We can therefore does not reproduce the measured temperature dependence of the muon precession frequencies. The rotation tests are shown in Supplementary Section~S8.

The $\Gamma_4$ candidate is also the magnetic structure predicted for ZnReO$_4$ by the first-principles effective Hamiltonian. However, point-dipole fields calculated using the predicted physical moments are smaller than measured at the candidate muon sites in both compounds. For ZnReO$_4$, matching the frequencies requires approximately $0.22\,\mu_{\mathrm B}$ per Re, compared with the predicted $0.12\,\mu_{\mathrm B}$. For MgReO$_4$, the effective-Hamiltonian $\mathbf{k}=(1/2,1/2,0)$ structure with a moment of $0.06\,\mu_{\mathrm B}$ gives frequencies of approximately
$0.54$ and $1.01$~MHz, compared with the measured $4.229(9)$ and $7.86(7)$~MHz. Matching the observed field requires approximately $0.47\,\mu_{\mathrm B}$ per Re. The representation analysis, calculated fields and fitted moment directions are given in Supplementary Sections~S9 and S10.

\begin{figure}
\centering\includegraphics[width=\columnwidth]{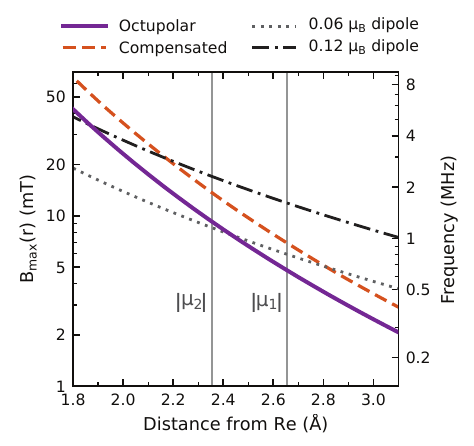}
\caption{\textbf{Finite local fields despite a vanishing integrated dipole.}
Illustrative atomic-field calculation for a Re$^{6+}$ $5d^1$ ion, including spin magnetization density and orbital current for $\langle r\rangle_{5d}=0.65$~\AA.  $B_{\max}(r)=\max_{\hat{\mathbf r}}|\mathbf B(r\hat{\mathbf r})|$ is the maximum single-ion field over directions on a sphere. The ideal octupolar state and the compensated $|3/2,+3/2\rangle$ state at $\gamma=1$ both have zero integrated magnetic dipole. The point-dipole curves use the dipole moments predicted by first-principles calculations: $0.06\,\mu_{\mathrm B}$ for MgReO$_4$ and $0.12\,\mu_{\mathrm B}$ for ZnReO$_4$. The vertical lines labelled $|\mu_2|$ and $|\mu_1|$ mark the calculated Re--muon distances for the candidate stopping sites in MgReO$_4$, $2.35$ and $2.65$~\AA. Model details and spatial field maps are given in Supplementary Section~S15.}
\label{fig:toy}
\end{figure}

To examine whether an octupolar state can generate a local magnetic field at a muon site even when its integrated dipole vanishes, we calculate the field of a single Re$^{6+}$ $5d^1$ ion, including spin magnetization density and orbital currents. Figure~\ref{fig:toy} shows the maximum field over directions as a function of distance from Re for an ideal octupolar state, point dipoles and a compensated reference state, whose integrated spin and orbital magnetic dipoles cancel. The point-dipole curves use the moments predicted by first-principles calculations for MgReO$_4$ and ZnReO$_4$. At the calculated Re--muon distances for the candidate stopping sites (Supplementary Section~S9), the octupolar field is comparable in scale to that of a $0.06\,\mu_{\mathrm B}$ point dipole. The octupolar state and the compensated reference state both produce finite local fields despite having zero integrated magnetic dipole moments. Related orbital-current fields have been considered for $5d^2$ osmates~\cite{Voleti2020}. The single-ion octupolar field remains smaller than
the local fields measured by muon spin spectroscopy. Model details and spatial field maps are given in Supplementary Section~S15, Fig.~S19.

\section*{Discussion}
\label{sec:discussion}
Our results support a crystal-field-selected, predominantly octupolar interpretation of magnetism in MgReO$_4$ and ZnReO$_4$. The central finding is that a spin–orbit-entangled Re$^{6+}$ ion can exhibit higher-rank magnetic degrees of freedom in a strongly distorted monoclinic environment. The crystal field selects the low-energy wavefunctions, which constrain the dipolar content of the ordering variable. Exchange then selects the ordered combination and its spatial arrangement. Within the calculated ordered states, the weak physical dipoles are therefore an incomplete description of what orders: an effective spin one-half captures the two-state low-energy manifold, but does not, by itself, identify the physical quantity that orders.

The distinction between pseudospin order and physical magnetization is essential here. In the specified $J=3/2$ tensor basis, the normalized ratios $m_1/m_3=0.07$ and $0.24$ quantify octupolar dominance in MgReO$_4$ and ZnReO$_4$, respectively. The physical dipole additionally depends on crystal-field mixing with the excited spin--orbit multiplet and on the orbital contribution, which is modified by covalency~\cite{Chen2010,Ahn2017,FioreMosca2025}. Consequently, a small dipolar component can accompany the dominant octupolar order without representing a separate phase. In this low-symmetry setting, the designation ``octupole-dominated'' describes the microscopic composition of the calculated order.

This interpretation extends, rather than replaces, the crystal-field route anticipated in models of $d^1$ double perovskites and identified microscopically for tetragonal Sr$_2$MgReO$_6$~\cite{Chen2010,FioreMosca2025}. The wolframites test that route in a monoclinic, edge-sharing network, where neither cubic degeneracy nor the double-perovskite exchange geometry is retained. Their calculated order is also distinct from the time-even charge-quadrupolar order identified in Ba$_2$MgReO$_6$~\cite{Soh2024}. Together with related work on rare-earth doublets and PrO$_2$~\cite{Huang2014,Khmelevskyi2024}, the present results emphasize that the microscopic operator content of the low-energy states, not their effective-spin label alone, is required to connect an exchange model to magnetic observables.

Comparing MgReO$_4$ and ZnReO$_4$ shows how similar local states can lead to different collective magnetic order. Their similar crystal-field splittings support an isolated-doublet description in both compounds, while their calculated exchange interactions select different magnetic structures. Further-neighbour interactions are more important in ZnReO$_4$, which also has the higher ordering temperature. The first-principles effective Hamiltonians predict collinear antiferromagnetic order in both compounds, with $\mathbf{k}=(1/2,1/2,0)$ and moments along $b$ in MgReO$_4$, and $\mathbf{k}=0$ with moments in the $ac$ plane in ZnReO$_4$. Their similar muon precession frequencies therefore do not necessarily imply similar magnetic structures. Additionally, in ZnReO$_4$, varying the orbital-reduction (covalency) factor $\gamma$ changes the $ac$-plane canting angle, while the dipole magnitude remains nearly unchanged.

The point-dipole calculations identify several magnetic structures that approximately reproduce the measured frequencies at the candidate muon sites when the moment magnitude is adjusted. For ZnReO$_4$, the dipolar-field analysis identifies the same $\Gamma_4$ structure as the effective Hamiltonian, but matching the frequencies requires approximately $0.22\,\mu_{\mathrm B}$ per Re, nearly twice the predicted $0.12\,\mu_{\mathrm B}$. For the MgReO$_4$ $\mathbf{k}=(1/2,1/2,0)$ structure predicted by the effective Hamiltonian, the required moment is approximately $0.47\,\mu_{\mathrm B}$ per Re, almost eight times the predicted $0.06\,\mu_{\mathrm B}$. Thus, the predicted dipole moments alone cannot reproduce the measured frequencies at the candidate muon sites.

The atomic spin-density and orbital-current model clarifies why the integrated dipole is an incomplete description of a local magnetic source. The octupolar
state and the compensated reference state both produce local magnetic fields despite their zero integrated magnetic dipole moments. Muon spin oscillations have also been observed in the $5d^2$ osmate Ba$_2$CaOsO$_6$~\cite{Thompson2014}, for which purely ferro-octupolar order has been proposed. In this description, orbital currents associated with octupolar order generate the local fields driving muon  recession~\cite{Voleti2020}. At the calculated Re--muon distances, the maximum octupolar field remains below the measured local fields, but is comparable to the field of a $0.06\,\mu_{\mathrm B}$ point dipole. A quantitative comparison requires the full spin magnetization density and orbital-current distribution of each compound in its ordered state, with their magnetic-field contributions summed as vectors over the crystal lattice at the candidate muon sites.

The electronic-structure results provide a complementary perspective on weak-moment magnetism. The HSE06+SOC calculations show that the Re spin and orbital
contributions nearly cancel in MgReO$_4$ and partially compensate in ZnReO$_4$, producing small local physical dipoles. Small static moments have also been reported for the double perovskite Sr$_2$ZnReO$_6$ within a canted-antiferromagnetic description~\cite{Isah2026}. These results illustrate that small dipole moments do not, by themselves, establish octupolar order. In this work, the prediction of predominantly octupolar order comes from solving the first-principles effective Hamiltonians.

The HSE06+SOC calculations predict semiconducting states for the relaxed structures, while our PBE calculations remain metallic for both the relaxed and
unrelaxed experimental structures. The contrast between PBE and HSE in our calculations highlights the influence of the treatment of electronic interactions on the predicted electronic state. A recent study by López-Moreno \textit{et al.} reports metallic MgReO$_4$ using PBEsol with SOC and HSE06+SOC~\cite{LopezMoreno2026}. Optical or transport measurements would help determine whether the compounds are metallic or semiconducting.

The observed distortion raises two questions: how it affects magnetic order and what causes it. The refined structures provide the static crystal field used in the magnetic Hamiltonian. Temperature-dependent diffraction, combined with calculations of the electronic energy gain and lattice energy cost associated with the ReO$_6$ distortions, could assess the role of Jahn--Teller coupling~\cite{Soh2024}. Temperature-dependent resonant inelastic X-ray scattering, interpreted using calculations that include coupling to lattice vibrations, could probe vibronic effects~\cite{Frontini2024}.

To establish the predominantly octupolar character of the magnetic order experimentally, further measurements are needed. Single-crystal neutron diffraction could separate reflections that overlap in powder patterns. The propagation vectors predicted by the first-principles effective Hamiltonians would guide targeted searches for weak magnetic reflections. Polarization analysis could also help separate weak magnetic signals from nuclear scattering. Comparing the momentum dependence of the measured intensities could distinguish dipolar and octupolar contributions~\cite{Shiina2007,FioreMosca2025}. Resonant X-ray scattering could provide a complementary test of the symmetry of the ordered electronic states through the polarization and azimuthal dependence of the intensity~\cite{Mannix2005}. Resonant inelastic X-ray scattering could measure the energies of local electronic excitations and test the predicted crystal-field splittings~\cite{Frontini2024}. 

In summary, our combined experimental and theoretical study supports predominantly octupolar magnetism in the monoclinic wolframites $\mathrm{MgReO}_4$ and $\mathrm{ZnReO}_4$. Magnetic transitions near 83 and 126~K, respectively, produce spontaneous local fields without a resolved magnetic neutron-diffraction contribution. The ab initio effective Hamiltonian provides a microscopic interpretation: a crystal-field splitting of approximately 200~meV isolates low-energy doublets whose exchange-selected order is dominated by octupolar components, accompanied by estimated physical dipoles of approximately $0.06$ and $0.12\,\mu_{\mathrm B}$ per Re. Differences in the calculated intersite interactions select distinct magnetic structures despite the similar muon precession frequencies. The single-ion atomic model further demonstrates that octupolar states can generate finite local magnetic fields even when their integrated magnetic dipole moment vanishes.

These findings extend the experimentally grounded search for octupolar magnetism beyond double perovskites to strongly distorted, edge-sharing wolframite oxides. They connect the proposed hidden order to crystal-field wavefunctions and exchange interactions, showing how the local crystal environment shapes the character of magnetic order. Analysing the dipolar and octupolar content of the low-energy states together with their intersite interactions provides a microscopic strategy for interpreting weak-moment magnets and identifying further candidates for multipolar order in low-symmetry materials.

\section*{Methods}\label{sec:methods}
\subsection*{Sample synthesis and handling}
Powder samples of $\mathrm{MgReO}_4$ and $\mathrm{ZnReO}_4$ were synthesized at the National Institute for Materials Science (NIMS), Ibaraki, Japan, using a high-pressure synthesis technique. The samples were held at $6$~GPa and $1300\,^{\circ}$C for one hour. Because the powders are sensitive to moisture, particularly $\mathrm{ZnReO}_4$, sample preparation was carried out in an argon glove box. Details of sample handling and the associated sample-quality assessments are given in Supplementary Secs.~S1 and S4.

\subsection*{AC magnetic susceptibility}
The real part of the AC magnetic susceptibility of $\mathrm{ZnReO}_4$ was measured using a Quantum Design PPMS DynaCool. The powder was compressed inside a capsule and sealed with Kapton tape. These measurements were used to assess the magnetic response and possible sample degradation. The susceptibility data and sample-handling qualifications are presented in Supplementary Sec.~S1.

\subsection*{Neutron powder diffraction}
Neutron powder diffraction patterns were collected using the high-resolution time-of-flight diffractometer iMATERIA at J-PARC, Ibaraki, Japan~\cite{Ishigaki2009}. Approximately $0.5$~g of each powder was sealed in a cylindrical vanadium can of $5$~mm diameter using aluminium screws and indium wire. The backscattering (BS) detector bank with a reported resolution of $\Delta d/d=0.15\%$ was used and refined using the \textsc{FullProf} suite~\cite{RodriguezCarvajal1993}. The low-angle LA35 bank was used to examine possible magnetic intensity.

The structural refinements used data collected at $7$~K. Patterns collected at $95$~K for $\mathrm{MgReO}_4$ and $140$~K for $\mathrm{ZnReO}_4$ provided comparisons above the respective magnetic transitions. Regions containing unindexed reflections were excluded from the nuclear refinements. The refinement parameters, excluded regions and magnetic-diffraction sensitivity tests are described in Supplementary Secs.~S2 and S3. Room-temperature X-ray diffraction measurements and their sample-handling limitations are documented separately in Supplementary Sec.~S4. The effective-Hamiltonian calculations use the neutron-refined $7$~K structures.

\subsection*{Muon spin spectroscopy and analysis}
Muon spin spectroscopy measurements were performed on the General Purpose Surface Muon (GPS) instrument at the Paul Scherrer Institute (PSI), Villigen, Switzerland. Approximately $200$~mg of each powder was sealed with Kapton tape in an aluminium--Mylar envelope of reported thickness approximately $50~\mu$m. The envelope was mounted on a low-background copper fork and inserted into a $^4$He flow cryostat with an operating range of $2$--$300$~K. Zero-field spectra were collected between $2$ and $100$~K for $\mathrm{MgReO}_4$ and between $2$ and $140$~K for $\mathrm{ZnReO}_4$. Weak-transverse-field measurements used $50$~G perpendicular to the initial muon spin polarization. Longitudinal-field measurements used $10$ and $20$~G parallel to the initial polarization performed above the magnetic ordering transition at $100$ and $140$~K for MgReO$_4$ and ZnReO$_4$, respectively. The data were analysed using \texttt{musrfit}~\cite{Suter2012}. The $\mathrm{MgReO}_4$ measurements were previously reported in Ref.~\cite{Nocerino2023} and are reanalyzed here.

The zero-field asymmetry was fitted using
\begin{align}
 A_{\mathrm{tot}}P_{\mathrm{ZF}}(t)={}&
 \sum_{j=\mathrm H,\mathrm L} A_j
 \cos(2\pi f_jt+\phi)e^{-\lambda_jt}\nonumber\\
 &+A_{\mathrm{KT}}G_{\mathrm{SGKT}}(t,\Delta_{\mathrm{KT}})
 e^{-\lambda_{\mathrm{KT}}t}\nonumber\\
 &+A_{\mathrm{tail}}e^{-\lambda_{\mathrm{tail}}t}
 +A_{\mathrm{bg}}e^{-\lambda_{\mathrm{bg}}t}.
 \label{eq:zf}
\end{align}
Here $A_j$, $f_j$ and $\lambda_j$ are the asymmetry, frequency and exponential relaxation rate of each
oscillatory component, and $\phi$ is their phase which was fitted commonly. $G_{\mathrm{SGKT}}$ is the static Gaussian Kubo--Toyabe function with width $\Delta_{\mathrm{KT}}$~\cite{Hayano1979}. The tail describes the non-precessing contribution in the ordered powder, and the background term accounts
for an additional non-ordering contribution. For the zero-field fits, $A_{\mathrm{bg}}$ was fixed using the
low-temperature weak-transverse-field analysis. Transition temperatures were obtained from sigmoid
fits to the temperature dependence of the weak-transverse-field asymmetry. The longitudinal-field
and weak-transverse-field models, parameter constraints and temperature-dependent fit results are given in
Supplementary Sections~S6--S8.

\subsection*{Muon-site candidates and dipolar fields}
Electrostatic-potential calculations were performed using the pseudopotential-based plane-wave method implemented in \textsc{Quantum ESPRESSO}~\cite{Giannozzi2009}. Candidate muon sites were identified under the assumption of a small local distortion induced by the implanted muon. The spherical average of the nuclear-field distribution width was evaluated for comparison with the paramagnetic Kubo--Toyabe width. Local dipolar fields for trial magnetic configurations were calculated using the Python package \texttt{muesr}~\cite{muesr_package}. The calculated muon sites, nuclear-width estimates and alternative magnetic structures and frequency-to-site mappings are presented in Supplementary Secs.~S9 and S10. 

\subsection*{Hybrid-functional electronic-structure calculations}
Electronic-structure calculations were performed using density functional theory (DFT) as implemented in the Vienna \emph{Ab initio} Simulation Package (VASP)~\cite{Kresse1996,KresseJoubert1999}. The experimentally determined $7$~K structures were used as the initial structures for PBE relaxation~\cite{PBE1996}. The resulting relaxed structures were used for the HSE and HSE06+SOC densities of states, local magnetic moments and charge/spin-density grids. Separate self-consistent PBE calculations on the unrelaxed experimental structures were used to assess the effect of structural relaxation on the spectra. HSE and HSE06+SOC calculations were restricted to the relaxed structures.

The screened Heyd--Scuseria--Ernzerhof hybrid functional (HSE06)~\cite{Heyd2003,Heyd2006,Krukau2006} was used with \texttt{AEXX=0.25} and \texttt{HFSCREEN=0.2}. Ionic cores and valence electrons were treated using the projector augmented-wave (PAW) method~\cite{Blochl1994,KresseJoubert1999}. A plane-wave cutoff of $600$~eV, a $\Gamma$-centred $5\times4\times5$ $\mathbf{k}$-point mesh~\cite{MonkhorstPack1976} and an electronic convergence criterion of $10^{-6}$~eV were employed. The force threshold for the preceding relaxation was $0.01~\mathrm{eV}\,\text{\AA}^{-1}$. The HSE06+SOC calculations were performed at fixed ionic positions, with \texttt{NSW=0} and \texttt{IBRION=-1}.

The magnetic calculations used the crystallographic unit cell containing two Re sites. Their spin moments were initialized antiparallel along Cartesian $z$, with no constraint on the converged moment magnitudes or directions. This cell represents $\mathbf{k}=0$ magnetic periodicity. SOC was enabled using \texttt{LSORBIT=.TRUE.}; no explicit \texttt{SAXIS} was specified. Local Re $d$-projected spin moments and orbital moments were obtained using \texttt{LORBIT=11} and \texttt{LORBMOM=.TRUE.}, respectively, and are reported in the default Cartesian frame. Band gaps were obtained from the valence-band maximum and conduction-band minimum, with the corresponding density of states used as a consistency check.

The detailed settings, PAW datasets, site-projected moment components and exploratory ferromagnetic--antiferromagnetic energy comparison are given in Supplementary Section~S11. The relaxed--unrelaxed PBE comparison and the real-space grid geometry are given in Supplementary Section~S16. The two-Re-cell calculation does not represent the larger-period MgReO$_4$ magnetic state predicted by the effective Hamiltonian, and the spin-density grids are not used as a quantitative prediction of the total field at the muon sites.

\subsection*{Ab initio effective Hamiltonian}
The effective exchange Hamiltonian  introduced in the Results contains the on-site crystal field $H_{\mathrm{cf}}$ and intersite couplings $V_{KK'}^{QQ'}(ij)$ between multipoles within the $J=3/2$ ground-state manifold. 
The $J=3/2$ label here specifies the four-state pseudospin basis used in the calculation rather than an exact angular-momentum quantum number of the monoclinic crystal.

All coupling constants in this Hamiltonian were calculated from first principles. First, the paramagnetic electronic structures of $\mathrm{MgReO}_4$ and $\mathrm{ZnReO}_4$ were calculated using DFT combined with dynamical mean-field theory (DFT+DMFT)~\cite{Georges1996,Anisimov1997,Lichtenstein1998} within the quasi-atomic Hubbard-I approximation~\cite{Hubbard1963}. The intersite exchange interactions within the $J=3/2$ manifold of the Re $d^1$ shell were subsequently evaluated using the FT-HI approach~\cite{Pourovskii2016}.

The DFT+DMFT calculations used the charge-self-consistent implementation of Refs.~\cite{Aichhorn2009,Aichhorn2011}, based on the full-potential linear augmented-plane-wave (LAPW) code \textsc{WIEN2k}~\cite{WIEN2k}, the DMFT implementation provided by the \textsc{TRIQS} library~\cite{Parcollet2015,Aichhorn2016}, and the Hubbard-I implementation in the \textsc{MagInt} code~\cite{MagInt}. The calculations employed $252$ $\mathbf{k}$ points in the full Brillouin zone, an LAPW basis cutoff $R_{\mathrm{mt}}K_{\max}=7$, the local-density approximation exchange--correlation potential~\cite{PerdewWang1992}, and the experimental $7$~K structures from the NPD refinements. The spin-orbit coupling was treated within the standard second-variational approach. Projective Wannier orbitals~\cite{Aichhorn2009} for the Re $5d$ shell were constructed from Kohn--Sham bands in the energy window $[-1.5,7]$~eV relative to the Kohn--Sham Fermi level, enclosing both the $t_{2g}$ and $e_g$ manifolds. The on-site Coulomb interaction for the full 5$d$ shell was specified by the Slater parameter $F^0=U=2.2$~eV and Hund's coupling $J_{\mathrm H}=0.5$~eV.
 The comparison of the resulting DFT+Hubbard-I spectral functions with the HSE06+SOC densities of states is presented in the Results and Supplementary Sec.~S12 as a cross-method consistency check. The double-counting correction was evaluated in the fully localized limit~\cite{Anisimov1991} using a fixed Re $5d$ occupancy of one electron, as appropriate for the Hubbard-I approximation~\cite{Pourovskii2007}.

During the self-consistent DFT+Hubbard-I calculations, the Boltzmann weights within the $J=3/2$ ground-state multiplet were averaged to suppress the DFT self-interaction contribution to $H_{\mathrm{cf}}$~\cite{Delange2017}. From the converged electronic structure, the intersite interactions $V_{KK'}^{QQ'}(ij)$ were evaluated using the FT-HI method implemented in \textsc{MagInt}, following previous applications to Re double perovskites~\cite{Soh2024,FioreMosca2024,FioreMosca2025}. The resulting Hamiltonian was solved in the single-site mean-field approximation using \textsc{McPhase}~\cite{Rotter2004} together with an in-house module to obtain transition temperatures and ordered magnetic structures.
The crystal-field parameters, basis conventions and exchange
matrices are presented in Supplementary Secs.~S12--S14. The ground-doublet projection is given in
Eq.~\eqref{eq:projection-main}.

\subsection*{Illustrative atomic magnetic-field model}
An atomic Re$^{6+}$ $5d^1$ model was used to calculate the single-ion fields of the ideal octupolar state,
$\lvert\psi_{\mathrm{oct}}\rangle= (\lvert3/2,+3/2\rangle+\lvert3/2,-3/2\rangle)/\sqrt{2}$,
and the compensated reference state $\lvert3/2,+3/2\rangle$ at $\gamma=1$. In the reference state, the integrated spin and orbital magnetic dipoles cancel. Both states have zero integrated magnetic dipole moment. The field calculation includes spin magnetization density and orbital currents, with a mean $5d$ radius $\langle r\rangle_{5d}=0.65$~\AA. Point-dipole fields were calculated using the moments
predicted by the first-principles effective Hamiltonians: $0.06\,\mu_{\mathrm B}$ for MgReO$_4$ and
$0.12\,\mu_{\mathrm B}$ for ZnReO$_4$.

For each distance $r$ from Re, we report the maximum single-ion field magnitude over directions on a sphere,
\begin{equation}
 B_{\max}(r)=\max_{\hat{\mathbf r}}
 \left|\mathbf B(r\hat{\mathbf r})\right|.
 \label{eq:methods-bmax}
\end{equation}
Model details and spatial field maps are given in Supplementary Section~S15.

\noindent{\bf \large{Acknowledgements}}
\vspace{4pt}

\noindent
We would like to thank A. Kentaro Inge from Stockholm University for his assistance with the X-ray diffraction measurements. NPD beamtime was performed at J-PARC using the iMATERIA instrument (Proposal Number:~2024A0387). We thank Takafumi Hawai for his support during the NPD measurements. Muon spin spectroscopy measurements were conducted at the General Purpose Surface Muon (GPS) instrument at Paul Scherrer Institute (PSI), Villigen, Switzerland (Experimental number:~20221283). We sincerely appreciate the invaluable support and expertise of the facility staff. L.V.P. is grateful to the CPHT computer team for support.

\bigskip
\vspace{4pt}
\bigskip

\noindent{\bf \large{Funding Information}}
\vspace{4pt}

\noindent
The research is funded by the Swedish Research Council, VR (Dnr. 2021-06157 and Dnr. 2022-03936), the Swedish Foundation for Strategic Research (SSF) within the Swedish national graduate school in neutron scattering (SwedNess), the Carl Tryggers Foundation for Scientific Research (CTS-22:2374), and the Knut and Alice Wallenberg Foundation through the grant 2021.0150. U.M. acknowledges funding from the KTH-SCI doctoral excellence program. L.V.P. acknowledges the support by CNRS through the Tremplin@Physique 2025 program
L060401-5. J.S. was supported by the Japan Society for the Promotion of Science (JSPS) KAKENHI Grant No. JP23H01840 and JP24H00042. O.K.F. is supported by the Swedish Research Council (VR) via a Grant 2022-06217 and the Foundation Blanceflor 2023 and 2024 fellow scholarships. E.N. acknowledges financial support from the SSF-Swedness grant SNP21-0004 and the Foundation Blanceflor 2024 fellow scholarship. S.S.P. and V.K. sincerely acknowledge the National Supercomputing Mission (NSM) for providing computational resources on ‘PARAM SEVA’ at IIT Hyderabad. S.S.P. acknowledges  DST-INSPIRE for a research fellowship. G V acknowledge the Institute of Eminence, University of Hyderabad (UOH-IOE-RC3–21–046) for funding.
%
Financial support from the Swedish Research Council (Vetenskapsrådet, VR; Grant No. 2024-04986), and the Knut and Alice Wallenberg Foundation (KAW; Grant No. 2022.0108), is acknowledged. 
The Wallenberg Initiative Materials Science for Sustainability (WISE) funded by the Knut and Alice Wallenberg Foundation and The Swedish e-Science Research Centre (SeRC) and are also acknowledged.
Some of the computations/data handling were enabled by resources provided by the National Academic Infrastructure for Supercomputing in Sweden (NAISS), partially funded by the Swedish Research Council through grant agreement no. 2022-06725.

\bigskip
\noindent{\bf{Author Contributions}}
\vspace{4pt}

\noindent
M.M., J.S. and Y.S. conceived the project. H.S. synthesized the MgReO$_4$
and ZnReO$_4$ samples. U.M., O.K.F., E.N., Y.G., F.E., H.L., R.P., C.W.,
M.M. and Y.S. conducted the muon spin spectroscopy measurements. U.M. and
O.K.F. analysed the muon data, and J.S., D.W.T., M.M. and Y.S. helped
them interpret the results. U.M. and T.I. performed the neutron powder
diffraction experiments, and U.M. analysed the resulting data. U.M.
conducted the X-ray diffraction experiments with assistance from M.S.
and analysed the resulting data. U.M. performed the AC magnetic
susceptibility measurements and analysis. L.V.P. performed the
\emph{ab initio} crystal-field, multipolar exchange and
effective-Hamiltonian calculations. S.S.P., V.K. and G.V. performed the
HSE06+SOC electronic-structure calculations. A.D. developed and
performed the illustrative atomic spin-density and orbital-current
magnetic-field calculations. M.M. and Y.S. supervised the project.
All authors contributed to the interpretation and discussion of the
results, reviewed and edited the manuscript, and approved the final
version.

\bigskip
\noindent{\bf \large{Data Availability and Analysis}}
\vspace{4pt}
\noindent
Data and their analysis are available from the corresponding authors upon reasonable request.

\bigskip
\noindent{\bf \large{Ethics declarations}}
\vspace{4pt}
\noindent The authors declare no competing interests.\\
LLM was used to assist with English-language editing, formatting, and improving the clarity and coherence of the manuscript. All suggested changes
were reviewed and verified by the authors, who retain full responsibility for the scientific content, interpretations, and conclusions.

\bibliographystyle{apsrev4-2}
\bibliography{refs}

\FloatBarrier  

\end{document}


\raggedright\raggedbottom
\begingroup\setstretch{1.0}\maketitle\endgroup

\setcounter{tocdepth}{1}
\begingroup\small\tableofcontents\endgroup
\clearpage
\setstretch{1.10}
\suppsection{sample}{Sample handling and AC magnetic susceptibility}
For AC magnetic susceptibility (ACMS) measurements, the powder was compressed in a capsule and sealed with Kapton tape. Since the samples are moisture sensitive, particularly \Zn, the transfer into the capsule was done under argon. The sample mass was not determined because no balance was available inside the glove box, and air exposure during transfer and mounting was minimized. Some degradation during transfer cannot be excluded.

The ACMS of \Zn was measured using the ACMS II option of a Quantum Design PPMS DynaCool in the KTH Nanolab, with an AC drive amplitude of approximately $5$~Oe at a frequency of $4$~kHz. $\chi'$ is the real part of the ACMS divided by the AC drive field, expressed in $\mathrm{emu/Oe}$. No normalization to sample mass was applied and no background subtraction conducted.

\begin{figure}[!htbp]
\centering\includegraphics[width=0.45\textwidth]{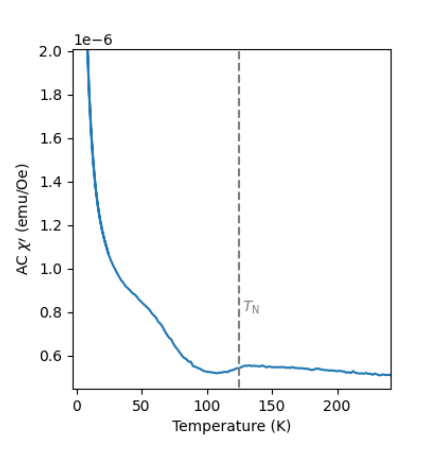}
\caption{Real component of the AC magnetic susceptibility (ACMS) of \Zn, measured on cooling with an AC drive of approximately $5$~Oe at $4.004$~kHz. The response is expressed per unit AC field. The dashed line marks the magnetic transition temperature obtained from the weak transverse field (TF) \musr analysis. The low-temperature upturn is treated as a possible impurity contribution.}
\label{fig:si-ac}
\end{figure}

The measurement was intended as a qualitative check that the expected magnetic transition remained detectable after sample handling, since the X-ray diffraction results available at the time did not provide a clear assessment of sample condition. Figure~\ref{fig:si-ac} shows a weak anomaly near $126$~K, consistent with the $125.67(10)$~K transition obtained from weak transverse-field (TF) \musr. The increase below approximately $100$~K is consistent with a contribution from paramagnetic impurities. The unknown sample mass, weak anomaly and background preclude a reliable extraction of an intrinsic magnetic moment from this ACMS data set alone. Earlier measurements on \Mg are described in Ref.~\cite{Nocerino2023}.

\suppsection{structure}{Low-temperature structural refinement and local coordination}
The $7$~K neutron powder diffraction (NPD) patterns were refined in space group $P2/c$ using \textsc{FullProf}~\cite{RodriguezCarvajal1993} using the backscattering (BS) detector bank. Table \ref{tab:si-lattice} shows the refined lattice parameters, cell volume and refinement $\chi^2$. The reduced $\chi^2$ values are $364.5$ for \Mg and $188.4$ for \Zn. The measurement time was very long and high counting statistics can make the $\chi^2$ very sensitive to small profile discrepancies~\cite{Toby2006}. The reported standard deviations are therefore scaled by $\sqrt{\chi^2}$ according to ref. \cite{Toby2006} to account for the inflated $\chi^2$. 

\begin{table}[!htbp]
\centering\small
\caption{Lattice parameters from the $7$~K neutron powder diffraction (NPD) refinements. Both compounds use the same monoclinic $P2/c$ structure, with $\alpha=\gamma=90^\circ$.
Parentheses give the standard uncertainties scaled by $\sqrt{\chi^2}$.}
\label{tab:si-lattice}
\begin{tabular}{lcc}
\toprule
Parameter & \Mg & \Zn\\
\midrule
$a$ (\AA) & 4.67746(7) & 4.68845(8)\\
$b$ (\AA) & 5.57642(9) & 5.60352(10)\\
$c$ (\AA) & 5.01176(7) & 5.02262(8)\\
$\beta$ (degrees) & 91.8834(13) & 91.2725(13)\\
$V$ (\AA$^3$) & 130.654(3) & 131.921(4)\\
$\chi^2$ & 364.5 & 188.4\\
\bottomrule
\end{tabular}
\end{table}
Table~\ref{tab:si-coordinates} shows the refined atomic coordinates and isotropic displacement parameters. In FullProf, the occupation factor is
\[
 o_{\mathrm{FP}} = f\,\frac{m}{M},
\]
where $f$ is the physical site occupancy, $m$ is the site multiplicity, and $M$ is the general-position multiplicity. For $P2/c$, $M=4$. Re occupies the $2e$ site and Mg or Zn occupies the $2f$ site, so full occupancy ($f=1$) corresponds to $o_{\mathrm{FP}}=2/4=0.5$. O1 and O2 occupy general $4g$ sites and therefore have
$o_{\mathrm{FP}}=1$. The resulting unit-cell contents are $2A+2\mathrm{Re}+8\mathrm{O}$,
with $A=\mathrm{Mg}$ or $\mathrm{Zn}$ and $Z=2$. All occupation factors were fixed during refinement.

\begin{table}[htbp]
\caption{Fractional atomic coordinates and isotropic displacement
parameters from $7$ K NPD refinements. The last column gives the FullProf occupation factor
$o_{\mathrm{FP}}$, rather than the physical site occupancy. Coordinates without uncertainties are fixed by symmetry. O1 and O2 share a single refined $B_{\rm iso}$ within each compound.
The Re $B_{\rm iso}$ in \Zn was refined separately and then fixed during final refinement.}
\label{tab:si-coordinates}
\centering\small
\begin{ruledtabular}
\begin{tabular}{llccccc}
Compound & Atom & $x$ & $y$ & $z$
& $B_{\rm iso}$ (\AA$^2$) & $o_{\mathrm{FP}}$\\
\Mg & Mg & 0.5000 & 0.6695(7) & 0.2500 & 0.66(7) & 0.5\\
 & Re & 0.0000 & 0.1725(4) & 0.2500 & 0.05(4) & 0.5\\
 & O1 & 0.2157(5) & 0.1101(4) & 0.9351(4) & 0.44(3) & 1\\
 & O2 & $-0.2639(4)$ & 0.6272(5) & 0.6045(4) & 0.44(3) & 1\\
\Zn & Zn & 0.5000 & 0.6825(8) & 0.2500 & 0.48(10) & 0.5\\
 & Re & 0.0000 & 0.1749(4) & 0.2500 & $0.07239^{\rm a}$ & 0.5\\
 & O1 & 0.2133(5) & 0.1089(5) & 0.9318(5) & 0.17(5) & 1\\
 & O2 & $-0.2642(5)$ & 0.6281(5) & 0.6087(5) & 0.17(5) & 1\\
\end{tabular}
\end{ruledtabular}
\par\smallskip
{\footnotesize $^{\rm a}$Fixed during refinement.}
\end{table}

Selected bond lengths and bond angles are listed in Table~\ref{tab:si-bonds}.
Each cation has six nearest oxygen neighbours, leading to three pairs of equal bond lengths. The bond lengths and angles deviate significantly from ideal octahedral coordination and show similar distortion patterns in both compounds. These structural results motivate the low-symmetry crystal-field treatment.

\begin{table}[htbp]
\caption{Bond lengths and selected bond angles calculated
from the 7K NPD refinements with $A=\mathrm{Mg}$ or $\mathrm{Zn}$. Each listed bond length has multiplicity two. The displayed digits indicate numerical precision rather than uncertainty estimates.}
\label{tab:si-bonds}
\centering\small
\begin{ruledtabular}
\begin{tabular}{lcc}
Bond & \Mg\ (\AA) & \Zn\ (\AA)\\
Re--O1 & 1.9326 & 1.9397\\
Re--O1 & 2.0740 & 2.0790\\
Re--O2 & 1.8007 & 1.7936\\
$A$--O1 & 2.0543 & 2.0153\\
$A$--O2 & 2.0745 & 2.1147\\
$A$--O2 & 2.1312 & 2.1887\\
\hline
Angle & \Mg\ (degrees) & \Zn\ (degrees)\\
O1--Re--O1 & 75.95, 81.07, 88.22 & 75.16, 80.18, 87.95\\
O1--Re--O2 & 88.47, 94.19, 98.65 & 88.61, 94.56, 98.93\\
O2--Re--O2 & 103.36 & 104.04\\
O1--$A$--O1 & 106.51 & 109.10\\
O1--$A$--O2 & 87.68, 90.89, 96.91 & 88.13, 92.24, 97.36\\
O2--$A$--O2 & 78.14, 83.46, 86.42 & 74.65, 82.38, 84.45\\
\end{tabular}
\end{ruledtabular}
\end{table}

\begin{figure}[!htbp]
\centering\includegraphics[width=0.60\textwidth]{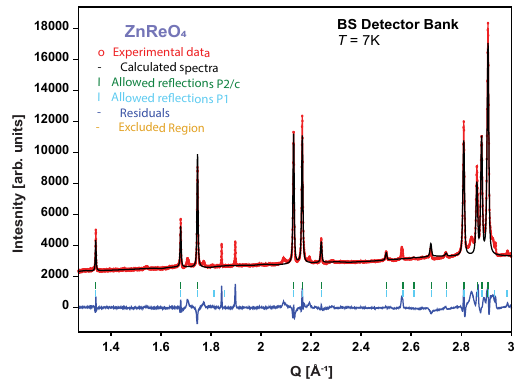}
\caption{\Zn NPD pattern at $7$~K in the backscattering bank, with the $P2/c$ refinement and allowed reflection positions for the tested $P1$ model. The excluded reflections are not accounted for by the displayed lower-symmetry test.}\label{fig:si-lower}
\end{figure}

To resolve the issue of unindexed NPD reflections, lower-symmetry models were tested. Figure~\ref{fig:si-lower} compares the \Zn pattern with allowed reflections of the $P2/c$ and triclinic $P1$ descriptions. The tested lower-symmetry models could not explain the unindexed peaks. This supports an impurity assignment within the tested structural models.

\begin{figure}[!htbp]
\centering\includegraphics[width=0.7\textwidth]{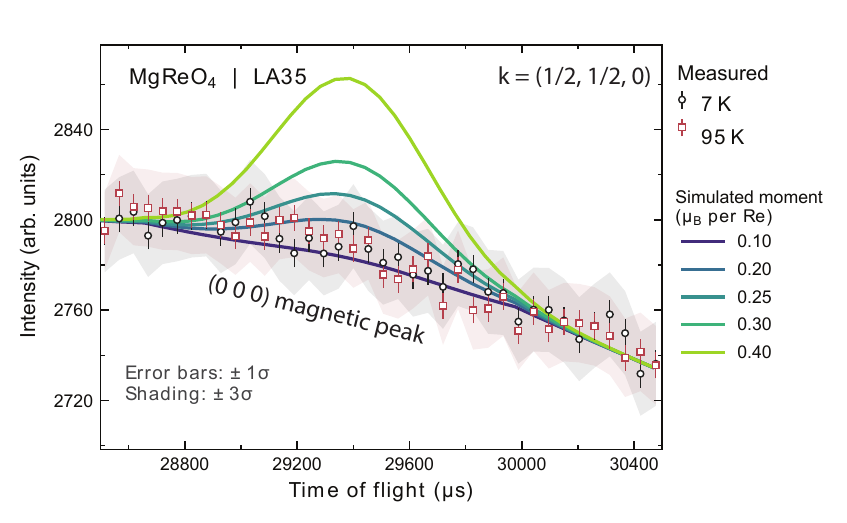}
\caption{Magnetic-diffraction sensitivity for the proposed \Mg $\mathbf{k}=(1/2,1/2,0)$ structure in the LA35 detector bank. The $7$ and $95$~K data are compared with simulated profiles for the ordered moments per Re for the magnetic $(0 0 0)$ peak. Error bars and shaded bands denote $\pm1\sigma$ and $\pm3\sigma$, respectively. The $0.30\,\mu_{\mathrm B}$ simulation exceeds the upper edge of the $7$~K band near the peak maximum.}\label{fig:si-detection}
\end{figure}

\suppsection{npdlimit}{Magnetic-diffraction sensitivity}
The sensitivity to magnetic neutron scattering was assessed for the proposed \Mg structure with propagation vector $\mathbf{k}=(1/2,1/2,0)$. Figure~\ref{fig:si-detection} compares the LA35 data at $7$ and $95$~K with simulated profiles for ordered moments of $0.10$, $0.20$, $0.25$, $0.30$ and $0.40\,\mu_{\mathrm B}$ per Re. The displayed magnetic reflection is the $(000)$ magnetic peak. For the simulated structure, a moment of $0.30\,\mu_{\mathrm B}$ per Re produces a peak clearly above the upper edge of the $7$~K $\pm3\sigma$ band. No corresponding magnetic peak is resolved in the measured data. We therefore adopt an approximate, model-dependent upper limit of $m_{\rm Re}\lesssim0.3\,\mu_{\mathrm B}$ for this structure. Inspection of the other detector banks gives a similar visual sensitivity.

The simulations use an Eu$^{2+}$ magnetic form factor, as there exists no tabulated value for Re$^{+6}$. Eu$^{2+}$ was chosen for having a large form factor to not underestimate the magnitude. No analogous upper limit is assigned to \Zn from this test, because the relevant magnetic reflections of the $\mathbf{k}=(0,0,0)$ model considered here overlap the nuclear peaks and make the assessment more difficult.

\suppsection{xrd}{Room-temperature X-ray diffraction}
Room-temperature powder X-ray diffraction (XRD) patterns were recorded using a Bruker D8 VENTURE diffractometer with a one-dimensional detector, Cu $K\alpha$ radiation and a G\"obel mirror in transmission geometry. The measurements used coupled $2\theta/\theta$ mode over a $2\theta$ range of $10$--$100^{\circ}$, with angular sampling of approximately $0.02^{\circ}$. The total acquisition times were $60$~h for \Mg\ and $36$~h for \Zn. Because of the high X-ray absorption of Re, the powder was mounted on a glass fibre coated with high-vacuum grease, placed inside a $0.3$~mm quartz capillary and sealed under argon. Refinements were performed using \textsc{FullProf}~\cite{RodriguezCarvajal1993}.

The NPD comparison raised concern that the XRD mounting procedure had altered the samples. The original analysis shows coordination-angle ranges of $73$--$111^{\circ}$ for \Mg\ and $62$--$136^{\circ}$ for \Zn\ in the XRD-refined structures. In particular, the refined oxygen coordinates imply much stronger distortions than the neutron results, especially for \Zn. As the samples are humidity sensitive, a possible reason is a reaction with the grease used to fix the powder to the glass capillary. However, a specific reaction pathway is not established. The XRD refinements are therefore retained as an example of sample-handling sensitivity and are not the structural input to the electronic, exchange or muon-site calculations.

\begin{figure}[!htbp]
\centering\includegraphics[width=0.91\textwidth]{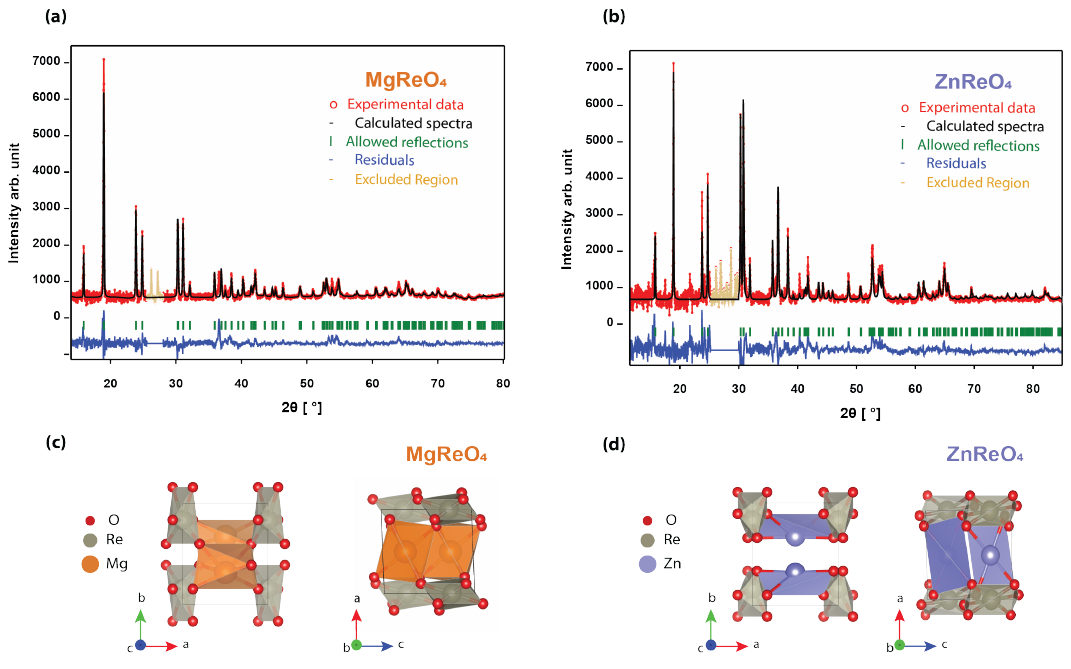}
\caption{Room-temperature XRD diagnostics: patterns and refinements for
(a) \Mg\ and (b) \Zn, with corresponding refined structures in (c) and
(d). The acquisition times were $60$~h for \Mg\ and $36$~h for \Zn.
Suspected sample degradation during mounting limits the interpretation of the oxygen coordinates. These XRD structures are not used for the magnetic calculations.}
\label{fig:si-xrd}
\end{figure}

The patterns can be largely indexed using the monoclinic wolframite structure, with strong unindexed reflections around $2\theta=25$--$30^{\circ}$. These regions were excluded from the refinements. Figure~\ref{fig:si-xrd} and Table~\ref{tab:si-xrd} show the structural refinement results. The plotted octahedra should not be interpreted as the intrinsic low-temperature distortion of the compounds. The powder used in these measurements was not used in the NPD measurements. 

\begin{table}[htbp]
\caption{Acquisition times, angular ranges and reported XRD refinement
parameters for the room-temperature diagnostic measurements. Both
refinements use $P2/c$. Uncertainties are those reported after
goodness-of-fit scaling and do not quantify possible degradation.}
\label{tab:si-xrd}
\centering\begin{ruledtabular}\setlength{\tabcolsep}{3pt}\footnotesize\begin{tabular}{lcc}
Parameter & \Mg & \Zn\\
Acquisition time (h)&60&36\\
Measured $2\theta$ range (degrees)&10--100&10--100\\
$\chi^2$ &12.16&24.49\\
Refinement exclusion intervals (degrees)&0--14, 25.5--28.0&0--14, 25.3--30.0\\
$a$ (\AA)&4.6822(3)&4.6990(2)\\
$b$ (\AA)&5.5750(3)&5.6130(4)\\
$c$ (\AA)&5.0071(3)&5.0270(4)\\
$\beta$ (degrees)&92.026(6)&91.260(6)\\
$V$ (\AA$^3$)&130.50(9)&132.40(10)\\
$A$ $(x,y,z)$&$(0.5,0.6768(4),0.25)$&$(0.5,0.6800(20),0.25)$\\
Re $(x,y,z)$&$(0,0.1633(1),0.25)$&$(0,0.1680(16),0.25)$\\
O1 $(x,y,z)$&$(0.1185(5),0.1606(5),0.9122(5))$&$(0.100(20),0.130(18),1.100(20))$\\
O2 $(x,y,z)$&$(-0.1962(5),0.6042(5),0.6027(5))$&$(0.000(18),0.710(10),0.470(12))$\\
\end{tabular}\end{ruledtabular}
\end{table}

\suppsection{bvs}{Bond-valence-sum analysis}
For stoichiometric $A$ReO$_4$ with divalent Mg or Zn and O$^{2-}$, charge balance requires an average Re oxidation state of $+6$ corresponding to a $5d^1$ configuration. A different valence distribution, including local mixed valence, could contribute to the reduced ordered moment. We therefore use
bond-valence-sum (BVS) analysis as a structural consistency check.

The BVS model~\cite{Brown2006,Brown2009} gives
\begin{equation}
 V=\sum_i s_i=\sum_i\exp\!\left(\frac{R_0-R_i}{b}\right),
 \qquad b=0.37~\text{\AA},
 \label{eq:si-bvs}
\end{equation}
where $R_i$ are cation--oxygen distances and $R_0$ is the distance corresponding to a bond valence of one for the selected ion pair.
All six oxygen neighbours are included. Parameter transferability depends on the coordination environments and bond-length distributions used in the calibration~\cite{Gagne2015}.

MnReO$_4$ was selected as the primary reference because of its related structure, distorted ReO$_6$ coordination and reported
Mn$^{2+}$--Re$^{6+}$ assignment~\cite{Bramnik2005}. Sr$_3$Re$_2$O$_9$ provides a second octahedral Re$^{6+}$ reference~\cite{Urushihara2021}. The experimental structures measured at 304 and 298~K, respectively, have Re--O distances spanning 1.8026--2.0530 and 1.7566--2.2247~\AA.

With $b$ fixed, requiring each reference Re site to have BVS~$=6$ gives
\begin{equation}
 R_0=b\ln\!\left[\frac{6}{\sum_i\exp(-R_i/b)}\right].
 \label{eq:si-bvs-calibration}
\end{equation}
The Mn-derived $R_0=1.92050$~\AA\ is adopted as the primary parameter (Table~\ref{tab:si-bvs}). Applying it to the two Sr reference sites gives BVS values approximately 3\% above six.

\begin{table}[!htbp]
\caption{Re--O calibration at $b=0.37$~\AA. Each fitted $R_0$ gives
site BVS~$=6$; the final column uses the common Mn-derived
$R_0=1.92050$~\AA.}
\label{tab:si-bvs}
\centering\small
\begin{tabular}{llcc}
\toprule
Compound & Re site & Fitted $R_0$ (\AA) & BVS\\
\midrule
MnReO$_4$ & Re1 & 1.92050 & 6.000\\
Sr$_3$Re$_2$O$_9$ & Re1 & 1.90863 & 6.196\\
 & Re2 & 1.91105 & 6.155\\
\bottomrule
\end{tabular}
\end{table}

An alternative Sr-derived $R_0=1.90984$~\AA\ is obtained by requiring the mean BVS of its two equal-multiplicity Re sites to be six.
The target results remain near six under both calibrations (Table~\ref{tab:si-bvs-targets}).

\begin{table}[!htbp]
\caption{Target Re BVS using the Mn- and Sr-derived parameters.
The approximate structural standard uncertainties $\sigma_{\rm str}$
apply to the Mn-column results at fixed $R_0$ and exclude calibration uncertainty.}
\label{tab:si-bvs-targets}
\centering\small
\begin{tabular}{lccc}
\toprule
Compound & BVS (Mn) & BVS (Sr) & Approx. $\sigma_{\rm str}$\\
\midrule
\Mg & 6.020 & 5.849 & 0.021\\
\Zn & 6.019 & 5.849 & 0.024\\
\bottomrule
\end{tabular}
\end{table}

Uncertainties were propagated from the cell and fractional-coordinate standard uncertainties, retaining symmetry constraints but neglecting
covariance between distinct refined parameters. The Mn-reference uncertainties give an approximate $\sigma(R_0)=0.0045$~\AA, contributing
approximately 0.073 to each target BVS. Combining this with the target structural uncertainties in quadrature gives 0.076 and 0.077 for
\Mg\ and \Zn, respectively. These estimates exclude uncertainty in $b$, systematic structural errors and calibration transferability.

The results support the average Re$^{6+}$ assignment, although local mixed valence cannot be excluded. We therefore examine
crystal-field effects on the assumed $5d^1$ state as a possible origin of the suppressed ordered moment.

\begin{figure}[!htbp]
\centering\includegraphics[width=0.81\textwidth,height=0.77\textheight,keepaspectratio]{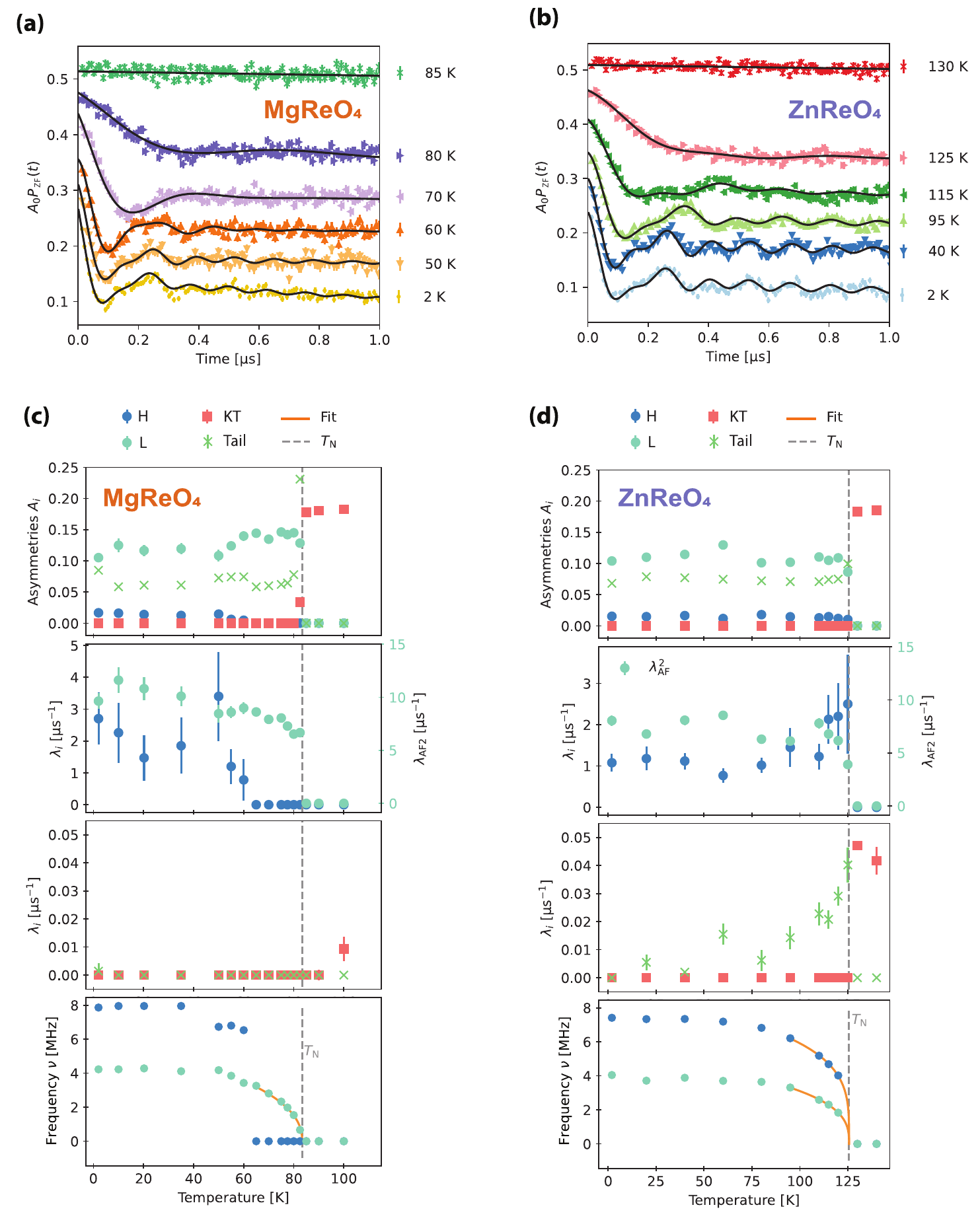}
\caption{zero-field (ZF) data and fit-parameter panels. (a,b) Spectra for \Mg and \Zn with fitted curves according to equation \eqref{eq:si-zf-full}; (c,d) asymmetries, relaxation rates and frequencies from fits. Zero-valued parameters denote unresolved or unused components, not measured fitted zero values.}\label{fig:si-zf}
\end{figure}

\suppsection{zf}{Zero-field fitting, powder amplitudes and longitudinal-field decoupling}
 The zero-field (ZF) $\mu^+\mathrm{SR}$ time spectra of $\mathrm{ZnReO}_4$ and $\mathrm{MgReO}_4$ are presented in Fig.~\ref{fig:si-zf}~(a)-(b). The ZF time spectra at low temperatures reveal two oscillations in both samples. The presence of oscillations indicates magnetic ordering, and multiple oscillations suggest multiple muon stopping sites inside the crystal structure. In a powder, two thirds of the internal magnetic field components are expected to be perpendicular to the initial muon spin polarization (causing oscillations), whereas one third is expected to be parallel to it (causing an exponentially decaying tail). 

As the temperature increases above $T_\mathrm{N}$, the oscillations evolve to a static Gaussian Kubo-Toyabe (KT) multiplied by an exponential type depolarization \cite{Hayano1979}, describing muon spin relaxation in a system of randomly oriented static spins within the muon lifetime. The static Gaussian KT is described as~\cite{Hayano1979,muSRbook_1}
\begin{equation}
 G_{\mathrm{SGKT}}(t,\Delta)=\frac13+\frac23(1-\Delta^2t^2)\exp\!\left(-\frac{\Delta^2t^2}{2}\right).\label{eq:si-sgkt}
\end{equation}
Here $\Delta$ has inverse-time units and describes the field distribution width of nuclear moments. 

The two oscillations are denoted by H and L. H denotes the original high-frequency component ($\sim 8$MHz) and L denotes the low-frequency component ($\sim 4$MHz). The complete ZF fitting panels are shown in Fig.~\ref{fig:si-zf}, including the asymmetries and relaxation rates omitted from the main paper figure. 

The ZF polarization function used for fitting the ZF \musr spectra is:
\begin{align}
 A_{\mathrm{tot}}P_{\mathrm{ZF}}(t)={}&\sum_{j=\mathrm H,\mathrm L}A_j\cos(2\pi f_jt+\phi)e^{-\lambda_jt}\nonumber\\
 &+A_{\mathrm{KT}}G_{\mathrm{SGKT}}(t,\Delta_{\mathrm{KT}})e^{-\lambda_{\mathrm{KT}}t}\nonumber\\
 &+A_{\mathrm{tail}}e^{-\lambda_{\mathrm{tail}}t}+A_{\mathrm{bg}}e^{-\lambda_{\mathrm{bg}}t}.\label{eq:si-zf-full}
\end{align}

where $P_\mathrm{ZF}(t)$ is the muon polarization in ZF, $A$ the corresponding asymmetry, $\lambda$ the damping, $\nu$ and $\phi$ the frequency and phase of the oscillations, and $\Delta_\mathrm{KT}$ the Gaussian distribution width of the KT and $\lambda_\mathrm{KT}$ the relaxation rate of the KT. In addition to the H, L and KT contributions, the polarization has two exponentially decaying terms; a very slowly decaying tail, $\lambda_\mathrm{tail}$ coming from the magnetic moments parallel to the initial muon spin polarization, and a decaying signal $\lambda_\mathrm{bg}$. The background asymmetry is fixed to the paramagnetic background found at base temperature from TF (See Sec. \siref{tf}). A similar background phase has been found in both compounds and it has previously been speculated for $\mathrm{MgReO}_4$ to arise from Re impurities including sample holder background contribution~\cite{Nocerino2023}. Furthermore, the phase of AF oscillations is theoretically $\phi = 0$. However, uncertainty in the muon implantation time, wide internal field distributions, and incommensurate (IC) order can cause $\phi \neq 0$, and therefore, it was treated as a fitted parameter. We confirm, however, that the fitted value remains close to zero at base temperature, and the phase is constrained to be the same for the two muon sites, H and L.

\begin{figure}[!htbp]
\centering\includegraphics[width=0.48\textwidth]{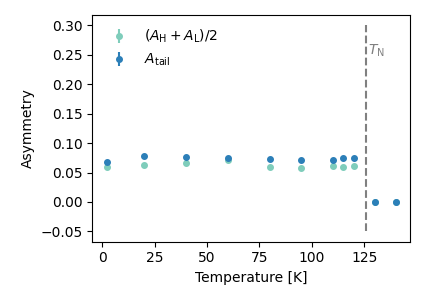}
\caption{Powder-amplitude consistency test for \Zn: one half of the summed oscillatory asymmetry compared with the slowly relaxing tail.}\label{fig:si-tail}
\end{figure}

We can verify the origin of the tail component in the ZF polarization function by checking that the sum of AFM asymmetries ($\frac{2}{3}$ of the signal) divided by two corresponds to the tail signal. Fig.~\ref{fig:si-tail} shows that these quantities align well in our case below $T_\mathrm{N}$ for $\mathrm{ZnReO}_4$, implying that the tail component indeed arises due to spins parallel to the initial muon spin polarization.

\begin{figure}[!htbp]
\centering\includegraphics[width=0.43\textwidth]{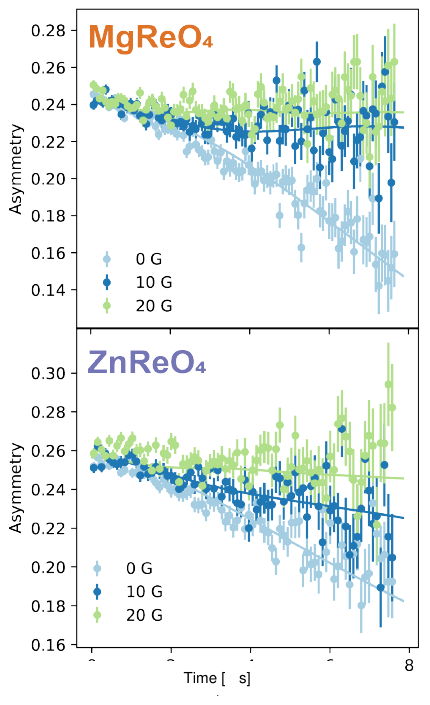}
\caption{Longitudinal-field spectra above the ordering temperatures for \Mg and \Zn at the fields indicated. The curves show the measured fitted relaxation using a static Gaussian Kubo--Toyabe function with an additional relaxation.}\label{fig:si-lf}
\end{figure}

Above the N\'eel temperature, spectra measured with $B_{\rm LF}=0$, $10$ and $20$~G at $100$~K for \Mg and $140$~K for \Zn show decoupling of the weak relaxation by a field of approximately $20$~G (Fig.~\ref{fig:si-lf}), consistent with a quasistatic
contribution. The LF spectra were fitted simultaneously over $0$--$8~\mu\mathrm{s}$ using a static Gaussian KT function multiplied by an exponential, together with an
additional relaxation contribution. Fits with this additional contribution fixed to the low-temperature weak-TF residual and nonrelaxing were compared with fits allowing its amplitude and relaxation rate to vary. The extracted Gaussian width
depends on this choice, while the finite time window limits the observation of the full KT evolution. The resulting nuclear-width comparison is presented in Sec.~\siref{muonsites}.

\begin{figure}[!htbp]
\centering\includegraphics[width=0.46\textwidth]{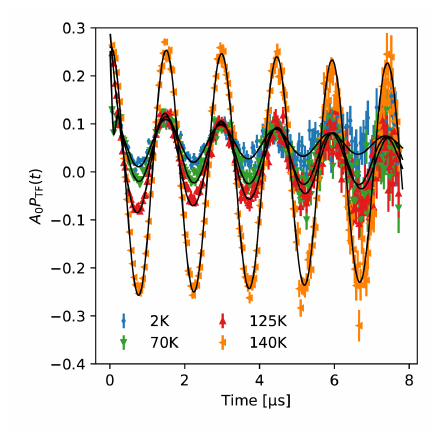}
\caption{Representative $50$~G weak-TF spectra of ZnReO$_4$ at a selected set of temperatures with the experimental error bars. Black lines show the fitted polarisation function with equation~\eqref{eq:si-tf}.}\label{fig:si-tf-spectra}
\end{figure}

\suppsection{tf}{Weak-transverse-field analysis and transition temperatures}
The weak-TF \musr measurements for both \Mg and \Zn used an applied field of $B=50$~G. Representative \Zn spectra at selected temperatures are shown in Fig.~\ref{fig:si-tf-spectra}.
The data are described by the polarization function: 
\begin{align}
 A_0P_{\rm TF}(t)={}&A_{\rm TF}\cos(2\pi f_{\rm TF}t+\phi_{\rm TF})\exp(-\lambda_{\rm TF}t)\nonumber\\
 &+A_{\rm AF}\cos(2\pi f_{\rm AF}t+\phi_{\rm AF})\exp(-\lambda_{\rm AF}t)
 +A_{\rm tail}\exp(-\lambda_{\rm tail}t).
 \label{eq:si-tf}
\end{align}

where $P_\mathrm{TF}(t)$ is the muon polarization in TF and $A_0$ the total asymmetry, which for our experimental setup is $A_0~\approx~0.25$. The polarization function is made up of three contributions; two relaxed oscillating terms coming from the applied TF and inner AFM field, and a slowly decaying magnetic tail arising from the internal magnetic moments which are parallel to the muon spin. In the polarization function, $A$ denotes the corresponding asymmetry, $\lambda$ the damping, and $f$ and $\phi$ the frequency and phase of the oscillations. In theory, we have $\phi = 0$, but uncertainties in implantation times, broad internal field distributions, and IC orders can cause $\phi \neq 0$, so the parameter was fitted. However, we confirm that the value is close to 0 at the base temperature.

\begin{figure}[!htbp]
\centering\includegraphics[width=0.80\textwidth]{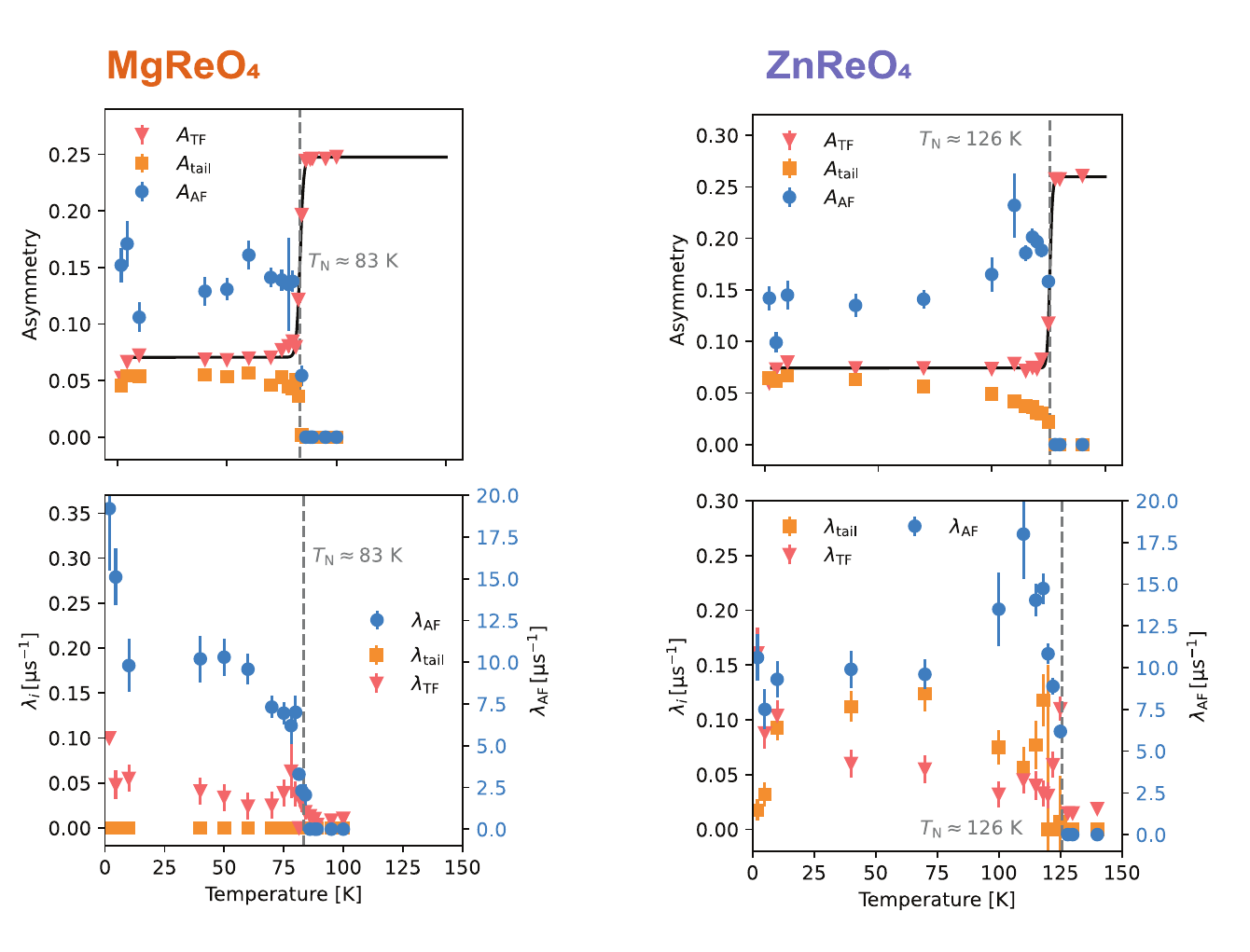}
\caption{Temperature-dependent weak-TF fitting coefficients for \Mg (left) and \Zn (right). The sigmoid fits to the TF amplitudes give $83.35(2)$ and $125.67(10)$~K, respectively. The labels near $83$ and $126$~K in the graphic are rounded.}\label{fig:si-tf-coeff}
\end{figure}

Fitting the temperature dependence of the TF coefficients [Fig.~\ref{fig:si-tf-coeff}] with a sigmoid function
\begin{equation}
 A_{\mathrm{TF}}(T)
 = c+\frac{L}{1+\exp[-k(T-T_\mathrm{N})]},
 \label{eq:wtf-sigmoid}
\end{equation}
where $c$ is the baseline asymmetry, $L$ is the asymmetry change across the transition and $T_\mathrm{N}$ is the N\'eel temperature. Fitting the asymmetry temperature evolution to this function allows us to extract $T_\mathrm{N}$ for $\mathrm{ZnReO}_4$ as $T_\mathrm{N}^\mathrm{Zn} = 125.67(10)$ and $T_\mathrm{N}^\mathrm{Mg} = 83.35(2)$ for $\mathrm{MgReO}_4$, where the error represents the variance in temperature from the fitting results. The TF asymmetry is not completely suppressed below $T_\mathrm{N}$ and stays $\approx 20\%$ for both compounds until base temperature. Part of this is due to sample environment background and the rest is likely due to paramagnetic impurities in the samples. 
 
We observe changes in the asymmetry distribution between the tail and AF component close to $T_\mathrm{N}$ for $\mathrm{ZnReO}_4$. This indicates that the two components become more difficult to distinguish closer to $T_\mathrm{N}$ and could be the cause of the relaxation rate dip before transition. 

\suppsection{temperature}{Temperature-dependent frequencies and restricted canting tests}
In \Mg the low-amplitude H oscillation becomes unresolved near $60$~K. Constraining the ratio between H and L in the fitting did not result in two separately resolvable frequencies above $60$~K. A possible reason for the unresolved H frequency was proposed to be due to spin canting~\cite{Nocerino2023}. To examine the possibility of the canting scenario in $\mathrm{MgReO}_4$ we let the spins cant within the $ac$ plane using the $\Gamma_3$ and $\Gamma_4$ dipolar candidates and follow the local field at each muon stopping site. The muon stopping sites can be found in Sec. \siref{muonsites}. This scenario is not viable for the first-principles predicted (1/2, 1/2, 0) structure, in which the dipole moments point strictly along $b$.

\begin{figure}[!htbp]
\centering\includegraphics[width=0.98\textwidth]{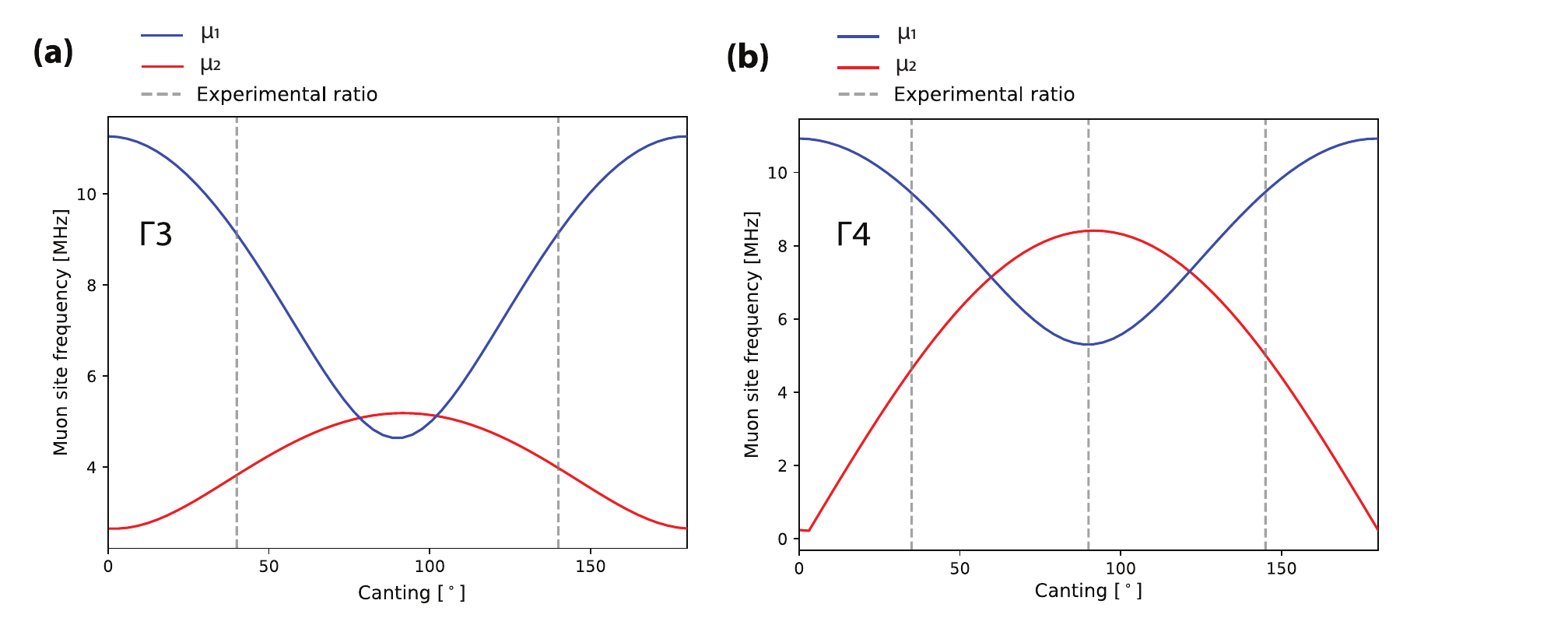}
    \caption{Dipolar-field muon precession frequency simulation for \Mg: (a) $\Gamma_3$, $\mathbf k=(0,1/2,0)$; (b) $\Gamma_4$, $\mathbf k=0$. The dashed line indicates where the experimentally observed H/L frequency ratio is realized. In both cases there are canting angles for which the two stopping sites experience the same local dipolar field.}\label{fig:si-canting}
\end{figure}

Fig.~\ref{fig:si-canting} shows the evolution of the two muon site frequencies as a function of canting angle. We see that there are canting angles for which the two site frequencies cross and experience the same local dipolar magnetic field. However, in a canting case, we should see a sharp increase in the lower frequency. In our experiments, we see that the higher frequency H is no longer distinguishable, but we do not see a drastic increase in the lower frequency L, which would be expected if the cause of the reduction in frequencies would be spin canting. Alternatively, we would have a sudden drastic decrease in the magnetic moment exactly coinciding with the canting, causing the lower L frequency to not change noticeably. As we do not observe any other signatures or structural transitions, we deem this scenario unlikely. Another observation here is that for $\Gamma_4$, the muon stopping sites assignment can be inverted and be associated with a different canting angle. 

The temperature dependence of the precession frequencies was fitted to the power-law parametrization
\begin{equation}
f(T)=\alpha\left(1-\frac{T}{T_{\rm N}}\right)^\beta,
\label{eq:si-powerlaw}
\end{equation}
with $T_{\rm N}$ fixed to the transition temperature obtained from the weak-TF analysis. For \Mg, the fit of the lower-frequency component using seven temperature points between $60$ and $82.5$~K, with $T_{\rm N}=83.366$~K fixed, gives $\beta=0.440(25)$. For \Zn, separate fits to the frequency components at $95$, $110$, $115$, and $120$~K, with $T_{\rm N}=125.67$~K fixed, give $\beta_{\rm H}=0.260(5)$ and $\beta_{\rm L}=0.349(8)$. Parentheses denote formal one-standard-error estimates obtained as the square roots of the fitted covariance-matrix diagonal elements. They exclude uncertainty in the fixed $T_{\rm N}$. With only a few temperature points included, the fits are treated as phenomenological descriptions rather than determinations of a universality class which would require many more measurement points closer to the transition~\cite{dor2004}.

\suppsection{muonsites}{Muon-site candidates, geometry and nuclear widths}
The electrostatic-potential calculation identifies two candidate muon stopping sites (see Table ~\ref{tab:si-sites}) in each compound under the assumption of negligible muon-induced distortion.  The measured H and L frequencies are kept separate from the site labels $\mu_1$ and $\mu_2$, since a unique frequency-to-site assignment is not established by the electrostatic calculation and depending on canting angle we see that the assignments can differ (Sec.~\siref{temperature}).

\begin{table}[htbp]
\caption{Candidate muon stopping-site coordinates, nearest Re
distances and calculated nuclear widths, using the $7$~K NPD
structures without local relaxation around the implanted muon.
For $\Delta_{\mathrm C}$, the additional amplitude was fixed
to the low-temperature weak-TF background value
($0.0644$ for \Mg and $0.0562$ for \Zn), with its relaxation
rate fixed to zero. For $\Delta_{\mathrm U}$, both its amplitude
and relaxation rate were fitted. The experimental widths
describe the whole KT contribution, not individual sites.
All widths are in $\mu\mathrm{s}^{-1}$.}
\label{tab:si-sites}
\centering
\begin{ruledtabular}
\begin{tabular}{llccccc}
Compound & Site & $(x,y,z)$ & Nearest Re (\AA)
& $\Delta_{\mathrm{calc}}$ & $\Delta_{\mathrm C}$
& $\Delta_{\mathrm U}$\\
\Mg & $\mu_1$ & $(0,0.33,0.75)$ & 2.655
& 0.139 & 0.0953(21) & 0.271(13)\\
& $\mu_2$ & $(0.5,0.22,0.25)$ & 2.354
& 0.155 & &\\
\Zn & $\mu_1$ & $(0,0.30,0.75)$ & 2.607
& 0.143 & 0.0869(22) & 0.290(13)\\
& $\mu_2$ & $(0.5,0.22,0.25)$ & 2.358
& 0.153 & &\\
\end{tabular}
\end{ruledtabular}
\end{table}

For the distance and nuclear-width calculations, we use the monoclinic metric with Cartesian lattice vectors
\begin{equation}
 \mathbf a=(a,0,0),\qquad\mathbf b=(0,b,0),\qquad
 \mathbf c=(c\cos\beta,0,c\sin\beta).
 \label{eq:si-cell}
\end{equation}

Assuming uncorrelated nuclear moments, the calculated nuclear width
for a powder sample is~\cite{muSRbook_1}
\begin{equation}
 \Delta_{\rm calc}^{2}=
 \frac{2\gamma_\mu^2}{3}
 \left(\frac{\mu_0\hbar}{4\pi}\right)^2
 \sum_j\frac{1}{r_j^6}
 \sum_\nu p_{j\nu}\gamma_{j\nu}^{2}I_{j\nu}(I_{j\nu}+1).
 \label{eq:si-nuclear}
\end{equation}
with $j$ labeling the nuclear sites, $r_j$ is the distance to the atom,
and $\nu$ is the isotopes with abundance $p_{j\nu}$, spin $I_{j\nu}$ and
angular gyromagnetic ratio $\gamma_{j\nu}$. Natural abundances are used:
$0.374$ for $^{185}$Re, $0.626$ for $^{187}$Re, $0.100$ for $^{25}$Mg,
$0.041$ for $^{67}$Zn and $0.00038$ for $^{17}$O. All these magnetic
isotopes have $I=5/2$. 

The nuclear gyromagnetic ratios are in $\mathrm{rad\,s^{-1}\,T^{-1}}$, and the unit conversion to $\mu s^{-1}$ uses $\gamma_\mu=851.616~\mathrm{rad\,\mu s^{-1}\,T^{-1}}$. The calculation was summed over translation indices $[-12,12]$ along each lattice direction. Increasing this range to $[-24,24]$ changed every calculated width by less than $9\times10^{-7}~\mu\mathrm{s}^{-1}$.

The fitted Gaussian width depends strongly on the treatment of the additional relaxation contribution, with similar
behaviour in both compounds. The finite measurement window also limits the observation of the full KT evolution. The calculated widths lie between the estimates obtained with the two fitting procedures. We report these estimates separately to show the dependence on the fitting assumptions
and use the electrostatic positions as candidate muon sites for the subsequent magnetic-field calculations.

\suppsection{dipolar}{Dipolar candidate structures and limits of the field comparison}
In testing dipolar candidate structures we considered propagation vectors $\mathbf{k}=(0,0,0)$, $(0,1/2,0)$ and $(1/2,0,0)$, selected based on studies of other magnetic wolframites~\cite{Fabelo2024,Prosnikov2017,Lautenschlager1993}. Re occupies the $2e$ site rather than the $2f$ magnetic site found in the related wolframites, so their allowed magnetic basis vectors cannot simply be transferred without a site-specific symmetry analysis. The search used \textsc{BasIreps} within \textsc{FullProf}~\cite{RodriguezCarvajal1993} to identify allowed components.

Table~\ref{tab:si-ir} shows the four irreducible representations (IR) found for the aforementioned $k$-vectors and their moment-direction restrictions. The rotational-part headings are abbreviated. The labels $\Gamma_3$ and $\Gamma_4$ in other parts of this work refer to these particular IRs describing the magnetic structures.
\begin{table}[htbp]
\caption{Character pattern and allowed moment components in the adopted representation labeling. The headings denote the rotational parts of the identity, twofold, inversion and glide operations. The full translation and magnetic-basis convention belongs to the propagation-vector-specific calculation.}\label{tab:si-ir}
\centering\setlength{\tabcolsep}{4pt}\begin{ruledtabular}\begin{tabular}{lccccc}
IR & $1$ & $2_y$ & $\bar 1$ & $m_y$ & Allowed components\\
$\Gamma_1$ &1&1&1&1&$(0,v,0)$\\
$\Gamma_2$ &1&1&$-1$&$-1$&$(0,v,0)$\\
$\Gamma_3$ &1&$-1$&1&$-1$&$(u,0,w)$\\
$\Gamma_4$ &1&$-1$&$-1$&1&$(u,0,w)$\\
\end{tabular}\end{ruledtabular}
\end{table}

\begin{figure}[!htbp]
\centering
\includegraphics[width=0.95\textwidth]{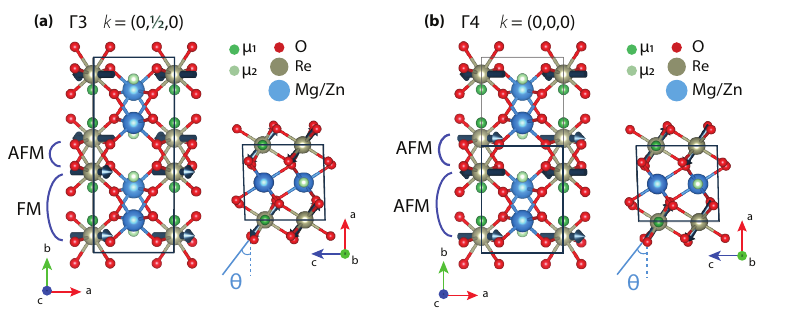}
\caption{Dipolar candidate structures: (a) $\Gamma_3$ with
$\mathbf{k}=(0,1/2,0)$ and (b) $\Gamma_4$ with $\mathbf{k}=0$.
Re moments lie in the $ac$ plane. Green markers indicate candidate muon
stopping positions.}
\label{fig:si-candidates}
\end{figure}

The candidates compared below are $\Gamma_3$ with $\mathbf{k}=(0,1/2,0)$, $\Gamma_4$ with $\mathbf{k}=(0,0,0)$, and the collinear
Mg candidate with $\mathbf{k}=(1/2,1/2,0)$ and moments along $b$. The $\Gamma_3$ and $\Gamma_4$ Re moments lie in the $ac$-plane
(Fig.~\ref{fig:si-candidates}); $\theta$ is measured from $a$ within this plane, so $90^\circ$ denotes moments along $c$. 

The candidate positions in the original chemical cell are similar for the two compounds with just a small shift along the $b$-axis (See Table \ref{tab:si-dipolar-summary}). The calculated precession frequencies and the defined low-to-high frequency ratio are
\begin{equation}
 f_i=\frac{\gamma_\mu}{2\pi}|\mathbf B_{{\rm dip},i}|,
 \qquad \frac{\gamma_\mu}{2\pi}=135.54~\mathrm{MHz\,T^{-1}},
 \qquad \rho=\frac{f_{\rm L}}{f_{\rm H}}.
 \label{eq:si-dipolar-frequency}
\end{equation}
Table~\ref{tab:si-dipolar-summary} lists the moment amplitude per Re, canting angle, calculated field vectors and frequency assignment for every candidate. The two $\Gamma_4$ angle choices exchange their H and L frequency assignments.

\begin{table*}[!htbp]
\caption{Calculated dipolar fields and comparison with the measured frequencies. Each model heading gives the propagation vector, input Re moment amplitude $\mu$, direction and frequency ratio $\rho=f_{\rm L}/f_{\rm H}$. H and L denote the high- and low-frequency
assignments of the two sites. }
\label{tab:si-dipolar-summary}
\label{tab:si-ppt-fields}
\centering\small
\setlength{\tabcolsep}{4pt}
\begin{tabular}{lccrrr}
\toprule
Site & Fractional position & $\mathbf B_{\rm dip}$ (T) & $|\mathbf B_{\rm dip}|$ (T)
 & $f_{\rm calc}$ (MHz) & $f_{\rm meas}$ (MHz)\\
\midrule
\multicolumn{6}{l}{\Mg, $\Gamma_3$: $\mathbf{k}=(0,1/2,0)$,
 $\mu=0.255\,\mu_{\rm B}$, $\theta=41^\circ$, $\rho=0.53405$}\\
 $\mu_1$ (L) & $(0.00,0.33,0.75)$ & $(-0.01201,0,0.02826)$ & 0.030708 & 4.162 & $4.229(9)$\\
$\mu_2$ (H) & $(0.50,0.22,0.25)$ & $(0.05407,0,-0.01956)$ & 0.057501 & 7.794 & $7.86(7)$\\
\addlinespace[5pt]
\multicolumn{6}{l}{\Mg, $\Gamma_4$: $\mathbf{k}=(0,0,0)$,
 $\mu=0.255\,\mu_{\rm B}$, $\theta=35^\circ$, $\rho=0.54942$}\\
 $\mu_1$ (L) & $(0.00,0.33,0.75)$ & $(-0.00313,0,0.03161)$ & 0.031761 & 4.305 & $4.229(9)$\\
$\mu_2$ (H) & $(0.50,0.22,0.25)$ & $(0.05461,0,-0.01895)$ & 0.057807 & 7.835 & $7.86(7)$\\
\addlinespace[5pt]
\multicolumn{6}{l}{\Mg, $\Gamma_4$: $\mathbf{k}=(0,0,0)$,
 $\mu=0.255\,\mu_{\rm B}$, $\theta=90^\circ$, $\rho=0.56399$}\\
 $\mu_1$ (H) & $(0.00,0.33,0.75)$ & $(-0.00172,0,0.05756)$ & 0.057584 & 7.805 & $7.86(7)$\\
$\mu_2$ (L) & $(0.50,0.22,0.25)$ & $(-0.00040,0,-0.03247)$ & 0.032476 & 4.402 & $4.229(9)$\\
\addlinespace[5pt]
\multicolumn{6}{l}{\Zn, $\Gamma_3$: $\mathbf{k}=(0,1/2,0)$,
 $\mu=0.232\,\mu_{\rm B}$, $\theta=36.5^\circ$, $\rho=0.54515$}\\
 $\mu_1$ (L) & $(0.00,0.30,0.75)$ & $(-0.01006,0,0.02808)$ & 0.029825 & 4.042 & $4.05(5)$\\
$\mu_2$ (H) & $(0.50,0.22,0.25)$ & $(0.05233,0,-0.01597)$ & 0.054709 & 7.415 & $7.42(3)$\\
\addlinespace[5pt]
\multicolumn{6}{l}{\Zn, $\Gamma_4$: $\mathbf{k}=(0,0,0)$,
 $\mu=0.227\,\mu_{\rm B}$, $\theta=32^\circ$, $\rho=0.54793$}\\
 $\mu_1$ (L) & $(0.00,0.30,0.75)$ & $(-0.00353,0,0.02978)$ & 0.029984 & 4.064 & $4.05(5)$\\
$\mu_2$ (H) & $(0.50,0.22,0.25)$ & $(0.05233,0,-0.01602)$ & 0.054722 & 7.417 & $7.42(3)$\\
\addlinespace[5pt]
\multicolumn{6}{l}{\Zn, $\Gamma_4$: $\mathbf{k}=(0,0,0)$,
 $\mu=0.218\,\mu_{\rm B}$, $\theta=80^\circ$, $\rho=0.54774$}\\
 $\mu_1$ (H) & $(0.00,0.30,0.75)$ & $(-0.00170,0,0.05476)$ & 0.054784 & 7.425 & $7.42(3)$\\
$\mu_2$ (L) & $(0.50,0.22,0.25)$ & $(0.01009,0,-0.02826)$ & 0.030007 & 4.067 & $4.05(5)$\\
\addlinespace[5pt]
\multicolumn{6}{l}{\Mg, collinear: $\mathbf{k}=(1/2,1/2,0)$,
 $\mu=0.470\,\mu_{\rm B}$, $\mathbf m\parallel b$, $\rho=0.54004$}\\
 $\mu_1$ (L) & $(0.00,0.33,0.75)$ & $(0,0.03144,0)$ & 0.031442 & 4.262 & $4.229(9)$\\
$\mu_2$ (H) & $(0.50,0.22,0.25)$ & $(-0.05818,0,-0.00231)$ & 0.058221 & 7.891 & $7.86(7)$\\
\bottomrule
\end{tabular}
\end{table*}

The $0.470\,\mu_{\rm B}$ moment used for the collinear \Mg candidate exceeds the approximately $0.06\,\mu_{\rm B}$ dipolar moment predicted
by the effective Hamiltonian. Likewise, the \Zn $\Gamma_4$ amplitudes of $0.218$--$0.227\,\mu_{\rm B}$ exceed its predicted $\sim0.12\,\mu_{\rm B}$. A frequency comparison after rescaling the moment thus does not establish consistency with the predicted physical
moment. The $\Gamma_3$ and $\Gamma_4$ amplitudes are below the representative $0.3\,\mu_{\rm B}$ NPD sensitivity scale, but the detectability of each candidate must be assessed from individual simulations. 

\suppsection{hse}{Hybrid-functional electronic structure and site-projected moments}
The HSE06+SOC results were obtained with \textsc{VASP} and PAW potentials~\cite{Kresse1996,KresseJoubert1999,Blochl1994}, using the screened HSE06 hybrid functional~\cite{Heyd2003,Heyd2006,Krukau2006}. The experimental $7$~K structures were the initial inputs for PBE relaxation~\cite{PBE1996}. The resulting relaxed structures and their self-consistent magnetic configurations were used for the DOS, local-moment and charge/spin-density calculations reported here. Thus, these outputs do not correspond directly to the unrelaxed experimental cells. HSE and HSE06+SOC were evaluated only for the relaxed structures. The PBE-optimized lattice parameters are compared with experiment in Table~\ref{tab:si-optimized}; a separate PBE calculation on the unrelaxed experimental structures is discussed in Sec.~\siref{versionaudit}.

\begin{table}[htbp]
\caption{Reported PBE-optimized lattice parameters compared with the rounded experimental parameters used in the original computational comparison. Both $\alpha$ and $\gamma$ are $90^{\circ}$. The more precise experimental values are given in Sec.~\siref{structure}.}\label{tab:si-optimized}
\centering\begin{ruledtabular}\begin{tabular}{llccccc}
Compound &Structure&$a$ (\AA)&$b$ (\AA)&$c$ (\AA)&$\beta$ (degrees)&$V$ (\AA$^3$)\\
\Mg &Experimental&4.677&5.576&5.011&91.88&130.65\\
 &PBE optimized&4.690&5.664&5.084&91.97&134.98\\
\Zn &Experimental&4.688&5.603&5.023&91.27&131.92\\
 &PBE optimized&4.724&5.696&5.089&92.31&136.82\\
\end{tabular}\end{ruledtabular}
\end{table}

\Needspace{8\baselineskip}
\subsection*{Computational settings and magnetic configuration}
The calculations employed a plane-wave cutoff of $600$~eV, a $\Gamma$-centred $5\times4\times5$ $\mathbf{k}$-point mesh and an electronic convergence criterion of $10^{-6}$~eV. The force threshold for the preceding PBE relaxations was $0.01~\mathrm{eV}\,\angstrom^{-1}$. The hybrid-functional settings were \texttt{LHFCALC=.TRUE.}, \texttt{AEXX=0.25}, \texttt{HFSCREEN=0.2}, \texttt{ALGO=ALL} and \texttt{TIME=0.4}. For HSE06+SOC, the ionic positions were fixed with \texttt{NSW=0} and \texttt{IBRION=-1}.

The PAW datasets were \texttt{PAW\_PBE Re 17Jan2003} and \texttt{PAW\_PBE O 08Apr2002} for both compounds, together with \texttt{PAW\_PBE Mg 05Jan2001} for \Mg or \texttt{PAW\_PBE Zn 06Sep2000} for \Zn. Spin--orbit coupling and orbital-moment output were enabled with \texttt{LSORBIT=.TRUE.} and \texttt{LORBMOM=.TRUE.}. The site- and angular-momentum-resolved projections used \texttt{LORBIT=11}. The tabulated quantities are specifically local Re $d$-projected moments.

The crystallographic unit cell contains two Re sites. The initial moments were specified as \texttt{MAGMOM = 0 0 14 0 0 -14 6*0 24*0}, placing the two initial Re spin moments antiparallel along Cartesian $z$. These values define the starting magnetic state, not constraints on the converged moments. No constrained-moment calculation was imposed. No explicit \texttt{SAXIS} was supplied, so the default spinor convention was used and the reported components are in the Cartesian $(x,y,z)$ frame. The periodicity of this two-Re-cell calculation is $\mathbf{k}=0$. It does not represent the $\mathbf{k}=(1/2,1/2,0)$ Mg magnetic state obtained from the effective Hamiltonian, which would require an enlarged magnetic cell.

\Needspace{8\baselineskip}
\subsection*{Electronic spectra and gap determination}
PBE without SOC gives metallic spectra for both compounds. HSE without SOC opens a gap of approximately $1$~eV, and HSE06+SOC retains the gapped electronic structure. The gaps quoted in the main paper are approximately $0.99$~eV for \Mg and $1.22$~eV for \Zn. They were determined from the separation between the valence-band maximum and conduction-band minimum; the DOS provides a consistent check for the absence of states in this energy interval. These are theoretical gaps, not experimental measurements. Their presence already without SOC means that the gap cannot be attributed specifically to switching on SOC. The predominantly O-derived occupied weight and Re-derived unoccupied weight are broadly similar to the DFT+Hubbard-I spectra in Sec.~\siref{hi}.

\Needspace{8\baselineskip}
\subsection*{Site-projected spin and orbital moments}
Table~\ref{tab:si-hse-moments} gives the converged Re spin and orbital components. Although the initial spin moments point along Cartesian $z$, the self-consistent Mg solution acquires substantial $x$ components without an imposed directional constraint. The Mg spin vectors nearly cancel when summed over the two Re sites, with a sum of approximately $(0.001,0.002,0)\,\mu_{\rm B}$; the two orbital vectors also nearly cancel. Separately, at a given Re site, opposing spin and orbital contributions strongly reduce the local residual moment. These are distinct cancellations: one concerns the two sites, and the other concerns spin and orbital contributions on a single site.

\begin{table}[htbp]
\caption{Local Re $d$-projected spin ($M$) and orbital ($L$) magnetic moments in $\mu_{\rm B}$ from HSE06+SOC calculations on the relaxed structures in the two-Re crystallographic cell. The tabulated $x$, $y$ and $z$ components refer to Cartesian
axes, with $y$ parallel to the crystallographic $b$ axis.}\label{tab:si-hse-moments}
\centering\begin{ruledtabular}\begin{tabular}{llrrrr}
Compound &Component&$M$(Re1)&$M$(Re2)&$L$(Re1)&$L$(Re2)\\
\Mg &$x$&$-0.615$&$+0.616$&$+0.603$&$-0.603$\\
 &$y$&$+0.001$&$+0.001$&$-0.000$&$-0.000$\\
 &$z$&$+0.268$&$-0.268$&$-0.216$&$+0.216$\\
\Zn &$x$&$-0.140$&$+0.050$&$+0.199$&$-0.130$\\
 &$y$&$-0.052$&$+0.053$&$+0.003$&$-0.003$\\
 &$z$&$+0.103$&$-0.085$&$+0.015$&$-0.033$\\
\end{tabular}\end{ruledtabular}
\end{table}

For Mg Re1, the sum of the printed spin and orbital components is $(-0.012,0.001,0.052)\,\mu_{\rm B}$, with a magnitude of approximately $0.053\,\mu_{\rm B}$. This is a residual Re moment in \Mg, not a moment on the Mg ion or a bulk magnetization. In \Zn, the two Re spin vectors are non-antiparallel and sum to $(-0.090,0.001,0.018)\,\mu_{\rm B}$. This describes a noncollinear Re-spin solution in the two-Re, $\mathbf{k}=0$ cell. It does not, by itself, establish that the spin-plus-orbital magnetic moments have the same directions or symmetry as the physical moments of the effective-Hamiltonian $\Gamma_4$ state.

Finite Re spin and orbital contributions have also been analysed in other Re-containing magnetic oxides~\cite{Vaitheeswaran2005,Vaitheeswaran2006,Vaitheeswaran2008,Winkler2009}. Those double-perovskite examples provide context for Re spin--orbital physics, not independent evidence for an octupolar order parameter in the present wolframites. The HSE results are used here as complementary evidence for the calculated electronic structure and local spin--orbital compensation; the octupolar assignment follows from the effective-Hamiltonian and multipolar-operator analysis.

A field estimate based only on the spin-magnetization density from a CHGCAR file omits the large orbital contribution in this compensated system. The site-integrated orbital moments in the OUTCAR output are also insufficient to reconstruct the spatial orbital-current field at an interstitial muon. Accordingly, no spin-density-only estimate is presented as the total local field, and the HSE results are used as complementary electronic-structure and compensation evidence rather than as an independent determination of multipolar order.

\subsection*{Functional comparison and orbital-resolved spectral weight}
\begin{figure}[!htbp]
\centering\includegraphics[width=0.94\textwidth]{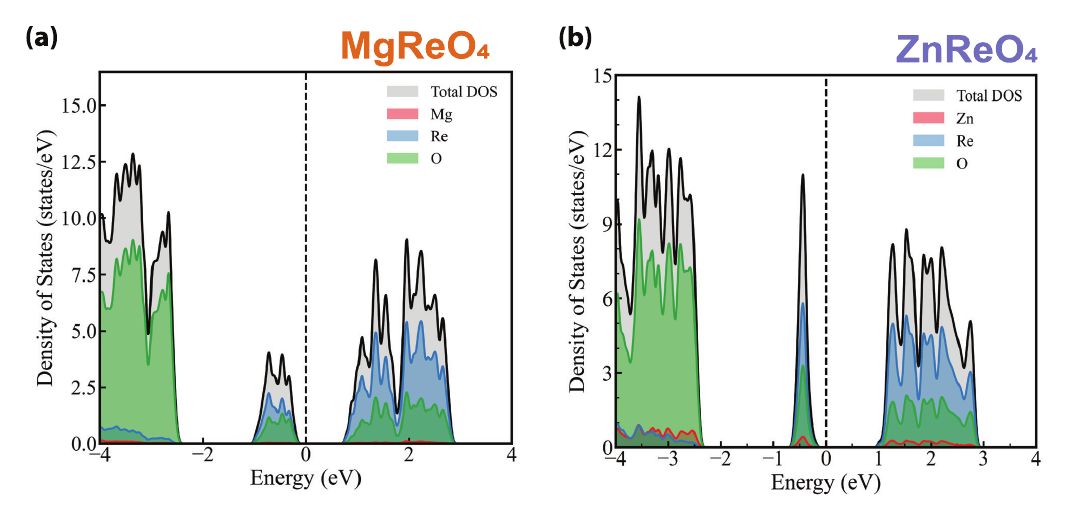}
\caption{HSE06+SOC total and atom-projected densities of states for (a) \Mg and (b) \Zn. The calculations use the relaxed structures and the two-Re crystallographic magnetic cell. Their gaps are theoretical predictions; spin and orbital compensation does not independently identify the multipolar rank.}\label{fig:si-hse-soc}
\end{figure}

\begin{figure}[!htbp]
\centering\includegraphics[width=0.88\textwidth]{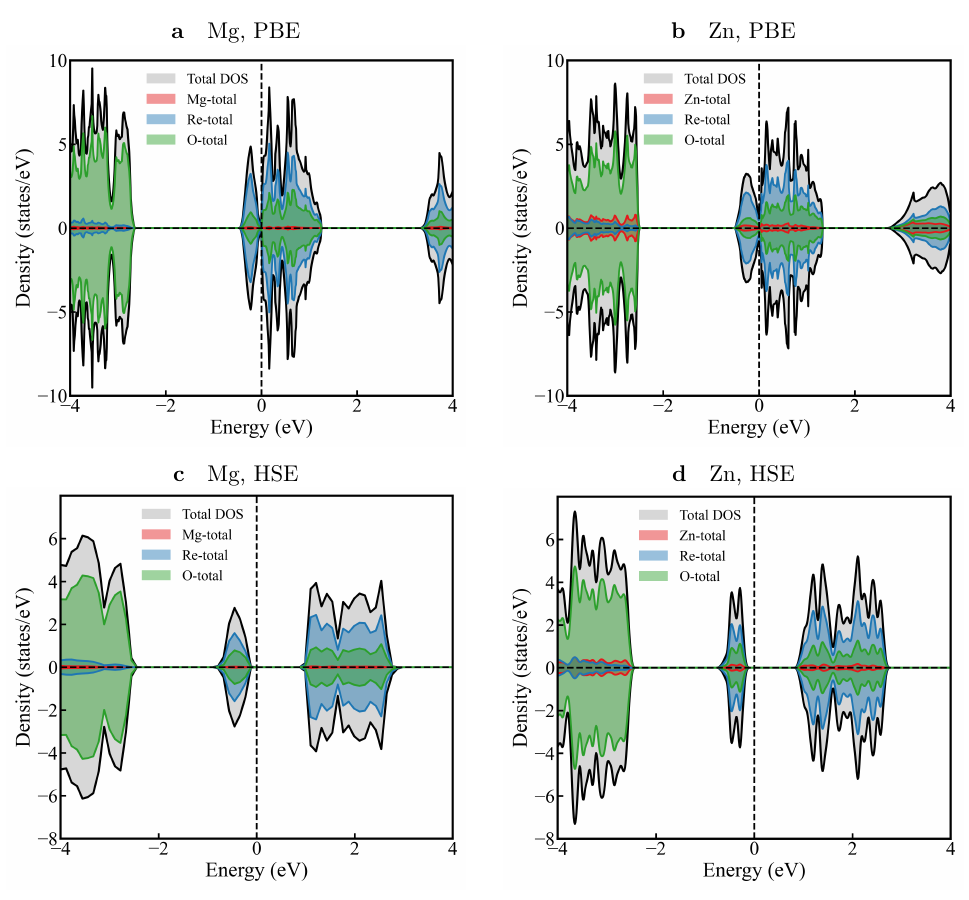}
\caption{Total and projected densities of states without SOC: PBE for (a) Mg and (b) Zn, and HSE for (c) Mg and (d) Zn. These panels use the relaxed structures; the separate PBE geometry comparison is given in Fig.~\ref{fig:si-pbe-relaxation}. PBE is metallic, whereas HSE is already gapped without SOC. Mirrored curves are the original spin-resolved no-SOC output and should not be read as spin channels of the noncollinear SOC calculation.}\label{fig:si-functional}
\end{figure}

\begin{figure}[!htbp]
\centering\includegraphics[width=0.99\textwidth]{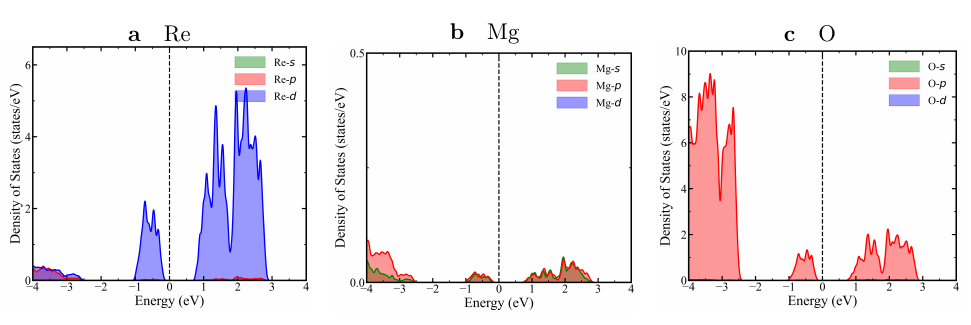}
\caption{ HSE06+SOC orbital-resolved densities of states for \Mg: (a) Re, (b) Mg and (c) O contributions. These exports use the relaxed structure. The original vertical scales are retained. These exports resolve atom-centred $s$, $p$ and $d$ contributions; they are not a measurement of a magnetic order parameter.}\label{fig:si-orbitalmg}
\medskip
\includegraphics[width=0.99\textwidth]{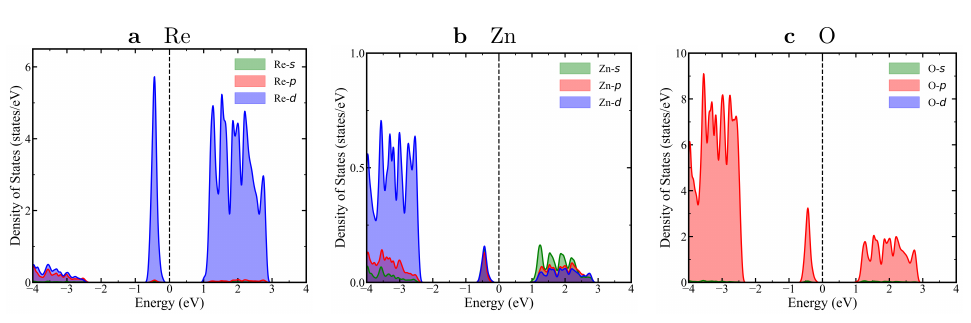}
\caption{HSE06+SOC orbital-resolved densities of states for \Zn: (a) Re, (b) Zn and (c) O contributions. These exports use the relaxed structure. The substantial occupied Zn-$d$ spectral weight is distinct from the low-energy Re-derived magnetic degrees of freedom.}\label{fig:si-orbitalzn}
\end{figure}

Figure~\ref{fig:si-functional} preserves the total and projected DOS from the semilocal and hybrid calculations without SOC. The mirrored positive/negative curves are the spin-resolved output of those calculations; they must not be interpreted as separate spin channels of the noncollinear SOC calculation. Figures~\ref{fig:si-orbitalmg} and \ref{fig:si-orbitalzn} give the separately supplied orbital-resolved weight. Re-$d$ and O-$p$ states dominate the respective manifolds, with smaller Mg-$p$ and Zn-$d$ contributions. Figure~\ref{fig:si-hse-soc} retains the HSE06+SOC comparison, also shown in the main text.

\Needspace{8\baselineskip}
\subsection*{Ferromagnetic–antiferromagnetic energy comparison}
The ferromagnetic--antiferromagnetic energy comparison used spin-polarized, collinear PBE calculations without SOC, with a plane-wave cutoff of $650$~eV and Gaussian smearing of $0.05$~eV. The moments were allowed to converge self-consistently without moment constraints.

The energy difference is defined as $\Delta E_{\mathrm{FM-AFM1}}=E_{\mathrm{FM}}-E_{\mathrm{AFM1}}$, using the same relaxed crystal structure and identical computational settings for the two configurations of each compound. The reported difference is per crystallographic cell containing two Re sites and is not divided further by the number of Re atoms. A positive value therefore places AFM1 below FM among the two tested configurations. The differences are $19.56$~meV for \Mg and $0.10\,\mu\mathrm{eV}=1.0\times10^{-7}$~eV for \Zn.

For \Zn, the very small splitting was re-examined with an electronic convergence criterion of $10^{-8}$~eV and a denser $7\times6\times7$ $\mathbf{k}$-point mesh, compared with $10^{-6}$~eV and $5\times4\times5$ in the original calculation. The reported difference remained approximately $0.10\,\mu\mathrm{eV}$. It is retained as an exploratory estimate of near-degeneracy between the two tested collinear configurations, not a quantitative ordering-energy scale. This comparison is distinct from the unconstrained HSE06+SOC local-moment calculation above. It does not establish the canted ground state or identify an octupolar order parameter.

\suppsection{hi}{DFT+Hubbard-I electronic structure and crystal field splitting}

\begin{figure}[!htbp]
\centering\includegraphics[width=0.88\textwidth]{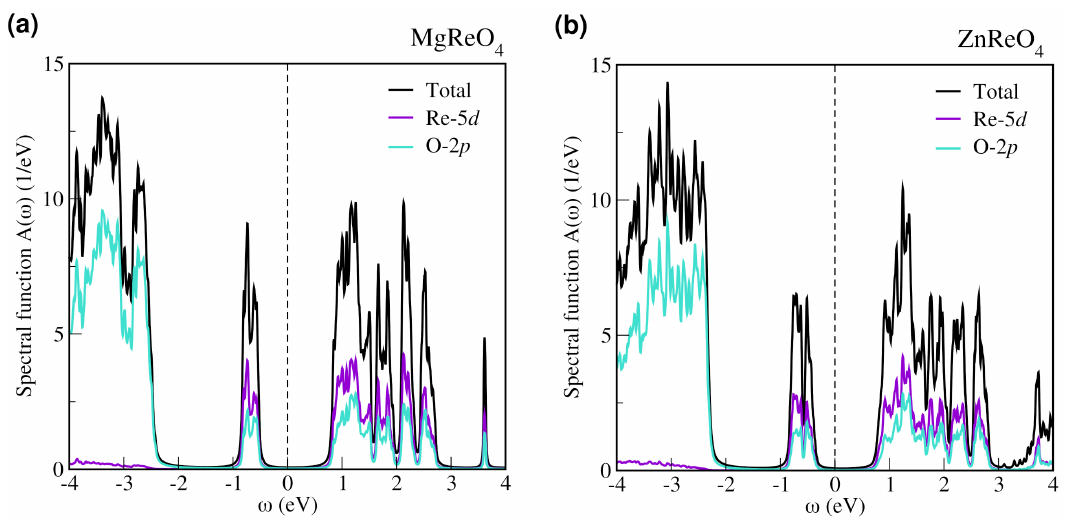}
\caption{Paramagnetic DFT+Hubbard-I one-electron spectral functions for (a) \Mg and (b) \Zn, calculated with $U=2.2$~eV and $J_{\rm H}=0.5$~eV. The relative band positions and gap scales can be compared with the HSE06+SOC results in the main text.}\label{fig:si-hi}
\end{figure}

Figure~\ref{fig:si-hi} shows the one-electron spectral functions of \Mg and \Zn calculated in DFT+Hubbard-I. Their gap scales and relative band positions agree well with the HSE06+SOC densities of states in the main text. Some differences are expected because DFT+Hubbard-I is evaluated in the paramagnetic phase, whereas the HSE result is a magnetic solution, and because the two methods treat correlations and multiplet effects differently. 

The structural inputs also differ: DFT+Hubbard-I uses the neutron-refined $7$~K structures, whereas the HSE06+SOC spectra use the PBE-relaxed structures described in Sec.~\siref{hse}.

The two lowest crystal-field doublets are predicted to be split by $196$~meV in \Mg and $190$~meV in \Zn. The excited predominantly $J=1/2$ spin--orbit doublet lies about $680$~meV above the ground doublet. The approximately $200$~meV splitting is appreciable compared with this spin--orbit scale, allowing a non-negligible admixture of the excited doublet into the formally $J=3/2$ four-state subspace. The level scheme is shown in the main electronic-structure figure.


\begin{table}[htbp]
\caption{Calculated spin--orbit coupling $\lambda_{\rm SO}$ and crystal-field parameters  transcribed in the calculation's $A_k^q\langle r^k\rangle$ convention (in meV). 
}\label{tab:si-cf}
\centering\begin{ruledtabular}\begin{tabular}{lrr}
Parameter &\Mg&\Zn\\
$\lambda_{\rm SO}$&290&286\\
$A_2^0\langle r^2\rangle$&$-75$&$-75$\\
$A_2^1\langle r^2\rangle$&859&1186\\
$A_2^2\langle r^2\rangle$&583&582\\
$A_4^0\langle r^4\rangle$&$-184$&$-175$\\
$A_4^1\langle r^4\rangle$&$-5986$&$-5884$\\
$A_4^2\langle r^4\rangle$&1574&1497\\
$A_4^3\langle r^4\rangle$&$-13162$&$-12595$\\
$A_4^4\langle r^4\rangle$&$-3336$&$-3255$\\
\end{tabular}\end{ruledtabular}
\end{table}

The calculated crystal-field parameters are listed in Table~\ref{tab:si-cf}. These parameters, the exchange matrices and the order parameters use one common global frame with $x||a$, $y||b$, and $z||x\times y$. This global frame does not coincide with a local point-group frame at each Re site, consequently, the crystal-field parameters with odd $q$ can be nonzero in this global frame. Most crystal-field parameters are close between the two materials, while $A_2^1\langle r^2\rangle$ differs more substantially and changes the composition of the crystal-field doublets. 

\suppsection{basis}{Multipolar operators and the low-energy pseudospin basis}
The operators $O_{KQ}$ are Hermitian spherical tensors for $J=3/2$, defined with the convention of Eq.~(10) of Ref.~\cite{Santini2009} and normalized such that $\Tr(O_{KQ}O_{K'Q'})=\delta_{KK'}\delta_{QQ'}$. Ranks $K=1,2,3$, which correspond to dipole, quadrupoles, octupoles, respectively, with $Q=-K, -K+1,...,K$ host the fifteen traceless operators of the four-state manifold. The rank labels describe this specified pseudospin representation, not directly a multipole expansion of a spatial magnetic field.

Because of the appreciable crystal-field mixing, a suitable pseudospin basis is constructed by diagonalizing the physical $d$-shell operator $\hat j_z=\hat l_z+\hat s_z$ in the subspace of the two lowest doublets. The resulting eigenvalues are close to, but not exactly, $\pm3/2$ and $\pm1/2$; in \Mg the reported values are $\pm1.478$ and $\pm0.482$. These eigenstates are labeled $\ket{\widetilde{3/2};M}$, with $M=\pm3/2,\pm1/2$ as pseudospin labels. Their phases are chosen so that the adjacent $\hat j_+$ matrix elements are positive and real.

In this basis the ground-state doublets are
\begin{align}
 \ket{\pm0}_{\rm Mg}={}&\pm0.587\ket{\widetilde{3/2};\pm3/2}
 -0.548\ket{\widetilde{3/2};\pm1/2}\nonumber\\
 &\quad\pm0.569\ket{\widetilde{3/2};\mp1/2}
 -0.177\ket{\widetilde{3/2};\mp3/2},\label{eq:si-mg-ground}\\
 \ket{\pm0}_{\rm Zn}={}&\pm0.699\ket{\widetilde{3/2};\pm3/2}
 -0.405\ket{\widetilde{3/2};\pm1/2}\nonumber\\
 &\quad\pm0.579\ket{\widetilde{3/2};\mp1/2}
 -0.112\ket{\widetilde{3/2};\mp3/2}.
 \label{eq:si-zn-ground}
\end{align}
The upper and lower signs label the two Kramers partners. The excited doublets are
\begin{align}
\ket{\pm1}_{\rm Mg}={}&0.072\ket{\widetilde{3/2};\pm3/2}
 \mp 0.518\ket{\widetilde{3/2};\pm1/2}\nonumber\\
 &\quad-0.328\ket{\widetilde{3/2};\mp1/2}
 \pm 0.787\ket{\widetilde{3/2};\mp3/2},\label{eq:si-mg-excited}\\
\ket{\pm1}_{\rm Zn}={}&\pm0.698\ket{\widetilde{3/2};\pm3/2}+0.406\ket{\widetilde{3/2};\pm1/2}\nonumber\\
&\mp0.579\ket{\widetilde{3/2};\mp1/2}-0.112\ket{\widetilde{3/2};\mp3/2}.\label{eq:si-zn-excited}
\end{align}

With the coefficients in the wavefunctions above rounded to three digits, orthonormality is preserved to the same precision.

Together with inter-doublet splittings of $196$ and $190$~meV for \Mg and \Zn, respectively, the ground-doublet and excited state wavefunctions (\ref{eq:si-mg-ground}-\ref{eq:si-zn-excited}) specify the crystal-field term $H_{\mathrm{cf}}$  in the four-state subspace. The ground state doublets (\ref{eq:si-mg-ground}) and (\ref{eq:si-zn-ground}) define the projection $P$ to the ground-state doublet space, eq.~3 of the main text.

\suppsection{exchange}{Intersite multipolar exchange interactions}
Intersite exchange interactions between Re $d^1$ pseudospins were evaluated using the FT-HI method~\cite{Pourovskii2016}. Starting from the paramagnetic DFT+Hubbard-I electronic structure, the method considers small trace-conserving fluctuations of the on-site ground-manifold density matrices. A fluctuation of a matrix element is written $\epsilon\rho_{M_1M_2}$, where $\epsilon$ is small and the fluctuation is restricted to the relevant low-energy manifold. The resulting exchange matrix element is
\begin{equation}
 \bra{M_1^i M_3^j}H_{\rm IEI}\ket{M_2^i M_4^j}
 =\Tr\left[G_{ij}\frac{\delta\Sigma_j}{\delta\rho^j_{M_3M_4}}
 G_{ji}\frac{\delta\Sigma_i}{\delta\rho^i_{M_1M_2}}\right],
 \label{eq:si-ft-hi}
\end{equation}
with the trace 
taken over the Matsubara frequencies and the correlated shell spin and orbital indices. $G_{ij}$ is the one-electron propagator between the Re shells and $\Sigma_i$ is the local Hubbard-I self-energy. The self-energy derivatives are evaluated analytically in the Hubbard-I approximation. Once these matrix elements are obtained, the exchange tensor $V^{QQ'}_{KK'}(ij)$ follows by expansion in the orthonormal multipolar operators.

\begin{figure}[!htbp]
\centering\includegraphics[width=0.94\textwidth]{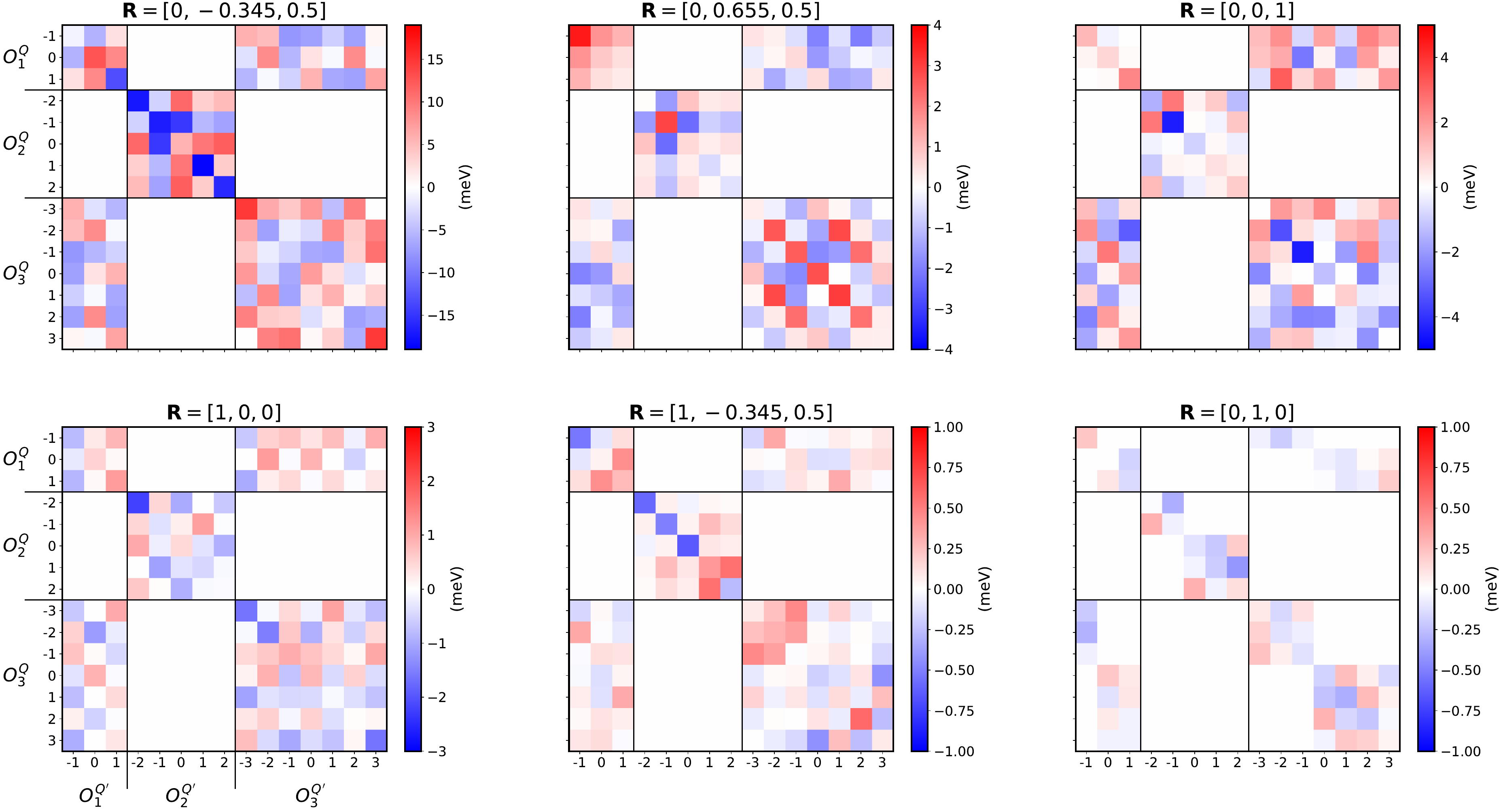}
\caption{Calculated intersite exchange tensors for \Mg. Bond vectors in fractional lattice units are shown above the matrices; color scales are in meV. The operators are ordered from top to bottom in order of increasing rank: dipole ($K=1$), quadrupole ($K=2$), and octupole ($K=3$), and, within each block, by increasing projection $Q=-K,\ldots,K$.} \label{fig:si-exchange-mg}
\vspace{1em}
\includegraphics[width=0.94\textwidth]{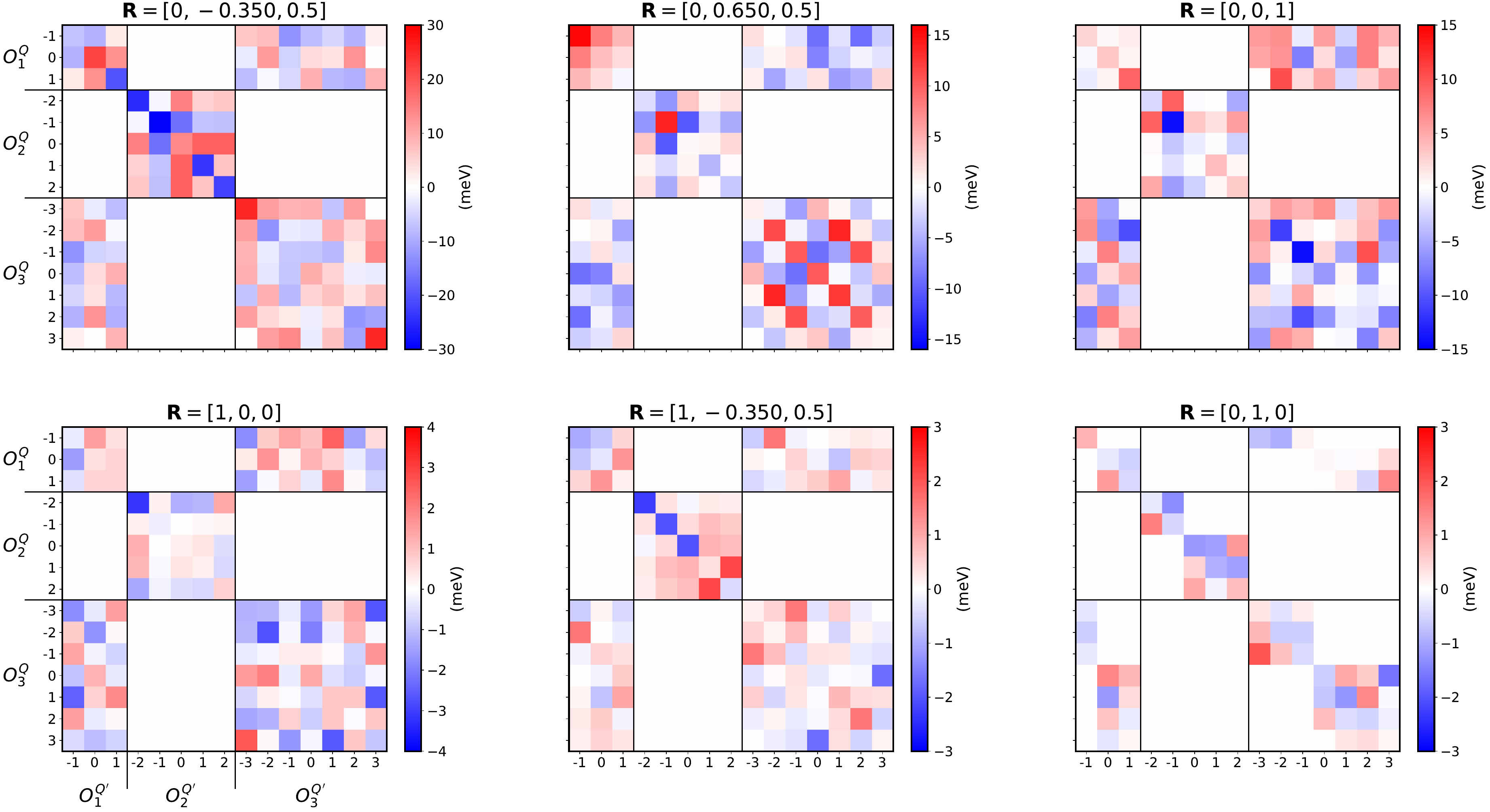}
\caption{Corresponding intersite exchange tensors for \Zn. The operator ordering and conventions are the same as in Fig.~\ref{fig:si-exchange-mg}. 
}\label{fig:si-exchange-zn}
\end{figure}

The calculation includes the first twelve neighbors, grouped into six coordination shells. Each bond has a real $15\times15$ exchange matrix. The matrix for the opposite vector $-\mathbf R$ is the transpose of that for $\mathbf R$. Time-reversal symmetry of the paramagnetic reference state forbids bilinear couplings between time-even quadrupoles and time-odd dipoles or octupoles, while the low spatial symmetry allows many other components to be nonzero. The calculated matrices are displayed as color maps in Figs.~\ref{fig:si-exchange-mg} and \ref{fig:si-exchange-zn}; their essential matrix elements are listed in Tables~\ref{tab:MgReO4-exchange} and ~\ref{tab:ZnReO4-exchange}. 
The exchange matrices are symmetric for some lattice vectors and non-symmetric for others; asymmetry arises from  asymmetric Dzyaloshinskii-Moriya-type exchange terms for the bonds for which they are allowed. All matrix elements are calculated independently using FT-HI eq.~\ref{eq:si-ft-hi}, hence, the effect of bond symmetry is not imposed in calculations, but appears nevertheless in the resulting exchange matrices.   

The overall magnitudes of the dipole and octupole exchange blocks are comparable. Octupolar dominance in the ordered state must therefore not be attributed simply to negligible dipolar exchange. The strongest coupling in each material is associated with the shortest Re--Re bond joining edge-sharing octahedra. In \Mg it is approximately four times the second-neighbor scale, whereas the interactions decay less rapidly with distance in \Zn. Beyond the fourth coordination shell, the terms are roughly an order of magnitude smaller than the strongest nearest-neighbor terms. This exchange hierarchy and the different longer-range interactions distinguish the two effective Hamiltonians.

\begin{table}[p]
\centering
\caption{Intersite exchange couplings of MgReO$_4$ (meV). Rows and columns follow $(K,Q)$ and $(K^{\prime},Q^{\prime})$, respectively, with $Q=-K,\ldots,K$. Blank entries have $|V|<0.05$ meV; rows below this cutoff for every bond are omitted. Values are rounded to two decimals.}
\label{tab:MgReO4-exchange}
\begingroup
\fontsize{7}{8}\selectfont
\setlength{\tabcolsep}{1.8pt}
\renewcommand{\arraystretch}{0.95}
\begin{minipage}[t]{0.49\linewidth}\centering\vspace{0pt}
\begin{tabular}{@{}rrrrrrrrrr@{}}
\toprule
$K$ & $Q$ & $K^{\prime}$ & $Q^{\prime}$ & $R_1$ & $R_2$ & $R_3$ & $R_4$ & $R_5$ & $R_6$ \\
\midrule
\multicolumn{10}{c}{Dipole--dipole} \\
$1$ & $-1$ & $1$ & $-1$ & $-0.84$ & $3.59$ & $1.38$ & $-0.80$ & $-0.53$ & $0.22$ \\
$1$ & $-1$ & $1$ & $0$ & $-5.66$ & $1.69$ & $-0.23$ & $0.28$ & $-0.10$ &  \\
$1$ & $-1$ & $1$ & $1$ & $2.36$ & $1.17$ &  & $0.86$ & $0.13$ &  \\
$1$ & $0$ & $1$ & $-1$ & $-5.66$ & $1.69$ & $0.23$ & $-0.28$ & $-0.10$ &  \\
$1$ & $0$ & $1$ & $0$ & $12.72$ & $0.85$ & $0.80$ & $0.51$ & $0.05$ &  \\
$1$ & $0$ & $1$ & $1$ & $8.77$ & $0.51$ & $0.10$ & $0.08$ & $0.44$ & $-0.17$ \\
$1$ & $1$ & $1$ & $-1$ & $2.36$ & $1.17$ &  & $-0.86$ & $0.13$ &  \\
$1$ & $1$ & $1$ & $0$ & $8.77$ & $0.51$ & $0.10$ & $0.08$ & $0.44$ & $0.10$ \\
$1$ & $1$ & $1$ & $1$ & $-13.14$ & $0.33$ & $2.35$ & $1.15$ & $0.27$ & $-0.15$ \\
\midrule
\multicolumn{10}{c}{Dipole--octupole} \\
$1$ & $-1$ & $3$ & $-3$ & $5.82$ & $0.41$ & $1.29$ & $-0.64$ & $-0.16$ & $-0.06$ \\
$1$ & $-1$ & $3$ & $-2$ & $4.95$ & $0.23$ & $2.17$ & $0.54$ & $0.34$ & $-0.20$ \\
$1$ & $-1$ & $3$ & $-1$ & $-7.78$ & $-0.50$ & $-0.77$ & $0.69$ &  &  \\
$1$ & $-1$ & $3$ & $0$ & $-7.03$ & $-1.92$ & $1.78$ & $0.31$ &  &  \\
$1$ & $-1$ & $3$ & $1$ & $-3.63$ & $-0.49$ & $-0.80$ & $0.75$ & $0.07$ &  \\
$1$ & $-1$ & $3$ & $2$ & $-7.08$ & $-2.00$ & $2.38$ & $-0.18$ &  &  \\
$1$ & $-1$ & $3$ & $3$ & $0.66$ & $-0.83$ & $1.75$ & $0.94$ & $0.10$ &  \\
$1$ & $0$ & $3$ & $-3$ & $-2.27$ & $-0.31$ & $1.15$ &  &  &  \\
$1$ & $0$ & $3$ & $-2$ & $8.59$ & $0.15$ & $1.66$ & $1.16$ &  &  \\
$1$ & $0$ & $3$ & $-1$ & $-5.45$ & $0.59$ & $-2.64$ & $-0.05$ & $0.13$ &  \\
$1$ & $0$ & $3$ & $0$ & $2.09$ & $-1.60$ & $0.27$ & $0.89$ & $-0.13$ & $-0.06$ \\
$1$ & $0$ & $3$ & $1$ & $-0.54$ & $-0.82$ & $-1.78$ &  & $-0.12$ & $-0.10$ \\
$1$ & $0$ & $3$ & $2$ & $8.49$ & $-0.13$ & $1.92$ & $-0.52$ & $0.11$ &  \\
$1$ & $0$ & $3$ & $3$ & $-0.56$ & $-0.36$ & $0.38$ &  & $0.14$ & $0.09$ \\
$1$ & $1$ & $3$ & $-3$ & $-5.47$ & $0.34$ & $-0.66$ & $-1.00$ & $-0.13$ &  \\
$1$ & $1$ & $3$ & $-2$ & $-0.53$ & $-1.31$ & $3.19$ & $0.23$ & $-0.09$ &  \\
$1$ & $1$ & $3$ & $-1$ & $-3.36$ & $-0.49$ & $0.82$ & $0.46$ & $0.12$ &  \\
$1$ & $1$ & $3$ & $0$ & $5.56$ & $0.53$ & $1.89$ & $-0.05$ &  &  \\
$1$ & $1$ & $3$ & $1$ & $-6.52$ & $-1.34$ & $-0.29$ & $0.44$ & $0.33$ & $-0.10$ \\
$1$ & $1$ & $3$ & $2$ & $-7.08$ & $-1.21$ & $0.29$ &  & $0.06$ & $-0.06$ \\
$1$ & $1$ & $3$ & $3$ & $6.82$ & $0.36$ & $2.00$ & $0.30$ &  & $0.20$ \\
\midrule
\multicolumn{10}{c}{Quadrupole--quadrupole} \\
$2$ & $-2$ & $2$ & $-2$ & $-17.27$ &  & $-1.52$ & $-2.20$ & $-0.59$ &  \\
$2$ & $-2$ & $2$ & $-1$ & $-3.28$ & $-1.56$ & $2.64$ & $0.49$ & $0.06$ & $-0.31$ \\
$2$ & $-2$ & $2$ & $0$ & $11.23$ & $0.94$ & $0.24$ & $-0.99$ &  &  \\
$2$ & $-2$ & $2$ & $1$ & $3.69$ & $0.34$ & $1.03$ &  &  &  \\
$2$ & $-2$ & $2$ & $2$ & $5.14$ & $0.43$ & $-1.34$ & $-0.65$ &  &  \\
$2$ & $-1$ & $2$ & $-2$ & $-3.28$ & $-1.56$ & $2.64$ & $0.49$ & $0.06$ & $0.31$ \\
$2$ & $-1$ & $2$ & $-1$ & $-16.79$ & $2.93$ & $-4.45$ & $-0.35$ & $-0.50$ & $-0.06$ \\
$2$ & $-1$ & $2$ & $0$ & $-14.77$ & $-2.30$ & $0.05$ & $0.20$ & $0.05$ &  \\
$2$ & $-1$ & $2$ & $1$ & $-5.33$ & $-0.73$ & $-0.22$ & $1.11$ & $0.26$ &  \\
$2$ & $-1$ & $2$ & $2$ & $-6.78$ & $-0.99$ & $1.10$ &  & $0.14$ &  \\
$2$ & $0$ & $2$ & $-2$ & $11.23$ & $0.94$ & $-0.24$ & $0.99$ &  &  \\
$2$ & $0$ & $2$ & $-1$ & $-14.77$ & $-2.30$ & $-0.05$ & $-0.20$ & $0.05$ &  \\
$2$ & $0$ & $2$ & $0$ & $5.55$ & $0.60$ & $-0.89$ & $0.46$ & $-0.65$ & $-0.11$ \\
$2$ & $0$ & $2$ & $1$ & $10.14$ & $0.29$ & $0.13$ & $-0.32$ & $0.10$ & $-0.21$ \\
$2$ & $0$ & $2$ & $2$ & $12.02$ & $0.49$ & $-0.31$ & $-0.94$ & $0.07$ & $0.20$ \\
$2$ & $1$ & $2$ & $-2$ & $3.69$ & $0.34$ & $-1.03$ &  &  &  \\
$2$ & $1$ & $2$ & $-1$ & $-5.33$ & $-0.73$ & $0.22$ & $-1.11$ & $0.26$ &  \\
$2$ & $1$ & $2$ & $0$ & $10.14$ & $0.29$ & $0.13$ & $-0.32$ & $0.10$ &  \\
$2$ & $1$ & $2$ & $1$ & $-18.64$ & $-0.57$ & $0.60$ & $-0.48$ & $0.40$ & $-0.20$ \\
$2$ & $1$ & $2$ & $2$ & $3.77$ & $0.11$ & $0.30$ & $-0.07$ & $0.56$ & $-0.41$ \\
$2$ & $2$ & $2$ & $-2$ & $5.14$ & $0.43$ & $1.34$ & $0.65$ &  &  \\
$2$ & $2$ & $2$ & $-1$ & $-6.78$ & $-0.99$ & $-1.10$ &  & $0.14$ &  \\
$2$ & $2$ & $2$ & $0$ & $12.02$ & $0.49$ & $-0.31$ & $-0.94$ & $0.07$ & $0.31$ \\
$2$ & $2$ & $2$ & $1$ & $3.77$ & $0.11$ & $0.30$ & $-0.07$ & $0.56$ &  \\
$2$ & $2$ & $2$ & $2$ & $-15.97$ & $-0.46$ & $0.98$ & $-0.06$ & $-0.27$ & $0.13$ \\
\bottomrule
\end{tabular}
\end{minipage}\hfill
\begin{minipage}[t]{0.49\linewidth}\centering\vspace{0pt}
\begin{tabular}{@{}rrrrrrrrrr@{}}
\toprule
$K$ & $Q$ & $K^{\prime}$ & $Q^{\prime}$ & $R_1$ & $R_2$ & $R_3$ & $R_4$ & $R_5$ & $R_6$ \\
\midrule
\multicolumn{10}{c}{Octupole--dipole} \\
$3$ & $-3$ & $1$ & $-1$ & $5.82$ & $0.41$ & $1.29$ & $-0.64$ & $-0.16$ & $-0.21$ \\
$3$ & $-3$ & $1$ & $0$ & $-2.27$ & $-0.31$ & $-1.15$ &  &  &  \\
$3$ & $-3$ & $1$ & $1$ & $-5.47$ & $0.34$ & $0.66$ & $1.00$ & $-0.13$ &  \\
$3$ & $-2$ & $1$ & $-1$ & $4.95$ & $0.23$ & $2.17$ & $0.54$ & $0.34$ & $-0.30$ \\
$3$ & $-2$ & $1$ & $0$ & $8.59$ & $0.15$ & $-1.66$ & $-1.16$ &  &  \\
$3$ & $-2$ & $1$ & $1$ & $-0.53$ & $-1.31$ & $-3.19$ & $-0.23$ & $-0.09$ &  \\
$3$ & $-1$ & $1$ & $-1$ & $-7.78$ & $-0.50$ & $-0.77$ & $0.69$ &  & $-0.06$ \\
$3$ & $-1$ & $1$ & $0$ & $-5.45$ & $0.59$ & $2.64$ & $0.05$ & $0.13$ &  \\
$3$ & $-1$ & $1$ & $1$ & $-3.36$ & $-0.49$ & $-0.82$ & $-0.46$ & $0.12$ &  \\
$3$ & $0$ & $1$ & $-1$ & $-7.03$ & $-1.92$ & $-1.78$ & $-0.31$ &  &  \\
$3$ & $0$ & $1$ & $0$ & $2.09$ & $-1.60$ & $0.27$ & $0.89$ & $-0.13$ & $0.21$ \\
$3$ & $0$ & $1$ & $1$ & $5.56$ & $0.53$ & $1.89$ & $-0.05$ &  & $0.09$ \\
$3$ & $1$ & $1$ & $-1$ & $-3.63$ & $-0.49$ & $0.80$ & $-0.75$ & $0.07$ &  \\
$3$ & $1$ & $1$ & $0$ & $-0.54$ & $-0.82$ & $-1.78$ &  & $-0.12$ & $-0.11$ \\
$3$ & $1$ & $1$ & $1$ & $-6.52$ & $-1.34$ & $-0.29$ & $0.44$ & $0.33$ & $0.10$ \\
$3$ & $2$ & $1$ & $-1$ & $-7.08$ & $-2.00$ & $-2.38$ & $0.18$ &  &  \\
$3$ & $2$ & $1$ & $0$ & $8.49$ & $-0.13$ & $1.92$ & $-0.52$ & $0.11$ & $0.08$ \\
$3$ & $2$ & $1$ & $1$ & $-7.08$ & $-1.21$ & $0.29$ &  & $0.06$ &  \\
$3$ & $3$ & $1$ & $-1$ & $0.66$ & $-0.83$ & $-1.75$ & $-0.94$ & $0.10$ &  \\
$3$ & $3$ & $1$ & $0$ & $-0.56$ & $-0.36$ & $0.38$ &  & $0.14$ &  \\
$3$ & $3$ & $1$ & $1$ & $6.82$ & $0.36$ & $2.00$ & $0.30$ &  &  \\
\midrule
\multicolumn{10}{c}{Octupole--octupole} \\
$3$ & $-3$ & $3$ & $-3$ & $14.83$ & $0.30$ &  & $-1.62$ & $0.08$ & $0.09$ \\
$3$ & $-3$ & $3$ & $-2$ & $6.25$ & $-0.23$ & $1.98$ & $-0.09$ & $0.24$ & $-0.15$ \\
$3$ & $-3$ & $3$ & $-1$ & $4.05$ & $-1.20$ & $1.20$ & $0.46$ & $0.46$ & $0.12$ \\
$3$ & $-3$ & $3$ & $0$ & $7.66$ & $0.94$ & $2.31$ & $-0.14$ & $-0.09$ &  \\
$3$ & $-3$ & $3$ & $1$ & $-4.80$ & $0.14$ & $-0.21$ & $1.09$ & $0.17$ &  \\
$3$ & $-3$ & $3$ & $2$ & $9.43$ & $-0.84$ & $0.65$ & $-0.29$ & $-0.08$ &  \\
$3$ & $-3$ & $3$ & $3$ & $0.05$ &  & $1.53$ & $-0.77$ &  &  \\
$3$ & $-2$ & $3$ & $-3$ & $6.25$ & $-0.23$ & $1.98$ & $-0.09$ & $0.24$ & $0.17$ \\
$3$ & $-2$ & $3$ & $-2$ & $-7.08$ & $2.64$ & $-3.41$ & $-1.50$ & $0.31$ & $-0.11$ \\
$3$ & $-2$ & $3$ & $-1$ & $-1.37$ & $-0.31$ & $0.65$ & $0.65$ & $0.37$ & $-0.07$ \\
$3$ & $-2$ & $3$ & $0$ & $-2.77$ & $-1.40$ & $-0.21$ & $-0.92$ &  &  \\
$3$ & $-2$ & $3$ & $1$ & $8.66$ & $2.87$ & $1.33$ & $0.34$ & $-0.06$ &  \\
$3$ & $-2$ & $3$ & $2$ & $3.79$ & $0.33$ & $1.64$ & $-0.54$ &  &  \\
$3$ & $-2$ & $3$ & $3$ & $9.47$ & $-0.83$ & $-1.03$ & $0.44$ & $-0.07$ &  \\
$3$ & $-1$ & $3$ & $-3$ & $4.05$ & $-1.20$ & $1.20$ & $0.46$ & $0.46$ & $0.24$ \\
$3$ & $-1$ & $3$ & $-2$ & $-1.37$ & $-0.31$ & $0.65$ & $0.65$ & $0.37$ & $0.07$ \\
$3$ & $-1$ & $3$ & $-1$ & $-3.23$ & $2.55$ & $-4.45$ & $0.98$ &  & $-0.11$ \\
$3$ & $-1$ & $3$ & $0$ & $-6.43$ & $-1.82$ &  & $0.70$ &  &  \\
$3$ & $-1$ & $3$ & $1$ & $-6.86$ & $-1.54$ & $-1.92$ & $0.46$ & $0.10$ &  \\
$3$ & $-1$ & $3$ & $2$ & $3.30$ & $2.26$ & $2.40$ & $0.10$ &  &  \\
$3$ & $-1$ & $3$ & $3$ & $10.61$ & $0.40$ & $-1.16$ & $1.01$ & $-0.15$ &  \\
$3$ & $0$ & $3$ & $-3$ & $7.66$ & $0.94$ & $-2.31$ & $0.14$ & $-0.09$ &  \\
$3$ & $0$ & $3$ & $-2$ & $-2.77$ & $-1.40$ & $0.21$ & $0.92$ &  &  \\
$3$ & $0$ & $3$ & $-1$ & $-6.43$ & $-1.82$ &  & $-0.70$ &  &  \\
$3$ & $0$ & $3$ & $0$ & $7.42$ & $2.74$ & $-1.22$ & $0.82$ & $-0.18$ & $-0.17$ \\
$3$ & $0$ & $3$ & $1$ & $2.36$ &  &  & $-0.42$ & $-0.12$ & $0.25$ \\
$3$ & $0$ & $3$ & $2$ & $-2.40$ & $-0.74$ & $-2.35$ & $0.52$ & $0.11$ & $0.06$ \\
$3$ & $0$ & $3$ & $3$ & $0.51$ & $0.85$ & $-0.38$ & $-0.42$ & $-0.43$ & $-0.15$ \\
$3$ & $1$ & $3$ & $-3$ & $-4.80$ & $0.14$ & $0.21$ & $-1.09$ & $0.17$ &  \\
$3$ & $1$ & $3$ & $-2$ & $8.66$ & $2.87$ & $-1.33$ & $-0.34$ & $-0.06$ &  \\
$3$ & $1$ & $3$ & $-1$ & $-6.86$ & $-1.54$ & $1.92$ & $-0.46$ & $0.10$ &  \\
$3$ & $1$ & $3$ & $0$ & $2.36$ &  &  & $-0.42$ & $-0.12$ & $-0.24$ \\
$3$ & $1$ & $3$ & $1$ & $5.80$ & $3.06$ & $0.95$ & $-0.08$ & $0.14$ & $-0.32$ \\
$3$ & $1$ & $3$ & $2$ & $0.91$ & $-0.32$ & $-0.38$ & $-0.39$ & $-0.07$ & $0.27$ \\
$3$ & $1$ & $3$ & $3$ & $3.62$ & $-0.91$ & $-0.30$ & $-0.71$ & $0.26$ & $0.06$ \\
$3$ & $2$ & $3$ & $-3$ & $9.43$ & $-0.84$ & $-0.65$ & $0.29$ & $-0.08$ &  \\
$3$ & $2$ & $3$ & $-2$ & $3.79$ & $0.33$ & $-1.64$ & $0.54$ &  &  \\
$3$ & $2$ & $3$ & $-1$ & $3.30$ & $2.26$ & $-2.40$ & $-0.10$ &  &  \\
$3$ & $2$ & $3$ & $0$ & $-2.40$ & $-0.74$ & $-2.35$ & $0.52$ & $0.11$ & $0.29$ \\
$3$ & $2$ & $3$ & $1$ & $0.91$ & $-0.32$ & $-0.38$ & $-0.39$ & $-0.07$ & $-0.16$ \\
$3$ & $2$ & $3$ & $2$ & $-7.12$ & $2.22$ & $-0.91$ &  & $0.58$ & $-0.23$ \\
$3$ & $2$ & $3$ & $3$ & $-5.94$ & $0.29$ & $-2.13$ & $0.10$ & $-0.25$ &  \\
$3$ & $3$ & $3$ & $-3$ & $0.05$ &  & $-1.53$ & $0.77$ &  &  \\
$3$ & $3$ & $3$ & $-2$ & $9.47$ & $-0.83$ & $1.03$ & $-0.44$ & $-0.07$ &  \\
$3$ & $3$ & $3$ & $-1$ & $10.61$ & $0.40$ & $1.16$ & $-1.01$ & $-0.15$ &  \\
$3$ & $3$ & $3$ & $0$ & $0.51$ & $0.85$ & $-0.38$ & $-0.42$ & $-0.43$ &  \\
$3$ & $3$ & $3$ & $1$ & $3.62$ & $-0.91$ & $-0.30$ & $-0.71$ & $0.26$ & $0.21$ \\
$3$ & $3$ & $3$ & $2$ & $-5.94$ & $0.29$ & $-2.13$ & $0.10$ & $-0.25$ & $0.18$ \\
$3$ & $3$ & $3$ & $3$ & $14.72$ & $0.26$ &  & $-1.64$ & $0.08$ &  \\
\bottomrule
\end{tabular}
\end{minipage}
\par\medskip
$R_1=(0,-0.345,0.5)$, $R_2=(0,0.655,0.5)$, $R_3=(0,0,1)$;\\
$R_4=(1,0,0)$, $R_5=(1,-0.345,0.5)$, $R_6=(0,1,0)$. Fractional coordinates.
\endgroup
\end{table}

\begin{table}[p]
\centering
\caption{Intersite exchange couplings of ZnReO$_4$ (meV). Rows and columns follow $(K,Q)$ and $(K^{\prime},Q^{\prime})$, respectively, with $Q=-K,\ldots,K$. Blank entries have $|V|<0.05$ meV; rows below this cutoff for every bond are omitted. Values are rounded to two decimals.}
\label{tab:ZnReO4-exchange}
\begingroup
\fontsize{7}{8}\selectfont
\setlength{\tabcolsep}{1.8pt}
\renewcommand{\arraystretch}{0.95}
\begin{minipage}[t]{0.49\linewidth}\centering\vspace{0pt}
\begin{tabular}{@{}rrrrrrrrrr@{}}
\toprule
$K$ & $Q$ & $K^{\prime}$ & $Q^{\prime}$ & $R_1$ & $R_2$ & $R_3$ & $R_4$ & $R_5$ & $R_6$ \\
\midrule
\multicolumn{10}{c}{Dipole--dipole} \\
$1$ & $-1$ & $1$ & $-1$ & $-7.18$ & $15.59$ & $2.40$ & $-0.29$ & $-1.00$ & $0.91$ \\
$1$ & $-1$ & $1$ & $0$ & $-8.71$ & $8.11$ & $0.37$ & $1.52$ & $-0.68$ &  \\
$1$ & $-1$ & $1$ & $1$ & $2.51$ & $4.38$ & $1.09$ & $0.49$ & $0.49$ &  \\
$1$ & $0$ & $1$ & $-1$ & $-8.71$ & $8.11$ & $-0.37$ & $-1.52$ & $-0.68$ &  \\
$1$ & $0$ & $1$ & $0$ & $21.90$ & $4.09$ & $3.32$ & $0.50$ & $-0.29$ & $-0.27$ \\
$1$ & $0$ & $1$ & $1$ & $13.07$ & $2.22$ & $0.77$ & $0.66$ & $1.23$ & $-0.56$ \\
$1$ & $1$ & $1$ & $-1$ & $2.51$ & $4.38$ & $-1.09$ & $-0.49$ & $0.49$ &  \\
$1$ & $1$ & $1$ & $0$ & $13.07$ & $2.22$ & $0.77$ & $0.66$ & $1.23$ & $1.17$ \\
$1$ & $1$ & $1$ & $1$ & $-19.95$ & $-0.58$ & $9.03$ & $0.67$ & $0.18$ & $-0.45$ \\
\midrule
\multicolumn{10}{c}{Dipole--octupole} \\
$1$ & $-1$ & $3$ & $-3$ & $6.61$ & $2.12$ & $5.94$ & $-1.79$ & $-0.60$ & $-0.73$ \\
$1$ & $-1$ & $3$ & $-2$ & $7.68$ & $0.14$ & $6.59$ & $0.81$ & $1.59$ & $-0.96$ \\
$1$ & $-1$ & $3$ & $-1$ & $-12.78$ & $-1.63$ & $-1.15$ & $1.43$ & $-0.13$ & $0.16$ \\
$1$ & $-1$ & $3$ & $0$ & $-7.65$ & $-8.99$ & $5.55$ & $0.95$ &  &  \\
$1$ & $-1$ & $3$ & $1$ & $-4.91$ & $-1.81$ & $-2.52$ & $2.46$ & $0.13$ &  \\
$1$ & $-1$ & $3$ & $2$ & $-8.73$ & $-9.05$ & $7.41$ & $-1.47$ & $0.27$ &  \\
$1$ & $-1$ & $3$ & $3$ & $1.74$ & $-3.11$ & $4.48$ & $0.57$ & $0.19$ &  \\
$1$ & $0$ & $3$ & $-3$ & $-2.56$ & $-1.41$ & $5.17$ & $0.36$ & $0.13$ &  \\
$1$ & $0$ & $3$ & $-2$ & $11.69$ & $0.66$ & $6.42$ & $1.70$ &  &  \\
$1$ & $0$ & $3$ & $-1$ & $-5.35$ & $1.84$ & $-7.40$ & $0.18$ & $0.50$ &  \\
$1$ & $0$ & $3$ & $0$ & $4.34$ & $-7.73$ & $2.01$ & $1.20$ & $-0.12$ & $0.07$ \\
$1$ & $0$ & $3$ & $1$ & $3.33$ & $-2.80$ & $-5.42$ & $0.72$ & $-0.70$ & $-0.06$ \\
$1$ & $0$ & $3$ & $2$ & $12.88$ & $-0.74$ & $7.47$ & $-0.31$ & $0.60$ & $0.06$ \\
$1$ & $0$ & $3$ & $3$ &  & $-1.69$ & $1.63$ & $-1.02$ & $0.51$ & $0.46$ \\
$1$ & $1$ & $3$ & $-3$ & $-7.76$ & $0.98$ & $0.14$ & $-1.50$ & $-0.45$ &  \\
$1$ & $1$ & $3$ & $-2$ & $-0.82$ & $-5.56$ & $10.33$ & $-0.12$ & $-0.23$ &  \\
$1$ & $1$ & $3$ & $-1$ & $-4.54$ & $-1.66$ & $2.19$ & $0.67$ & $0.36$ &  \\
$1$ & $1$ & $3$ & $0$ & $9.24$ & $1.81$ & $4.92$ & $-0.37$ & $0.59$ &  \\
$1$ & $1$ & $3$ & $1$ & $-8.41$ & $-6.11$ & $-2.32$ & $1.81$ & $1.05$ & $0.21$ \\
$1$ & $1$ & $3$ & $2$ & $-9.16$ & $-4.68$ & $2.63$ & $0.10$ & $-0.13$ & $-0.47$ \\
$1$ & $1$ & $3$ & $3$ & $8.94$ & $2.62$ & $5.65$ & $-0.70$ & $0.29$ & $1.42$ \\
\midrule
\multicolumn{10}{c}{Quadrupole--quadrupole} \\
$2$ & $-2$ & $2$ & $-2$ & $-25.08$ & $-2.17$ & $-2.33$ & $-3.14$ & $-2.28$ & $-0.26$ \\
$2$ & $-2$ & $2$ & $-1$ & $-1.01$ & $-6.58$ & $9.27$ & $0.24$ & $0.33$ & $-1.35$ \\
$2$ & $-2$ & $2$ & $0$ & $14.99$ & $3.55$ & $-0.30$ & $-1.25$ & $-0.11$ &  \\
$2$ & $-2$ & $2$ & $1$ & $5.58$ & $0.85$ & $-0.11$ & $-1.15$ & $0.25$ &  \\
$2$ & $-2$ & $2$ & $2$ & $6.49$ & $1.85$ & $-4.97$ & $1.35$ & $0.23$ &  \\
$2$ & $-1$ & $2$ & $-2$ & $-1.02$ & $-6.58$ & $9.27$ & $0.24$ & $0.33$ & $1.50$ \\
$2$ & $-1$ & $2$ & $-1$ & $-29.57$ & $13.87$ & $-14.24$ & $-0.26$ & $-2.03$ & $-0.48$ \\
$2$ & $-1$ & $2$ & $0$ & $-16.72$ & $-10.40$ & $3.39$ &  & $0.43$ &  \\
$2$ & $-1$ & $2$ & $1$ & $-6.86$ & $-2.25$ & $1.87$ & $0.10$ & $0.78$ &  \\
$2$ & $-1$ & $2$ & $2$ & $-7.33$ & $-5.23$ & $5.64$ & $0.20$ & $0.60$ &  \\
$2$ & $0$ & $2$ & $-2$ & $14.99$ & $3.55$ & $0.30$ & $1.25$ & $-0.12$ &  \\
$2$ & $0$ & $2$ & $-1$ & $-16.72$ & $-10.40$ & $-3.39$ &  & $0.43$ &  \\
$2$ & $0$ & $2$ & $0$ & $13.91$ & $0.49$ & $-1.09$ & $0.24$ & $-2.03$ & $-1.21$ \\
$2$ & $0$ & $2$ & $1$ & $18.57$ & $0.63$ & $-0.19$ & $0.42$ & $0.88$ & $-1.11$ \\
$2$ & $0$ & $2$ & $2$ & $18.62$ & $2.39$ & $-2.78$ & $-0.52$ & $0.79$ & $1.21$ \\
$2$ & $1$ & $2$ & $-2$ & $5.58$ & $0.85$ & $0.11$ & $1.15$ & $0.26$ &  \\
$2$ & $1$ & $2$ & $-1$ & $-6.86$ & $-2.25$ & $-1.87$ & $-0.10$ & $0.78$ &  \\
$2$ & $1$ & $2$ & $0$ & $18.57$ & $0.63$ & $-0.19$ & $0.42$ & $0.88$ & $0.51$ \\
$2$ & $1$ & $2$ & $1$ & $-23.63$ & $-4.59$ & $3.92$ & $0.27$ & $0.37$ & $-0.94$ \\
$2$ & $1$ & $2$ & $2$ & $6.87$ & $0.32$ & $0.57$ & $-0.61$ & $2.16$ & $-1.11$ \\
$2$ & $2$ & $2$ & $-2$ & $6.49$ & $1.85$ & $4.97$ & $-1.35$ & $0.23$ &  \\
$2$ & $2$ & $2$ & $-1$ & $-7.33$ & $-5.23$ & $-5.64$ & $-0.20$ & $0.60$ &  \\
$2$ & $2$ & $2$ & $0$ & $18.62$ & $2.39$ & $-2.78$ & $-0.52$ & $0.79$ & $1.02$ \\
$2$ & $2$ & $2$ & $1$ & $6.87$ & $0.32$ & $0.57$ & $-0.61$ & $2.16$ & $-0.14$ \\
$2$ & $2$ & $2$ & $2$ & $-22.09$ & $-3.40$ & $3.10$ & $0.73$ & $-0.43$ & $0.76$ \\
\bottomrule
\end{tabular}
\end{minipage}\hfill
\begin{minipage}[t]{0.49\linewidth}\centering\vspace{0pt}
\begin{tabular}{@{}rrrrrrrrrr@{}}
\toprule
$K$ & $Q$ & $K^{\prime}$ & $Q^{\prime}$ & $R_1$ & $R_2$ & $R_3$ & $R_4$ & $R_5$ & $R_6$ \\
\midrule
\multicolumn{10}{c}{Octupole--dipole} \\
$3$ & $-3$ & $1$ & $-1$ & $6.61$ & $2.11$ & $5.94$ & $-1.79$ & $-0.60$ & $-0.30$ \\
$3$ & $-3$ & $1$ & $0$ & $-2.57$ & $-1.41$ & $-5.17$ & $-0.37$ & $0.13$ &  \\
$3$ & $-3$ & $1$ & $1$ & $-7.76$ & $0.98$ & $-0.14$ & $1.50$ & $-0.45$ &  \\
$3$ & $-2$ & $1$ & $-1$ & $7.68$ & $0.14$ & $6.59$ & $0.81$ & $1.59$ & $-0.60$ \\
$3$ & $-2$ & $1$ & $0$ & $11.68$ & $0.66$ & $-6.42$ & $-1.70$ &  &  \\
$3$ & $-2$ & $1$ & $1$ & $-0.82$ & $-5.56$ & $-10.34$ & $0.12$ & $-0.23$ &  \\
$3$ & $-1$ & $1$ & $-1$ & $-12.78$ & $-1.63$ & $-1.15$ & $1.43$ & $-0.13$ & $-0.26$ \\
$3$ & $-1$ & $1$ & $0$ & $-5.35$ & $1.84$ & $7.40$ & $-0.18$ & $0.50$ &  \\
$3$ & $-1$ & $1$ & $1$ & $-4.54$ & $-1.66$ & $-2.19$ & $-0.67$ & $0.36$ &  \\
$3$ & $0$ & $1$ & $-1$ & $-7.65$ & $-8.99$ & $-5.55$ & $-0.95$ &  &  \\
$3$ & $0$ & $1$ & $0$ & $4.34$ & $-7.73$ & $2.00$ & $1.20$ & $-0.12$ & $1.43$ \\
$3$ & $0$ & $1$ & $1$ & $9.25$ & $1.81$ & $4.92$ & $-0.37$ & $0.59$ & $0.86$ \\
$3$ & $1$ & $1$ & $-1$ & $-4.91$ & $-1.81$ & $2.52$ & $-2.46$ & $0.12$ &  \\
$3$ & $1$ & $1$ & $0$ & $3.33$ & $-2.80$ & $-5.42$ & $0.72$ & $-0.70$ & $-1.20$ \\
$3$ & $1$ & $1$ & $1$ & $-8.41$ & $-6.11$ & $-2.32$ & $1.81$ & $1.05$ & $0.43$ \\
$3$ & $2$ & $1$ & $-1$ & $-8.73$ & $-9.05$ & $-7.41$ & $1.47$ & $0.28$ &  \\
$3$ & $2$ & $1$ & $0$ & $12.88$ & $-0.74$ & $7.47$ & $-0.31$ & $0.60$ & $0.70$ \\
$3$ & $2$ & $1$ & $1$ & $-9.17$ & $-4.68$ & $2.63$ & $0.10$ & $-0.13$ & $-0.26$ \\
$3$ & $3$ & $1$ & $-1$ & $1.74$ & $-3.11$ & $-4.48$ & $-0.57$ & $0.19$ &  \\
$3$ & $3$ & $1$ & $0$ &  & $-1.69$ & $1.63$ & $-1.02$ & $0.51$ & $-0.29$ \\
$3$ & $3$ & $1$ & $1$ & $8.94$ & $2.62$ & $5.65$ & $-0.70$ & $0.29$ & $0.10$ \\
\midrule
\multicolumn{10}{c}{Octupole--octupole} \\
$3$ & $-3$ & $3$ & $-3$ & $25.54$ & $0.88$ & $2.47$ & $-1.17$ & $0.18$ & $0.30$ \\
$3$ & $-3$ & $3$ & $-2$ & $11.41$ & $-0.78$ & $5.64$ & $-1.13$ & $0.50$ & $-0.31$ \\
$3$ & $-3$ & $3$ & $-1$ & $8.82$ & $-5.97$ & $4.46$ & $-0.33$ & $1.56$ & $0.23$ \\
$3$ & $-3$ & $3$ & $0$ & $9.26$ & $4.62$ & $6.51$ & $-1.57$ & $-0.33$ &  \\
$3$ & $-3$ & $3$ & $1$ & $-6.80$ & $0.61$ & $-1.80$ & $0.64$ & $0.58$ &  \\
$3$ & $-3$ & $3$ & $2$ & $11.32$ & $-3.56$ & $3.69$ & $1.40$ & $-0.21$ &  \\
$3$ & $-3$ & $3$ & $3$ & $0.13$ &  & $5.82$ & $-2.63$ &  &  \\
$3$ & $-2$ & $3$ & $-3$ & $11.41$ & $-0.78$ & $5.64$ & $-1.13$ & $0.50$ & $0.83$ \\
$3$ & $-2$ & $3$ & $-2$ & $-12.87$ & $11.45$ & $-11.15$ & $-2.71$ & $0.13$ & $-0.59$ \\
$3$ & $-2$ & $3$ & $-1$ & $-2.17$ & $-0.76$ & $1.02$ & $-0.14$ & $0.79$ & $-0.56$ \\
$3$ & $-2$ & $3$ & $0$ & $-3.02$ & $-4.99$ & $0.18$ & $-1.98$ & $0.07$ &  \\
$3$ & $-2$ & $3$ & $1$ & $9.25$ & $13.75$ & $1.47$ & $-0.25$ & $-0.48$ &  \\
$3$ & $-2$ & $3$ & $2$ & $4.54$ & $1.45$ & $3.99$ & $1.16$ & $0.13$ &  \\
$3$ & $-2$ & $3$ & $3$ & $11.43$ & $-3.57$ & $-6.24$ & $-0.10$ & $-0.19$ &  \\
$3$ & $-1$ & $3$ & $-3$ & $8.82$ & $-5.97$ & $4.46$ & $-0.33$ & $1.56$ & $2.01$ \\
$3$ & $-1$ & $3$ & $-2$ & $-2.17$ & $-0.76$ & $1.01$ & $-0.14$ & $0.79$ & $0.75$ \\
$3$ & $-1$ & $3$ & $-1$ & $-6.36$ & $10.37$ & $-14.23$ & $0.29$ & $-0.42$ & $-0.43$ \\
$3$ & $-1$ & $3$ & $0$ & $-6.58$ & $-8.78$ & $2.24$ & $0.29$ & $0.33$ &  \\
$3$ & $-1$ & $3$ & $1$ & $-8.34$ & $-5.63$ & $-5.15$ & $0.07$ & $0.28$ &  \\
$3$ & $-1$ & $3$ & $2$ & $2.44$ & $10.79$ & $10.21$ & $-0.70$ & $-0.25$ &  \\
$3$ & $-1$ & $3$ & $3$ & $13.84$ & $1.64$ & $-4.69$ & $1.65$ & $-0.34$ &  \\
$3$ & $0$ & $3$ & $-3$ & $9.27$ & $4.62$ & $-6.51$ & $1.57$ & $-0.33$ &  \\
$3$ & $0$ & $3$ & $-2$ & $-3.02$ & $-4.99$ & $-0.18$ & $1.98$ & $0.07$ &  \\
$3$ & $0$ & $3$ & $-1$ & $-6.58$ & $-8.78$ & $-2.24$ & $-0.29$ & $0.33$ &  \\
$3$ & $0$ & $3$ & $0$ & $9.81$ & $10.35$ & $-5.99$ & $1.37$ & $-0.27$ & $-0.58$ \\
$3$ & $0$ & $3$ & $1$ & $5.01$ & $-0.50$ & $0.49$ & $-0.48$ & $-0.07$ & $1.02$ \\
$3$ & $0$ & $3$ & $2$ & $-2.04$ & $-3.49$ & $-6.17$ & $-0.81$ & $-0.08$ & $0.62$ \\
$3$ & $0$ & $3$ & $3$ & $-2.28$ & $3.58$ & $0.11$ & $-0.14$ & $-1.73$ & $-1.64$ \\
$3$ & $1$ & $3$ & $-3$ & $-6.80$ & $0.61$ & $1.80$ & $-0.65$ & $0.58$ &  \\
$3$ & $1$ & $3$ & $-2$ & $9.25$ & $13.75$ & $-1.47$ & $0.25$ & $-0.48$ &  \\
$3$ & $1$ & $3$ & $-1$ & $-8.34$ & $-5.63$ & $5.16$ & $-0.07$ & $0.28$ &  \\
$3$ & $1$ & $3$ & $0$ & $5.01$ & $-0.50$ & $0.49$ & $-0.48$ & $-0.07$ & $-0.66$ \\
$3$ & $1$ & $3$ & $1$ & $7.36$ & $12.42$ & $-0.14$ & $0.86$ & $0.84$ & $-1.24$ \\
$3$ & $1$ & $3$ & $2$ & $3.50$ & $-2.01$ & $-1.20$ & $0.84$ & $0.41$ & $1.39$ \\
$3$ & $1$ & $3$ & $3$ & $7.18$ & $-5.24$ & $-0.44$ & $-2.60$ & $0.38$ & $-0.08$ \\
$3$ & $2$ & $3$ & $-3$ & $11.32$ & $-3.56$ & $-3.69$ & $-1.40$ & $-0.21$ &  \\
$3$ & $2$ & $3$ & $-2$ & $4.53$ & $1.45$ & $-4.00$ & $-1.16$ & $0.13$ &  \\
$3$ & $2$ & $3$ & $-1$ & $2.44$ & $10.79$ & $-10.21$ & $0.70$ & $-0.25$ &  \\
$3$ & $2$ & $3$ & $0$ & $-2.05$ & $-3.49$ & $-6.17$ & $-0.81$ & $-0.08$ & $0.79$ \\
$3$ & $2$ & $3$ & $1$ & $3.50$ & $-2.00$ & $-1.20$ & $0.84$ & $0.41$ & $-0.40$ \\
$3$ & $2$ & $3$ & $2$ & $-12.23$ & $10.24$ & $-1.61$ & $-0.09$ & $1.60$ & $-0.51$ \\
$3$ & $2$ & $3$ & $3$ & $-10.89$ & $1.19$ & $-7.27$ & $0.88$ & $-0.50$ & $-0.17$ \\
$3$ & $3$ & $3$ & $-3$ & $0.12$ &  & $-5.82$ & $2.63$ &  &  \\
$3$ & $3$ & $3$ & $-2$ & $11.43$ & $-3.57$ & $6.24$ & $0.10$ & $-0.19$ &  \\
$3$ & $3$ & $3$ & $-1$ & $13.84$ & $1.65$ & $4.69$ & $-1.65$ & $-0.34$ &  \\
$3$ & $3$ & $3$ & $0$ & $-2.28$ & $3.58$ & $0.11$ & $-0.14$ & $-1.73$ &  \\
$3$ & $3$ & $3$ & $1$ & $7.18$ & $-5.24$ & $-0.44$ & $-2.60$ & $0.38$ & $0.29$ \\
$3$ & $3$ & $3$ & $2$ & $-10.89$ & $1.19$ & $-7.27$ & $0.87$ & $-0.50$ & $0.41$ \\
$3$ & $3$ & $3$ & $3$ & $25.38$ & $0.65$ & $3.20$ & $-0.90$ & $0.15$ & $0.11$ \\
\bottomrule
\end{tabular}
\end{minipage}
\par\medskip
$R_1=(0,-0.350,0.5)$, $R_2=(0,0.650,0.5)$, $R_3=(0,0,1)$;\\
$R_4=(1,0,0)$, $R_5=(1,-0.350,0.5)$, $R_6=(0,1,0)$. Fractional coordinates.
\endgroup
\end{table}

\suppsection{toy}{Illustrative atomic spin-density and orbital-current field model}
The atomic comparison considers a Re$^{6+}$ $5d^1$ ion in idealized $J=3/2$ states. The pure-octupolar reference is
\begin{equation}
 \ket{\psi_{\rm oct}}=
 \frac{\ket{3/2,+3/2}+\ket{3/2,-3/2}}{\sqrt{2}},
 \label{eq:si-octstate}
\end{equation}
and the compensated reference state is $\ket{3/2,+3/2}$ at $\gamma=1$, for which the integrated spin and orbital magnetic dipoles cancel. The field calculation includes spin magnetization density and orbital currents, with a mean $5d$ radius $\langle r\rangle_{5d}=0.65$~\AA. These ideal atomic states differ from the crystal-field-mixed pseudospin states described in
Sec.~\siref{basis}.

\begin{figure}[!htbp]
\centering
\includegraphics[width=0.97\textwidth]{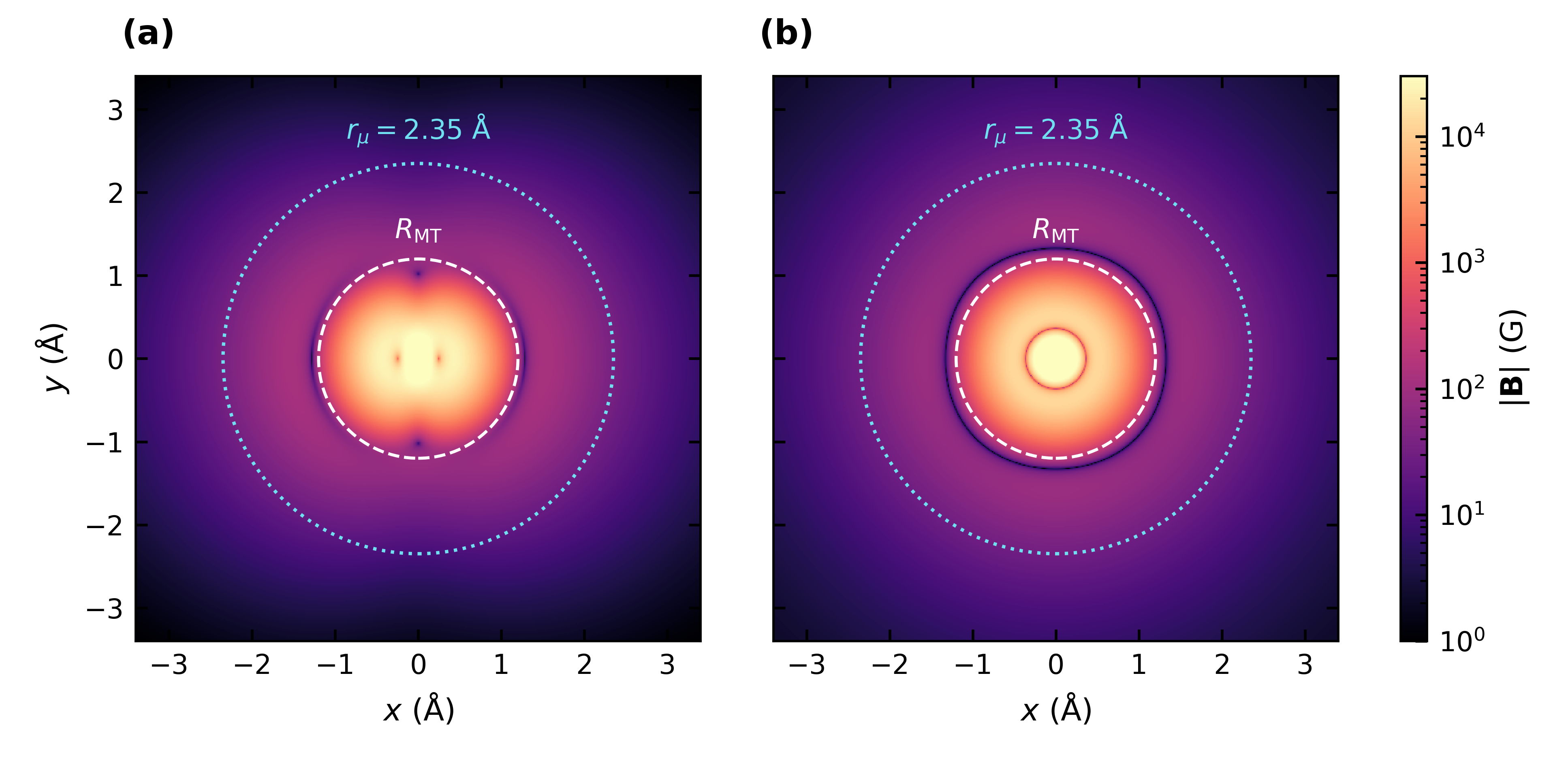}
\caption{Single-ion magnetic-field maps including spin density and orbital currents for (a) the ideal octupolar superposition in Eq.~\eqref{eq:si-octstate} and (b) the compensated $\ket{3/2,+3/2}$ state at $\gamma=1$. Both states have zero integrated magnetic dipole. The outer dotted circles mark the approximate Re--muon distance of $2.35$~\AA\ for the nearest candidate muon site. The inner dashed circles mark the reference radius $R_{\rm MT}$ used in the atomic model.}
\label{fig:si-toymaps}
\end{figure}

For each distance $r$ from Re, we calculate the maximum single-ion field magnitude over directions on a sphere,
\begin{equation}
 B_{\max}(r)=\max_{\hat{\mathbf r}}
 |\mathbf B(r\hat{\mathbf r})|,
 \label{eq:si-bmax}
\end{equation}
where the sphere is centred on Re. The ideal octupolar state and the compensated reference state both produce finite local fields despite their zero integrated magnetic dipole moments. The distance dependence of $B_{\max}(r)$ is shown in Fig.~5 of the main text. The corresponding spatial field maps are shown in
Fig.~\ref{fig:si-toymaps}.

\suppsection{versionaudit}{Structural provenance and complementary calculation inputs}
\subsection*{Parameter-set identity}
The magnetic results in the main paper and Sections~S12--S14 use $U=2.2$~eV and $J_{\mathrm H}=0.5$~eV, with common trace-normalized tensor operators. 

\subsection*{Geometry of the real-space spin-density files}
The charge and spin-density grids were generated for the same relaxed structures and self-consistent magnetic configurations as the DOS and local moments described in Sec.~\siref{hse}. The CHGCAR headers contain two formula units (two Re, two Mg/Zn and eight O atoms). Their lattice metrics, calculated directly from the header vectors, are given in Table~\ref{tab:si-gridmetric}. They match the PBE-relaxed cells in Sec.~\siref{hse}, not the experimental $7$~K cells. Four volumetric blocks with grids $72\times96\times80$ and $80\times96\times84$ are present for Mg and Zn, respectively, consistent with total charge plus three noncollinear spin-magnetization components in the VASP format \cite{VASPCHGCAR}. An orbital-current density is not supplied in these files.
\begin{table}[htbp]
\caption{Cell metrics obtained directly from the CHGCAR header vectors, compared with the refined experimental volumes. Header values are arithmetic checks, not new electronic-structure calculations.}\label{tab:si-gridmetric}
\centering\begin{tabular}{lcc}\toprule
Quantity&Mg grid&Zn grid\\\midrule
$a$ (\AA)&4.690285&4.723699\\
$b$ (\AA)&5.664579&5.696500\\
$c$ (\AA)&5.083640&5.088647\\
$\beta$ (degrees)&91.968789&92.314040\\
$V$ (\AA$^3$)&134.984911&136.816466\\
Experimental $7$~K $V$ (\AA$^3$)&130.655&131.921\\\bottomrule
\end{tabular}
\end{table}

The grids belong to the relaxed two-Re, $\mathbf{k}=0$ calculations with the default VASP spinor frame and HSE06+SOC settings specified in Sec.~\siref{hse}. They cannot be inserted directly into a muon-field calculation for the experimental structure without an explicit geometry mapping or a matched recalculation. In addition, their crystallographic-cell periodicity does not represent the $\mathbf{k}=(1/2,1/2,0)$ Mg state predicted by the effective Hamiltonian. The absence of a spatial orbital-current contribution is a separate limitation. These grids are therefore not used as a prediction of the full local field.
The atomic orbital moments are integrated quantities defined with respect to atomic centers. A bulk orbital magnetization obtained by the Berry-phase formulation includes contributions beyond an atomic partition, but is still not a spatially resolved orbital-current field at a muon. Adding a bulk orbital number to a local spin-density field is not a substitute for a consistent microscopic field calculation.

\Needspace{8\baselineskip}
\subsection*{PBE comparison of relaxed and experimental structures}
\begin{figure}[!htbp]
\centering
\includegraphics[width=0.78\textwidth]{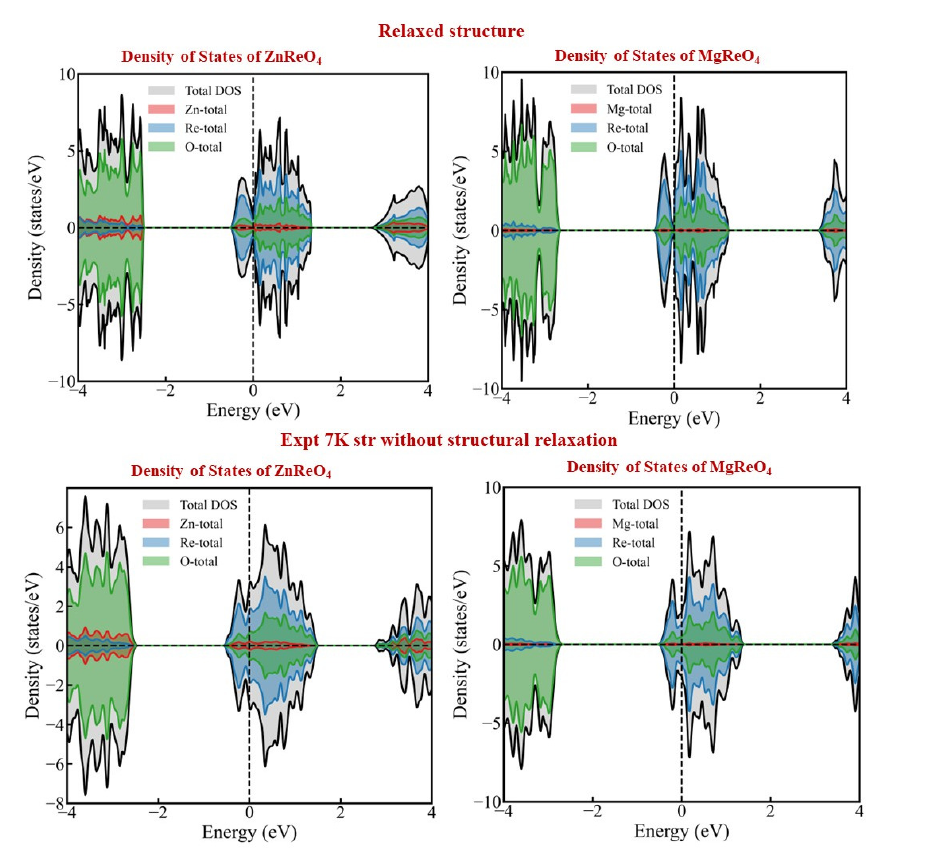}
\caption{PBE densities of states for relaxed structures and unrelaxed experimental $7$~K structures. The upper row shows the relaxed calculations and the lower row the unrelaxed experimental structures; ZnReO$_4$ is shown on the left and MgReO$_4$ on the right. Total and atom-projected DOS contributions are identified in the legends. The dashed vertical lines indicate the plotted Fermi-level reference, and mirrored curves represent the two spin channels of the calculations without SOC. The original panel scales and curves are retained. This is a PBE geometry comparison, not a test of the geometry dependence of the HSE06+SOC gaps.}
\label{fig:si-pbe-relaxation}
\end{figure}

To examine the effect of relaxation, additional self-consistent PBE calculations were performed for the unrelaxed experimental $7$~K structures and compared with the relaxed-structure results (Fig.~\ref{fig:si-pbe-relaxation}). Relaxation changes the positions and intensities of individual DOS features, while the overall electronic structure around the Fermi level remains qualitatively similar for each compound. PBE without SOC is metallic in the comparison. HSE and HSE06+SOC were subsequently applied only to the relaxed structures; no quantitative relaxed--unrelaxed comparison of their gaps or moments was performed.

\FloatBarrier
\begingroup\setstretch{1.0}\small
\bibliographystyle{naturemag}
\bibliography{refs}
\endgroup